\documentclass[10pt,letterpaper]{article}
\usepackage[top=0.85in,left=2.75in,footskip=0.75in]{geometry}

\usepackage{amsfonts,amsmath,amsthm,amsxtra,amssymb}
\usepackage{bm}
\usepackage{textcomp,marvosym}
\usepackage{nameref}
\usepackage[right]{lineno}
\usepackage{subcaption}
\usepackage{changepage} 

\usepackage{cite}

\usepackage{array}
\newcolumntype{+}{!{\vrule width 2pt}}
\newlength\savedwidth

\raggedright
\usepackage[aboveskip=1pt,labelfont=bf,labelsep=period,justification=raggedright,singlelinecheck=off]{caption}

\makeatletter
\renewcommand{\@biblabel}[1]{\quad#1.}
\makeatother

\usepackage{lastpage,fancyhdr,graphicx}
\usepackage{epstopdf}
\fancyheadoffset[L]{2.25in}
\fancyfootoffset[L]{2.25in}
\newlength{\offsetpage}
\newenvironment{widepage}{\begin{adjustwidth}{-5\offsetpage}{}\hsize=\linewidth%
    \addtolength{\textwidth}{2\offsetpage}}%
{\end{adjustwidth}}

\usepackage{wrapfig,graphics}
\usepackage{tabularx, multirow, fancybox, lscape,listings,float,rotating,url}
\usepackage{siunitx} 
\usepackage{color,hyperref}
\hypersetup{colorlinks,breaklinks, linkcolor=blue,urlcolor=blue,anchorcolor=lue,citecolor=blue}
\usepackage{enumerate}
\usepackage[normalem]{ulem}
\usepackage{booktabs}
\usepackage[table,xcdraw,dvipsnames]{xcolor}

\usepackage{color,hyperref}
\hypersetup{colorlinks,breaklinks,
linkcolor=blue,urlcolor=blue,
anchorcolor=lue,citecolor=blue}

\usepackage[nameinlink,capitalise]{cleveref}
\crefformat{figure}{#2Fig~#1#3}
\Crefformat{figure}{#2Fig~#1#3}

\usepackage{tikz}
\newcommand{\insertfigdefault}[3]{
\begin{tikzpicture}
\node[anchor=north west,inner sep=0pt] at (0,0){\includegraphics[width= #1 \linewidth]{#2}};
\node[font=\sffamily\bfseries\large] at (1ex,-1ex) {#3};
\end{tikzpicture}}

\newcommand{\insertfigclose}[3]{
\begin{tikzpicture}
\node[anchor=north west,inner sep=0pt] at (0,0){\includegraphics[width= #1 \linewidth]{#2}};
\node[font=\sffamily\bfseries\large] at (2ex,-1ex) {#3};
\end{tikzpicture}}

\newcommand{\insertfigcloser}[3]{
\begin{tikzpicture}
\node[anchor=north west,inner sep=0pt] at (0,0){\includegraphics[width= #1 \linewidth,trim={0.75cm 0cm 2cm 0cm}, clip]{#2}};
\node[font=\sffamily\bfseries\large] at (4ex,-1ex) {#3};
\end{tikzpicture}}

\def \W {\textit{Wolbachia~}}
\def \Wns {\textit{Wolbachia}}

\begin{document}

\vspace*{0.2in}

\begin{flushleft}
{\Large
\textbf\newline{Multistage spatial model for informing release of Wolbachia-infected mosquitoes as disease control} 
}
\newline
\\
Zhuolin Qu\textsuperscript{1*},
Tong Wu\textsuperscript{1}
\\
\bigskip
\textbf{1} Department of Mathematics, University of Texas at San Antonio, San Antonio, TX, United States of America
\\
\bigskip

* Corresponding author\\
Email: zhuolin.qu@utsa.edu (ZQ)

\end{flushleft}


\section*{Abstract}

\textit{Wolbachia} is a naturally occurring bacterium that can infect \textit{Aedes} mosquitoes and reduce the transmission of mosquito-borne diseases, including dengue fever, Zika, and chikungunya. Field trials have been conducted worldwide to suppress local epidemics. We introduce a novel partial differential equation model to simulate the spread of \textit{Wolbachia} infection in mosquito populations. Our model incorporates the intricate \textit{Wolbachia} maternal transmission cycle and detailed mosquito life stages, while also accounting for the spatial heterogeneity induced by mosquito dispersion across a two-dimensional domain. Prior modeling studies and field data indicate that a critical threshold of \textit{Wolbachia}-infected mosquitoes is necessary for infection to persist among the mosquito population. Through our spatial model, we identify a threshold condition, termed the ``critical bubble'', for having a self-sustainable \textit{Wolbachia} infection in the field. When releasing beyond this threshold, the model predicts a spatial wave of \textit{Wolbachia} infection. We further quantify how this threshold and infection wave velocity depend on the diffusion process and other parameters. We numerically study various intervention scenarios to inform efficient \textit{Wolbachia} release strategies. Our findings suggest that: (1) integrating \textit{Wolbachia} release with pre-release mitigations targeting the adult mosquitoes, rather than the aquatic stages, better reduces the threshold for \textit{Wolbachia} establishment; Habitats modification before the release may increase the threshold; (2) releases in the dry regions lower the threshold, though the infection waves may slow down or stall at the dry-wet interfaces due to the difference in carrying capacities; and (3) initiating releases just before the wet season further reduces the release threshold. 

\section*{Author summary}
In this study, we developed a new mathematical model to better understand how \textit{Wolbachia}, a naturally occurring bacterium, spreads in \textit{Aedes} mosquito populations. When mosquitoes carry \textit{Wolbachia}, they are much less likely to transmit harmful diseases like dengue fever, Zika, and chikungunya to humans. Our model looks at how \textit{Wolbachia} infection propagates through mosquito populations over time, taking into account both mosquito life cycles and their movement across different environments. Field studies show that for \textit{Wolbachia} to effectively control disease, a critical proportion of the mosquito population needs to be infected. We use our model to explore how this critical threshold can be achieved efficiently in real-world conditions. Our findings suggest that the timing and location of releases can significantly impact success. Releasing \textit{Wolbachia}-infected mosquitoes in dry areas or just before the wet season can lower the required release threshold, improving the chances of establishing self-sustaining \textit{Wolbachia} populations. Additionally, targeting adult mosquitoes rather than the aquatic stages before release can further lower this threshold. These insights could enhance \textit{Wolbachia} release programs, helping to refine strategies for more efficient and sustainable mosquito-borne disease prevention across diverse settings.

\section{Introduction}
Mosquitoes are among the deadliest creatures in the world due to the mosquito-borne diseases they can transmit. Traditional mosquito control methods, such as larvicides and indoor residual spraying, focus on reducing mosquito populations. However, these methods have limitations and may not offer reliable, long-term solutions for mosquito-borne diseases \cite{acheeAlternativeStrategiesMosquitoborne2019}. 
Novel biological controls are being developed to provide alternative strategies for managing mosquito populations and the spread of the diseases. These include \Wns-based methods, which can be used for either population suppression \cite{laven1967eradication,mcmeniman2009stable} or replacement \cite{walker2011wmel}, as well as the sterile insect technique \cite{alphey2010sterile,patterson1970suppression,lofgren1974release} and gene editing \cite{alphey2014genetica,kyrou2018crispra}. Among these approaches, \Wns-based controls have shown promise in field trials for \textit{Aedes} mosquitoes combat dengue epidemics worldwide \cite{hoffmann2011successful,hoffmann2014stability,ryan2020establishment,indriani2020reduced,gesto2021largescale,utarini2021efficacy}. Extensive laboratory work has also been conducted to introduce \W into \textit{Anopheles} mosquitoes as a potential malaria control strategy \cite{bian2013wolbachia,joshi2014wolbachia,gomes2018infection}.

The \Wns-based population replacement strategy aims to replace the wild-type mosquitoes with \Wns-infected ones by leveraging their reproductive advantages, achieved through a mechanism called cytoplasmic incompatibility (CI) \cite{laven1967eradication}. Moreover, unlike wild-type mosquitoes, \Wns-infected mosquitoes are less capable of transmitting diseases \cite{walker2011wmel}. This population replacement approach hinges on a threshold condition for \W persistence, which has been observed in both population cage experiments \cite{axford2016fitness} and field trials \cite{jiggins2017spread}. If the proportion of infected mosquitoes exceeds this threshold, \textit{Wolbachia} can persist and spread through the population; if not, the infection will eventually die out. Identifying this threshold is therefore critical for designing a successful field release program that achieves sustainable \W infection in target mosquito populations.

Mathematical modeling has become an essential tool in designing prevention and control measures for infectious diseases, including \Wns-based intervention. Modeling can identify the key \W features and entomological parameters that influence the success of mosquito release programs. It can quantify the critical threshold condition necessary for \W establishment, characterize the spatial spread of \W infection in the field, and offer insights into optimizing release scenarios that otherwise would be expensive and challenging to test in the field.

Previous modeling studies have identified the threshold infection frequency needed for \W establishment (e.g., \cite{koiller2014aedes,qu2018modeling,qu2022modeling,xue2016two,ndii2018mathematical,turelli2017deploying,ferguson2015modelinga,childs2020role}). This threshold depends on the trade-off between the fitness costs of \W infection (reduced lifespan and fecundity in female mosquitoes) and the reproduction advantage conferred by the CI effect of the \Wns-infection. Mathematical characterizations of this threshold typically use backward bifurcation results from ordinary differential equation (ODE) models \cite{childs2020role,xue2016two,qu2018modeling,koiller2014aedes,zhang2015models}. However, the threshold predicted by ODE models may not accurately capture the dynamics of \W invasion in field releases, as they overlook the spatial heterogeneity that naturally arises during the field releases, such as the compactness of initial release, spatial dispersion of the infected and uninfected mosquitoes, and other complexity comes from the surrounding environment. 

To address these limitations, a few spatial models have been developed at varying resolutions. Barton and Turelli \cite{bartonSpatialWavesAdvance2011} introduced a reaction-diffusion partial differential equation (PDE) model, incorporating the \Wns-induced CI effect and fitness costs. Their model used an idealized cubic approximation for maternal transmission of \W and tracked the fraction of infection among the mosquitoes using a single PDE. The model was later extended to include biologically realistic non-Gaussian dispersal kernels \cite{turelli2017deploying}, which accounts for a higher probability of long-distance mosquito flights. Both models identified a minimal \Wns-infection area as a threshold for \W establishment. Similar spatial models have been analyzed using different approximations to account for imperfect maternal transmission \cite{schofield2002spatially}. In Huang et al. \cite{huang2015wolbachia}, a reaction-diffusion two-PDE model was used to track infected and uninfected mosquito cohorts separately, and their analyses indicate spatial diffusion slightly increases the threshold. Duprez et al. \cite{duprez2021optimization} formulated a two-PDE model, which was then reduced to a one-PDE model for an optimal control problem to determine the best spatial strategy for \W release. 

The challenge of analyzing large PDE systems has led to proposals of simplified PDE spatial models to provide qualitative understandings of the \W dynamics, though these simplified models come with limitations. For instance, models that do not explicitly track the aquatic mosquito stages can not represent pre-release interventions, such as larviciding, that target these stages. Similarly, models that only track female mosquitoes often assume a 1:1 sex ratio, which may be inaccurate due to the difference in male and female lifespans. 

We developed a multistage spatial model to investigate the spatial spread of \Wns-infection in the field mosquitoes, which accounts for the detailed \W maternal transmission dynamics across mosquito life stages as well as spatial dispersion resulting from mosquito flights. This work is based on our previous ODE models \cite{florez2022modeling,qu2018modeling}, where we focused on analyzing threshold conditions and conducted a preliminary study on various \W release scenarios. In the present work, we extend our approach by constructing a reaction-diffusion spatial model to systematically identify key parameters influencing the \W establishment threshold and spatial propagation through a global sensitivity analysis. Additionally, we explore different field release scenarios, including the integration of \W with traditional mitigation approaches as pre-release mitigation, the selection of release sites in regions with disparities in carrying capacities, and the timing of \W releases under seasonal climate variations. Our numerical simulations provide insights into optimal strategies for efficient \W release programs. 

\section{Methods} \label{sec:methods}
Our spatial model accounts for both the complex \W transmission dynamics throughout the mosquito life cycle and the spatial heterogeneity. Key enhancements to our previous ODE model \cite{florez2022modeling} include the incorporation of spatial dispersion dynamics for adult-stage mosquitoes (\cref{sec:model}). The addition introduces nontrivial spatial effects, resulting in waves of infection that significantly complicate the representation of the threshold condition for \W invasion (\cref{sec:threshold}). We also made substantial advancements in model parameterization (\cref{sec:param}), incorporating detailed estimates from laboratory and field studies on wMel strain of \textit{Wolbachia} and \textit{Aedes aegypti} mosquitoes. Furthermore, we accounted for temporal heterogeneity due to seasonal variations in both temperature and rainfall, enhancing the model's relevance to real-world conditions.

\subsection{Model formulation}\label{sec:model}
\begin{figure}[ht!]
\centering
\includegraphics[width=0.9\textwidth]{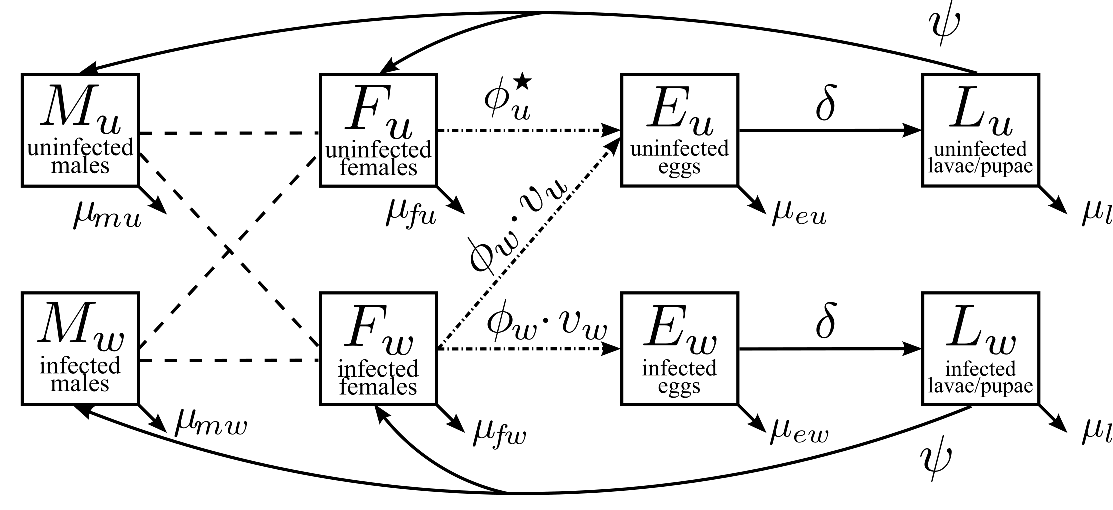}
\caption{Flow diagram for the \Wns-maternal transmission routes among mosquitoes, including the mating process between males and females (dashed arrows), females reproduction of eggs (dash-dotted arrows), and mosquito transitions through different life stages and deaths (solid arrows). Maternal transmission depends on the infection status of the parental male and female mosquitoes. Uninfected females ($F_u$) produce only uninfected eggs at a rate of $\phi_u^\bigstar = \phi_u \frac{M_u}{M_u+M_w}$, when mating with uninfected males ($M_u$). $\phi_u$ represents the per-capita egg-laying rate for the uninfected females. No offspring are reproduced when $F_u$ mate with $M_w$ due to cytoplasmic incompatibility. Infected females ($F_w$) lay eggs at a per-capita rate of $\phi_w$; among these, a fraction $v_w$ are infected eggs ($E_w$) due to the \W maternal transmission, while the remaining $1-v_w$ fraction are uninfected eggs ($E_u$).  \label{fig:diagram}}
\end{figure}

The mosquitoes population is tracked by different life stages (\cref{fig:diagram}), including eggs ($E_u$ and $E_w$), larvae/pupae combined ($L_u$ and $L_w$), males ($M_u$ and $M_w$), and females ($F_u$ and $F_w$). 
The \Wns-infected groups and wild-type groups are denoted with subscript $w$ and $u$, respectively.

The \W transmission from female mosquitoes to their offspring depends on the infection status of the male and female mosquitoes at mating \cite{walker2011wmel}. Infected females can pass the infection to their offspring at a per-capita rate of $v_w\phi_w$, where $\phi_w$ is the egg-laying rate for infected females, and $v_w$ is the maternal transmission rate, indicating the fraction of infected offspring. Uninfected females, when mating with uninfected males, produce uninfected offspring at a rate of $\phi_u \frac{M_u}{M_u+M_w}$, where $\phi_u$ is the egg-laying rate for uninfected females. The term $\frac{M_u}{M_u+M_w}$ models the probability of encountering an uninfected male, assuming equal mating competence between the infected and uninfected males \cite{segoli2014effect} and homogeneous mixing among males. When mating with infected males, the uninfected females produce nonviable offspring due to the \Wns-induced CI  \cite{walker2011wmel,laven1967eradication}.

Eggs hatch into larvae at a rate of $\delta$ and experience die at rates of $\mu_{eu}$ and $\mu_{ew}$ for uninfected and infected cohorts, respectively. The total larvae/pupae population, $L_u+L_w$, is constrained by a carrying capacity, $K_l$, to account for resource competition at these stages. Larvae then develop into adult mosquitoes at a rate of $\psi$ and die at a rate of $\mu_l$. Upon emergence, the probabilities of becoming male and female mosquitoes are given by $b_m$ and $b_f$, respectively. We assume equal progression rates $\delta$ and $\psi$ for infected and uninfected cohorts, as \W infection does not significantly affect these traits in \textit{Aedes aegypti} mosquitoes \cite{walker2011wmel}. The per-capita mortality rates for male mosquitoes are $\mu_{mu}$ and $\mu_{mw}$ for uninfected and infected individuals, respectively, and for females, they are $\mu_{fu}$ and $\mu_{fw}$.

Adult mosquitoes engage in local flights, characterized by random and unidirectional movements, which can be approximated by a diffusion
process \cite{takahashi2005mathematical}. To capture this behavior -- particularly relevant for field releases and the establishment of \W infection -- we incorporate diffusion terms for both male and female mosquitoes, with diffusion coefficients $D_i (i = 1,\cdots,4)$, representing the mean squared displacement of the mosquito flights per time unit. In contrast, juvenile-stage mosquitoes remain in
standing water until they develop into flying adults; thus, we assume no diffusion terms for these early life stages. 

Based on the assumptions above, we describe the \textit{Wolbachia} transmission dynamics among mosquitoes using the following PDE system:
\begin{equation}
\begin{aligned}\label{eq:PDE8}
(E_u)_t&= \phi_u\frac{M_u}{M_u+M_w} F_u + (1-v_w)\phi_wF_w - \delta E_u - \mu_{eu} E_u, \\
(E_w)_t&= v_w \phi_wF_w - \delta E_w - \mu_{ew} E_w,\\
(L_u)_t&= \delta E_u \left(1-\frac{L_u+L_w}{K_l}\right) - \psi L_u - \mu_l L_u,\\
(L_w)_t&= \delta E_w\left(1-\frac{L_u+L_w}{K_l}\right) - \psi L_w - \mu_l L_w,\\
(F_u)_t&= b_f \psi L_u -  \mu_{fu} F_u+D_1\Delta F_u,\\
(F_w)_t&=b_f\psi L_w - \mu_{fw}F_w+D_2\Delta F_w,\\
(M_u)_t&= b_m\psi L_u - \mu_{mu}M_u+D_3\Delta M_u,\\
(M_w)_t&= b_m \psi L_w - \mu_{mw} M_w+D_4\Delta M_w.
\end{aligned}
\end{equation}
The state variables are labeled as shown in \cref{fig:diagram}, and the descriptions of the parameters, along with their baseline values, are provided in \cref{tab:parameter_all}.

\begin{table}[H]
\centering
\caption{Model parameters and their baseline values for wMel strain of \textit{Wolbachia} in \textit{Aedes aegypti} mosquitoes. All rates are expressed in units of $day^{-1}$. We assume a perfect maternal transmission rate and complete cytoplasmic incompatibility \cite{walker2011wmel,hoffmann2014stability}. $^\dag$indicates \textit{per capita}. Note that different parameter ranges are used for sensitivity analysis, see \cref{tab:parameter_SA} in \cref{sec:appendixC}.}
\label{tab:parameter_all}
\begin{tabular}{@{}lllll@{}}
\toprule
\multicolumn{2}{l}{Parameters} & Baseline & Range & Reference\\ \midrule
$b_f$ & Female birth probability & 0.5& &\cite{walker2011wmel}\\
$b_m$ & Male birth probability $(=1-b_f)$& 0.5& &\cite{walker2011wmel}\\
$\phi_u$ & Egg-laying rate for uninfected females$^\dag$  & 3.7 & $1\sim 8$& \cite{hoffmann2014stability,styer2007mortality,yang2011follow}\\
$\phi_w$ &Egg-laying rate for infected females$^\dag$  &  3.5 & $0.9\sim 7.5$ &\cite{hoffmann2014stability,styer2007mortality} \\
$\delta$ & Hatching rate for eggs$^\dag$ & 1/2  & 1/4 $\sim$ 1 &\cite{soares-pinheiro2016eggs,foster2002mosquitoes} \\
$\psi$ & Development rate for larvae/pupae$^\dag$ & 1/10 & 1/30$\sim$1/8 & \cite{foster2002mosquitoes,walker2011wmel} \\
$\mu_{eu}$& Mortality rate for uninfected eggs$^\dag$ & 0.088& 0.08 $\sim$ 0.17 & \cite{hoffmann2014stability,soares-pinheiro2016eggs}\\
$\mu_{ew}$& Mortality rate for infected eggs$^\dag$ & 0.185& 0.17 $\sim$ 0.33 &\cite{hoffmann2014stability,soares-pinheiro2016eggs} \\
$\mu_{l}$& Mortality rate for uninfected larvae/pupae$^\dag$ & 0.12 & 0.06 $\sim$ 0.2& \cite{yang2011follow}\\
$\mu_{fu}$ &Mortality rate for uninfected females$^\dag$ & 1/17.5 & 1/21 $\sim$ 1/14 & \cite{qu2018modeling,zettel2009yellowa,nationalenvironmentagency2020aedes} \\
$\mu_{fw}$ &Mortality rate for infected females$^\dag$ & 1/15.8 & 1/19 $\sim$ 1/12.6 & \cite{walker2011wmel}\\
$\mu_{mu}$ &Mortality rate for uninfected males$^\dag$ &1/10.5 & 1/14 $\sim$ 1/7 &\cite{styer2007mortality} \\
$\mu_{mw}$ &Mortality rate for infected males$^\dag$ &1/10.5 & 1/14 $\sim$ 1/7& \cite{styer2007mortality}\\
$v_w$ & Maternal transmission rate & 1 & 0.89 $\sim$ 1 & \cite{walker2011wmel,hoffmann2014stability} \\
$K_l$ & Carrying capacity of larvae stage (/m$^2$)& $1$ & & Assumed\\ 
$D_i$ & Diffusion coefficients (m$^2$/day)& 200 & $100 \sim 300$& \cite{kay1998aedes,russell2005markreleaserecapture,turelli2017deploying}\\ \bottomrule
\end{tabular}
\end{table}

\subsection{Threshold condition for \W invasion}\label{sec:threshold}
\subsubsection{Existence of threshold}
For \Wns-infected mosquitoes to successfully invade and replace the wild-type mosquitoes in the field, a threshold condition on the initial release is required. Previous ODE models have shown that this condition leads to bistable dynamics. Specifically, when 
\begin{equation}\label{eq:R0}
4v_w(1-v_w)\le\mathcal{R}_0, \quad \mathcal{R}_0 = v_w \frac{\mu_{fu}\phi_w(\delta+\mu_{eu})}{\mu_{fw}\phi_u (\delta +\mu_{ew})},
\end{equation}
where $\mathcal{R}_0$ represents the basic reproduction number for \W transmission among mosquitoes, the system presents two distinct outcomes: if the infection level surpasses a critical threshold, \W infection can spread and stabilize at a high prevalence; otherwise, the infection will die out \cite{qu2018modeling}. Under baseline parameter values, this threshold \Wns-prevalence is around 25\%. 

\begin{figure}[ht!]
\centering
\begin{tikzpicture}
\node[anchor=north west,inner sep=0pt] at (0,0){\includegraphics[width= 0.305\textwidth]{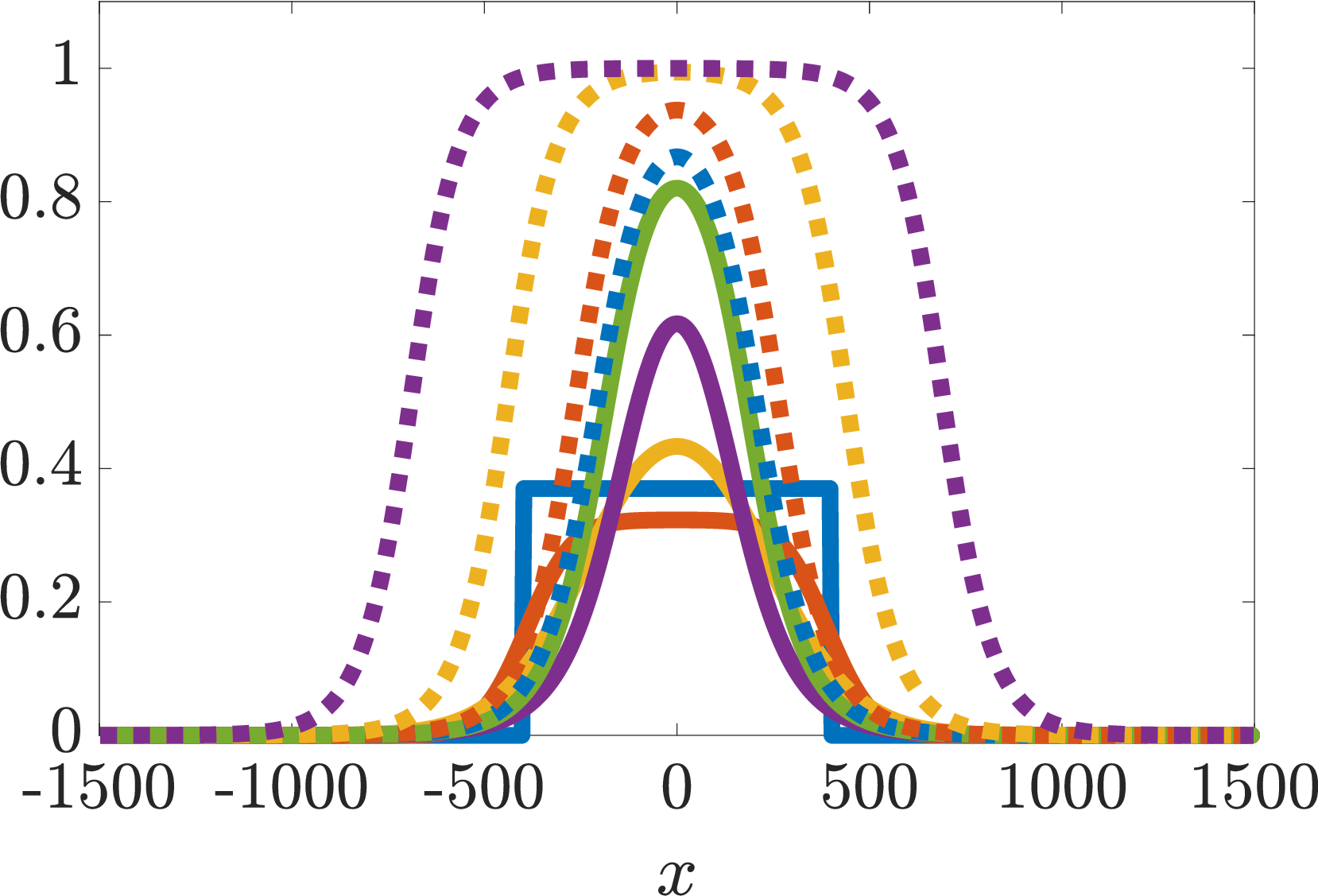}};
\node[font=\sffamily\bfseries\large] at (0,1ex) {A};
\end{tikzpicture}
\begin{tikzpicture}
\node[anchor=north west,inner sep=0pt] at (0,0){\includegraphics[width= 0.305\textwidth]{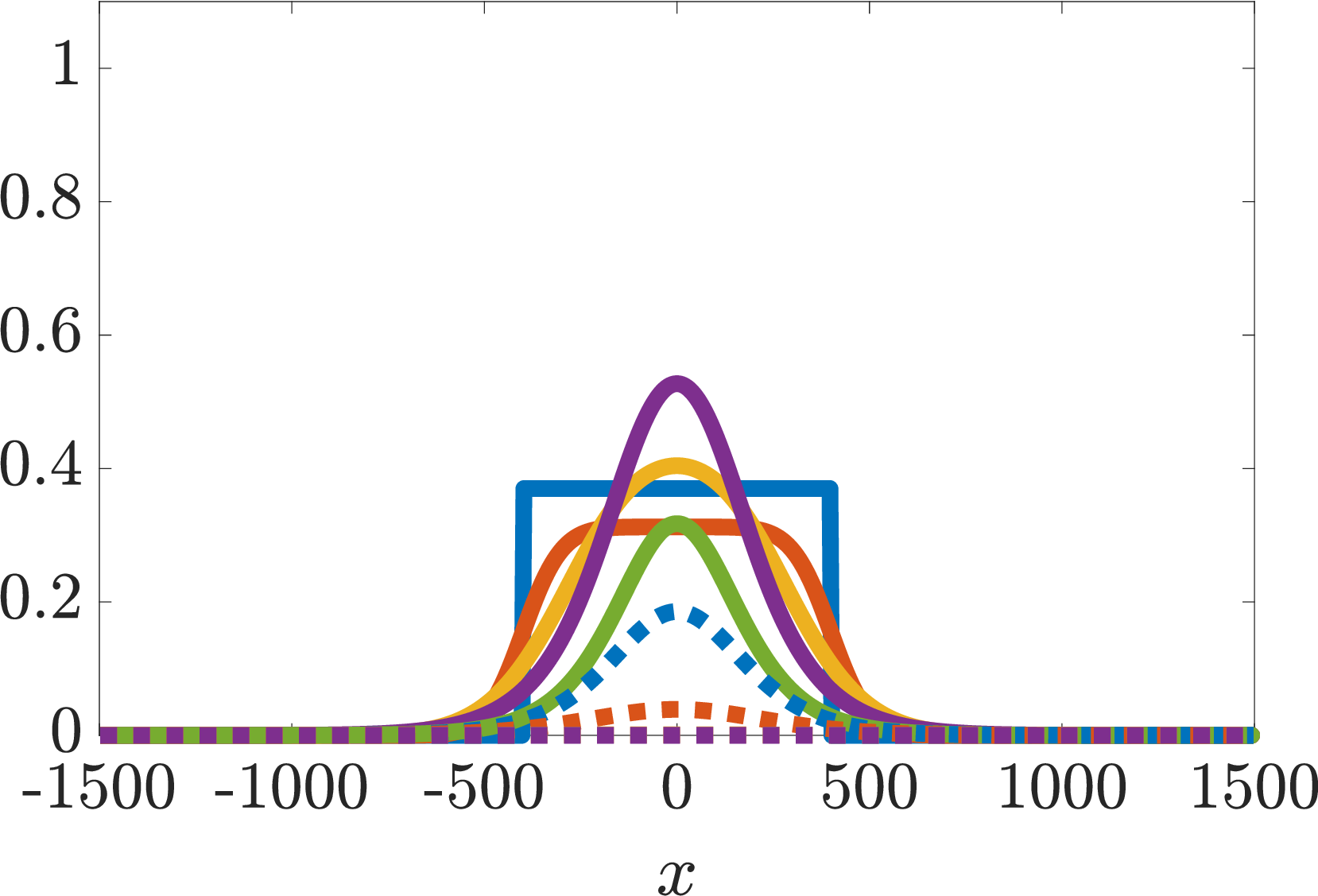}};
\node[font=\sffamily\bfseries\large] at (0,1ex) {B};
\end{tikzpicture}
\begin{tikzpicture}
\node[anchor=north west,inner sep=0pt] at (0,0){\includegraphics[width= 0.305\textwidth]{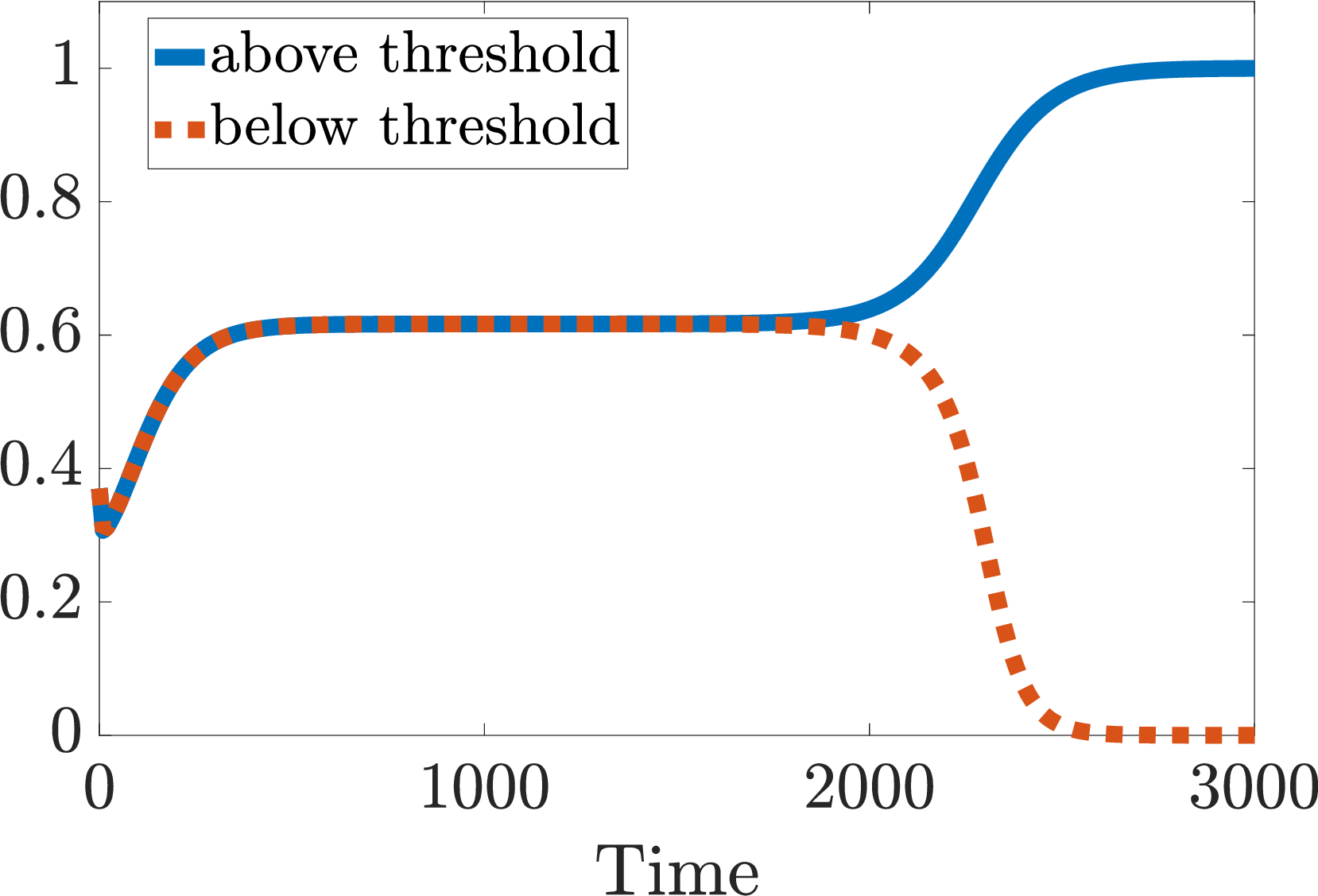}};
\node[font=\sffamily\bfseries\large] at (0,1ex) {C};
\end{tikzpicture}
\caption{Capturing the threshold for \W establishment. (A) The infection is established when a step-shaped initial release is above the threshold level. (B) The infection collapses when the release level is below the threshold level. (C) The time series of the infection fraction at the release center, $(x,y)=(0,0)$, for simulations in panels A and B, demonstrating the bifurcating dynamics of the system. \label{fig:step_releases}}
\end{figure}
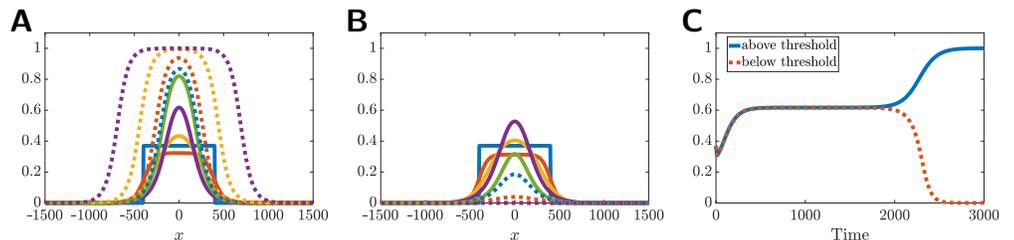

Our proposed PDE model retains these bistable dynamics but introduces new complexities due to spatial effects. Mosquito dispersal impacts the establishment of \W infection, the threshold level now depends on the distance from the release center, leading to a spatially heterogeneous threshold profile, referred to as a ``critical bubble'' \cite{qu2022modeling}. As illustrated in \cref{fig:step_releases}, if the infection is above the threshold (to be detailed later), it establishes locally near the release center and propagates into nearby areas (panel A); when the infection is below the threshold, the infection collapses (panel B).

\subsubsection{Capturing the threshold} \label{sec:threshold2}
To characterize the threshold ``critical bubble'', we adapt the numerical algorithm from \cite{qu2022modeling} and leverage the bistable behavior of the system. We design the following root-finding problem:
\begin{align}\label{eq:Jc}
\mathcal{J}(c) = 1-2\times \underbrace{\Big\{p_0(T;c)<p_0(T-\Delta t;c)\Big\}}_{\text{Condition I}}
 - 2\times \underbrace{\Big\{|p_0(T;c)|<10^{-4}\Big\}}_{\text{Condition II}}\\
 +2\times \underbrace{\Big\{|p_0(T;c)-p_{EE}|<10^{-4}\Big\}}_{\text{Condition III}}=0,
\end{align}
where $p_0$ is the fraction of infected females, $p_0=F_w/(F_u+F_w)$, at the release center $(x,y) = (0,0)$, $T$ is the final simulation time, $\Delta t$ is the time step for the numerical discretization, and $p_{EE}$ is the fraction of infected females at the endemic steady state, as derived in \cite[Eq. (9)]{florez2022modeling}. The algorithm iterates on the variable $c$, the initial release level, to find the threshold release level $c^*$, for which $\mathcal{J}(c^*)=0$. Each condition in \cref{eq:Jc} is evaluated to 1 if true or 0 false. For $c > c^*$, $p_0(t)$ approaches $p_{EE}$ from below (blue solid curve in \cref{fig:step_releases}C), resulting in $\mathcal{J}(c)>0$ (Condition I = 0, II = 0, and III = 0 or 1 depending on $T$); For $c < c^*$, $p_0(t)$ approaches zero from above (red dotted curve in \cref{fig:step_releases}C), leading to $\mathcal{J}(c)<0$ (Condition I = 1, II = 0 or 1 depending $T$, and III = 0). Although $\mathcal{J}(c)=0$ has no exact root, iterating on this root-finding problem provides a robust approximation of the threshold release level. The critical bubble can then be determined by identifying the stagnant phase of the transition before the bifurcation (plateau in the time series of \cref{fig:step_releases}C around $t=1000$).

Unless stated otherwise, our simulations use initial step-shaped releases of infection (male: female = 1:1), where $F_w(x,y,0) = M_w(x,y,0) = c$ within a circular domain ($x^2+y^2 \le r^2$, $r = 200$ by default). The threshold identification algorithm can be applied to different distributions of initial releases, and the resulting critical bubble remains comparable. We have included more details in \cref{sec:appendixB}. 

\subsection{Parameterization and numerical method}\label{sec:param}
All codes used to generate the results and figures will be made available on GitHub upon acceptance. Simulations were developed in MATLAB 2023a/2023b. Spatial derivatives were discretized using a fourth-order central difference scheme with free boundary conditions, and the time evolution was integrated using MATLAB \texttt{ODE45} solver. Initial conditions for all the simulations start at \Wns-free equilibrium, which is subsequently modified by any pre-release mitigation and then by the corresponding \Wns-release strategies. We assume an equal number of male and female mosquitoes are released within a circular domain. The parameters are summarized in \cref{tab:parameter_all} for the wMel strain of \W and \textit{Aedes aegypti} mosquitoes. For further details on the derivations of model parameters, see \cref{sec:param_detail}.

\section{Results}
We perform numerical simulations of the proposed model to gain insights into \Wns-based intervention strategies, with a focus on the scenarios relevant to the implementation of release programs. These results are intended to provide qualitative insights and establish guiding principles for designing efficient and effective release programs.

\subsection{Critical bubbles and sensitivity analysis}\label{sec:SA}
\begin{figure}[htbp]
\centering
\insertfigclose{0.43}{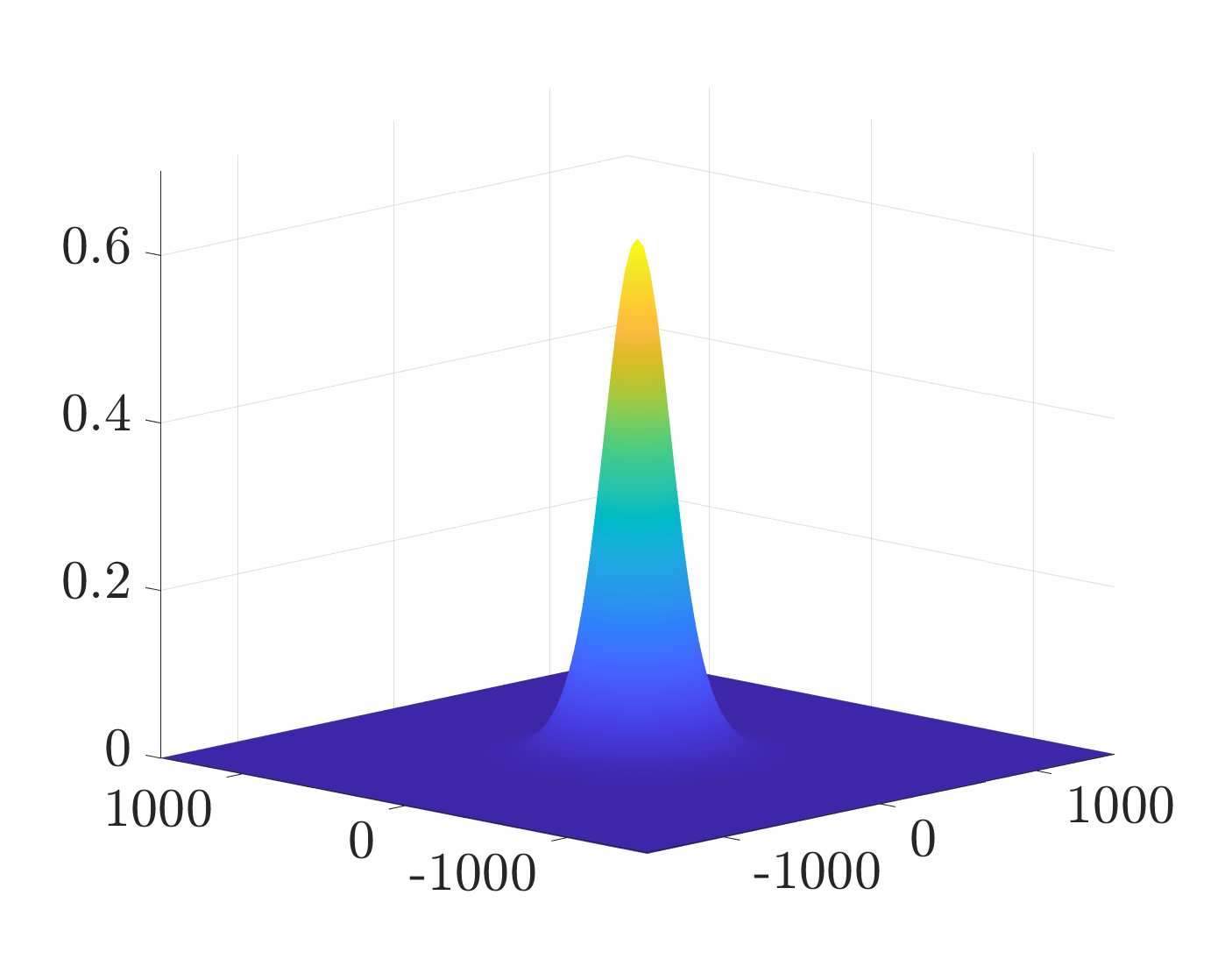}{A}
\insertfigclose{0.43}{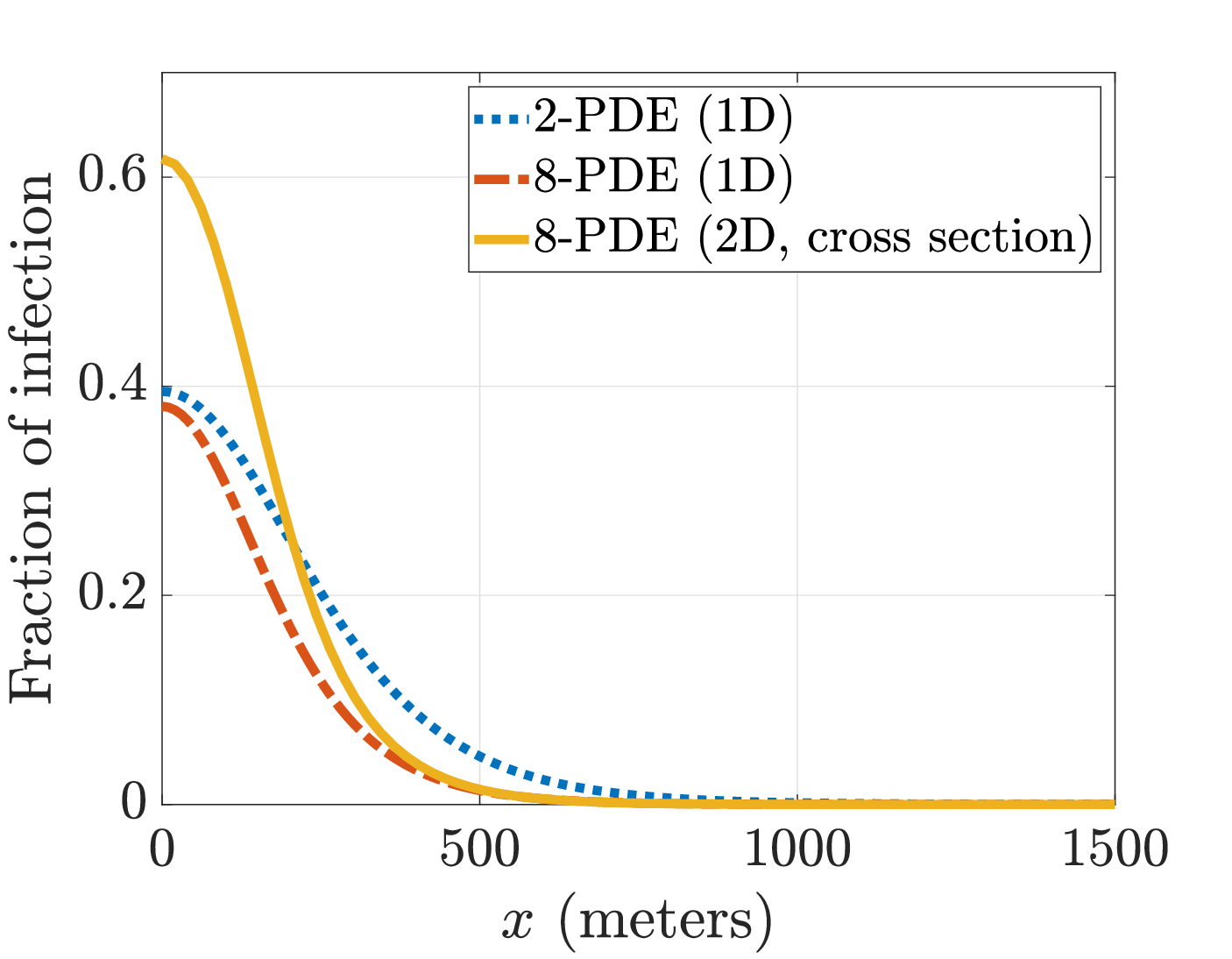}{B}
\caption{Threshold condition for sustainable \W releases. (A) Critical bubble, plotted as the fraction of infection in female mosquitoes, in a two-dimensional spatial field with the release center at $(x,y)=(0,0)$. (B) Comparison of the critical bubble (cross-sectional view from panel A) with the threshold identified using the 2-PDE model in \cite{qu2022modeling} and the current model with a one-dimensional spatial field. For biology relevance, all subsequent simulations in this study will use a two-dimensional spatial field. \label{fig:bubble}}
\end{figure}

We first visualize the critical bubble, which is the threshold release profile necessary to achieve a sustainable \W coverage among mosquitoes, at baseline parametrization in the 2-D field (\cref{fig:bubble}A). The threshold exhibits a bubble-shaped profile, requiring a significantly higher infection rate (approximately $60\%$) at the release center $(x,y) = (0,0)$ to support the initial local establishment of \Wns-infection and counteract the diffusion dynamics that may dilute the concentration of infection upon releases. In contrast, the required infection level decreases substantially as the distance from the release center increases.

For comparison, we calculate the results in the 1-D field (using the same diffusion coefficients) as well as using the 2-PDE model presented in \cite{qu2022modeling}, which tracks only female mosquitoes. The critical bubble derived from the current 8-PDE model in 1-D is in close agreement with its 2-PDE counterpart but is slightly lower and steeper (\cref{fig:bubble}B). This difference is likely due to the zero-diffusion assumption for the aquatic stages, which reduces the overall average diffusion in the system, thereby lowering the initial infection level needed for local \W establishment. Additionally, the bubble profile in the 2-D field (with a cross-sectional view in \cref{fig:bubble}B) is substantially higher than in the 1-D case, as the same diffusion coefficients represent a much greater amount of diffusion in 2-D compared to 1-D. Although the computational cost for 2-D simulations is much higher, we focus exclusively on 2-D spatial fields for the numerical results below, for their biological relevance for simulating field releases.

We then assess the importance of model parameters on the critical bubble and \W infection wave to identify the key factors in infection establishment and propagation. To do so, we conduct the global sensitivity analysis (SA) using the Latin Hypercube Sampling/Pearson Partial Rank Correlation Coefficient (LHS/PRCC) technique and analyze the quantities of interest (QOIs), including metrics of the critical bubble (threshold level, total infection, and width) and infection wave (wave velocity and width). To ensure biological relevance, we impose fitness costs on \Wns-infected mosquitoes, assuming reduced reproduction rate ($\phi_w<\phi_u$, \cite{hoffmann2014stability}) and higher death rates for \Wns-infected females ($\mu_{fw}>\mu_{fu}$, \cite{walker2011wmel}) and eggs ($\mu_{ew}>\mu_{eu}$, \cite{hoffmann2014stability}). Details on the QOI definitions and sampling methods are given in \cref{sec:appendixC}.

\begin{figure}[htbp]
\begin{widepage}
\centering
\insertfigclose{0.32}{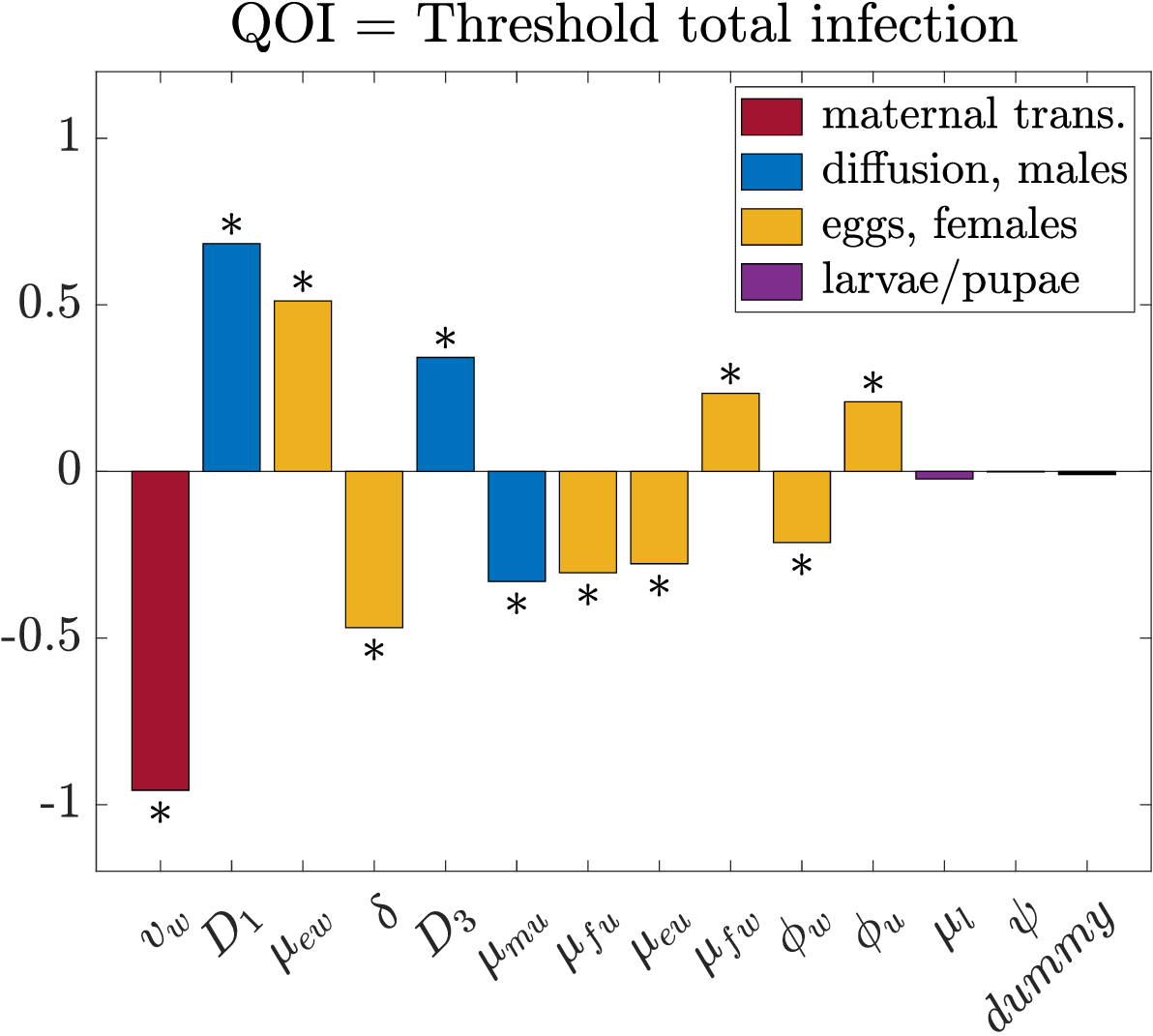}{A}\hfill
\insertfigclose{0.32}{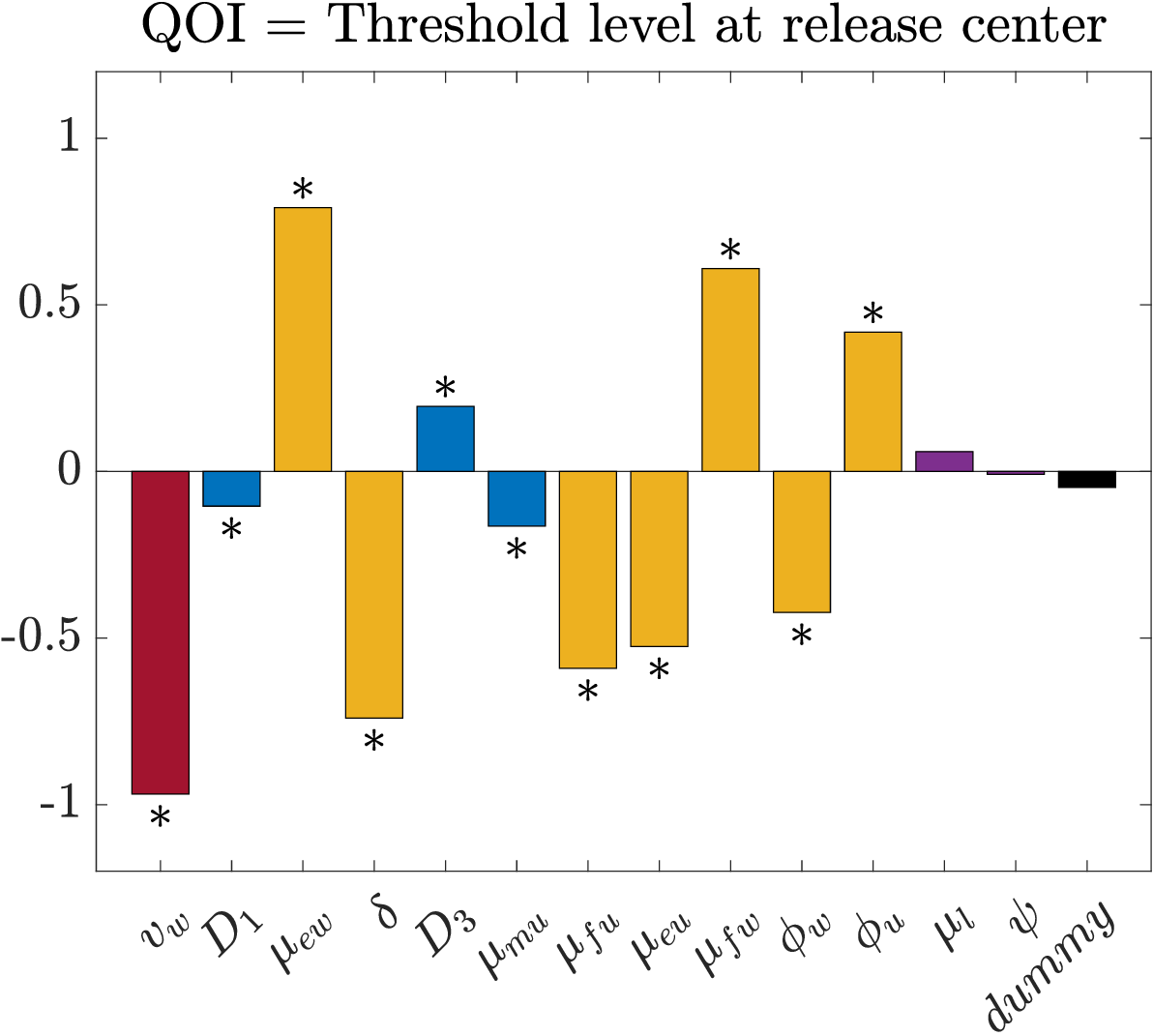}{B}\hfill
\insertfigclose{0.32}{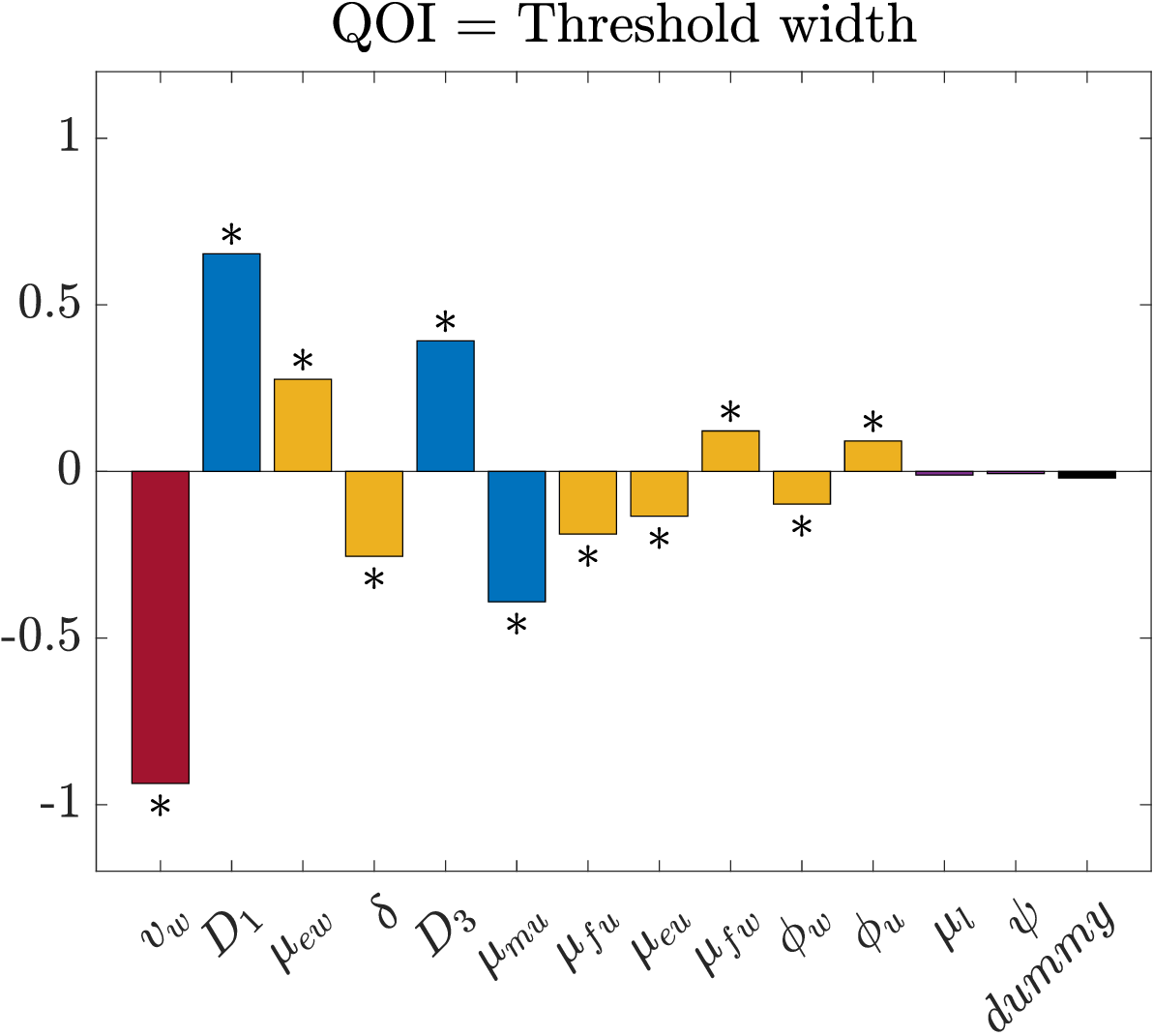}{C}\\
\insertfigclose{0.32}{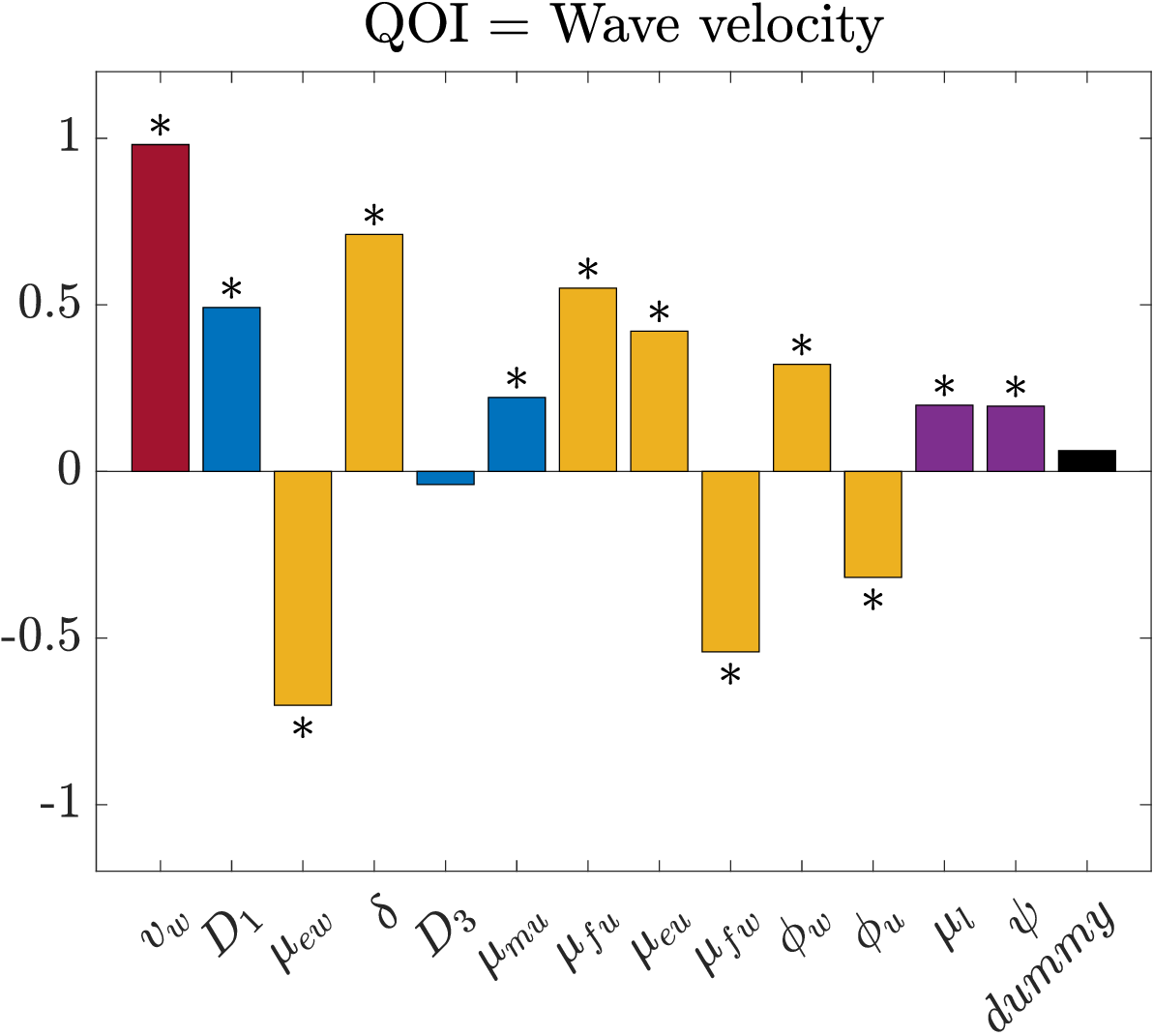}{D}\quad
\insertfigclose{0.32}{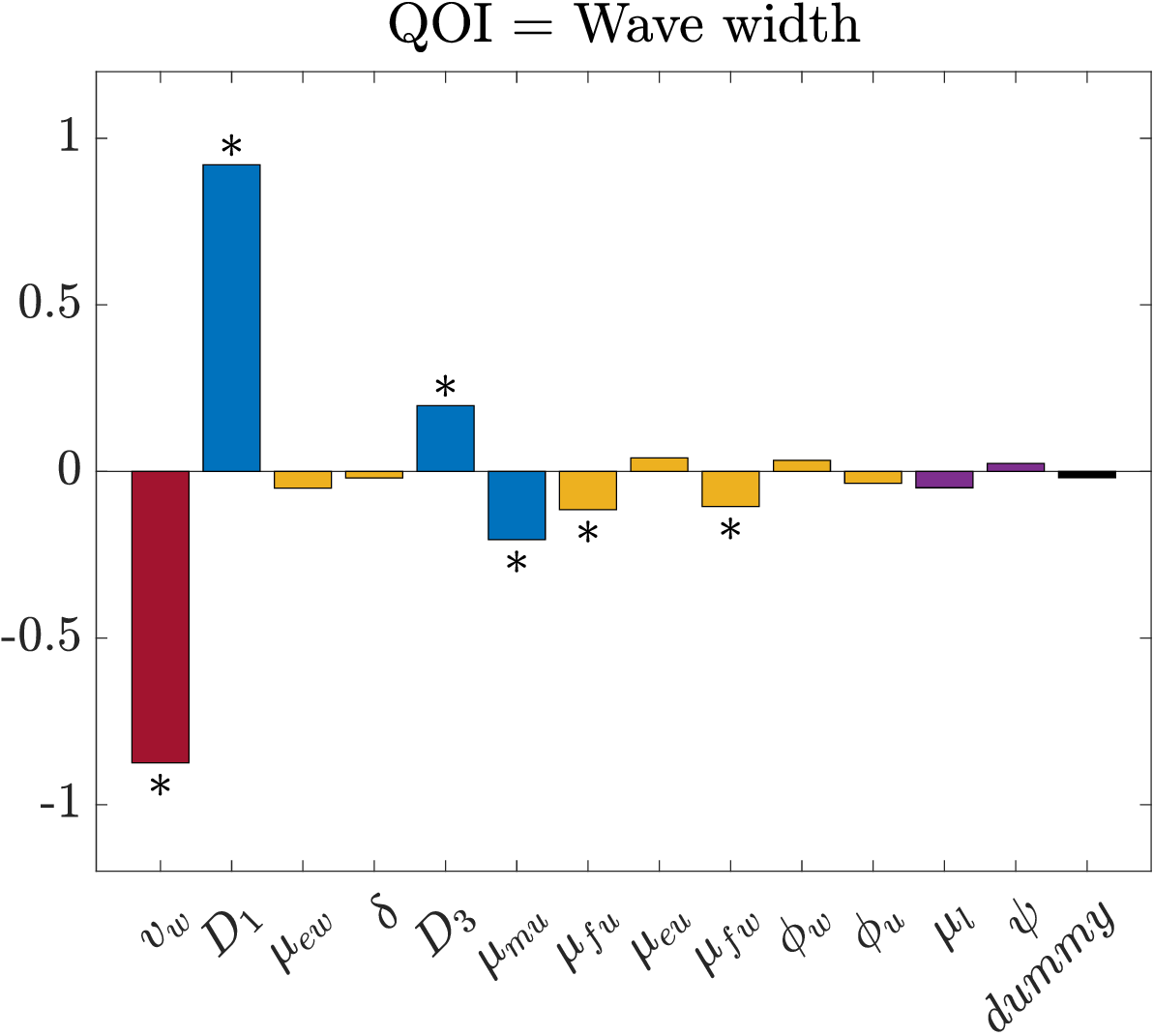}{E}
\caption{Global sensitivity analysis using PRCC on measurements of critical bubble profile and \W infection wave. QOI = quantity of interest. The colors of the bars indicate the categories of parameters, as shown in the legend, and a dummy parameter is included as a null comparison. Significance ($p<0.05$) is indicated with an asterisk (*). } \label{fig:SA_PRCC}
\end{widepage}
\end{figure}

The maternal transmission rate ($v_w$, red) has a dominating impact on all the QOIs considered (\cref{fig:SA_PRCC} A--E), which is consistent with previous modeling studies \cite{florez2022modeling,qu2018modeling}. Reproduction-related parameters (yellow), including egg survivorship (egg death rates $\mu_{eu},~\mu_{ew}$ and development rate $\delta$) and female fitness parameters (female death rates $\mu_{fu}, ~ \mu_{fw}$ and egg-laying rates $\phi_u, ~ \phi_w$) strongly influence the threshold level at the release center (\cref{fig:SA_PRCC}B) as well as the propagation of the infection wave (\cref{fig:SA_PRCC}D). These findings underscore the importance of eggs and female fitness in facilitating the population replacement process during both the establishment and propagation stages, highlighting the need to understand how much these traits could be altered by the \W infection for efficient \Wns-based programs.

Diffusion coefficients ($D_1,~ D_3$, blue) and male death rate ($\mu_{mu}$, blue) greatly impact the widths of the threshold bubble (\cref{fig:SA_PRCC}C) and the established wave (\cref{fig:SA_PRCC}E) and moderately influence wave velocity (\cref{fig:SA_PRCC}D). Notably, the female diffusion coefficient $D_1$ ($D_2=D_1$) has a dominating impact on the shape of the wave (\cref{fig:SA_PRCC}E, comparable to the impact of $v_w$. 

Larvae/pupae parameters ($\mu_l, \psi$) are not significant for the threshold quantities (Figs. \ref{fig:SA_PRCC}A--C). This may be because we have assumed that \Wns-infection does not impact the life traits at these two stages, thus they don't significantly contribute to the competition for population replacement. They have a modest impact on the wave velocity (\cref{fig:SA_PRCC}D).

\subsection{Pre-release mitigation}
Pre-release mitigation, such as larviciding and spraying, reduces the native mosquito population, thereby lowering the threshold for establishing a stable \textit{Wolbachia} infection. Starting from the \Wns-free equilibrium $(E_u^0, 0, L_u^0, 0, F_u^0, 0, M_u^0, 0)$, we model the pre-release mitigation by adjusting the mosquito abundance in the targeted stages. For example, a mitigation efficacy of $\theta$ applied to adult mosquitoes reduces uninfected females and males to $(1-\theta)F_u^0$ and  $(1-\theta)M_u^0$, respectively. We then examine threshold release numbers under varying release radius (where infected mosquitoes are released) and mitigation radius (where uninfected mosquitoes are reduced) to find the optimal combination of coverage.

\subsubsection{Pre-release mitigation against adult stages}
\begin{figure}[htbp]
\begin{widepage}
\centering
\insertfigclose{0.34}{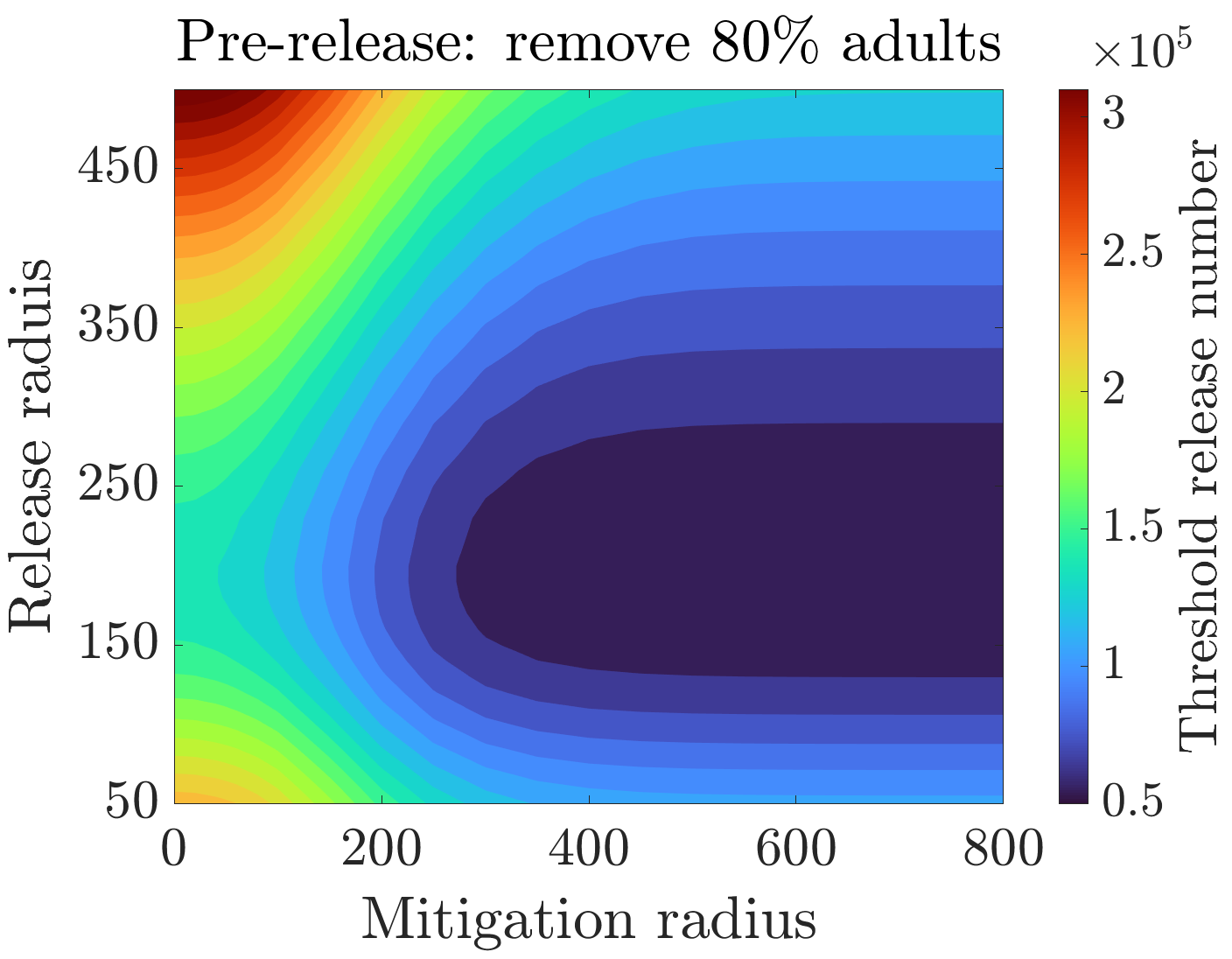}{A}\hfill
\insertfigdefault{0.31}{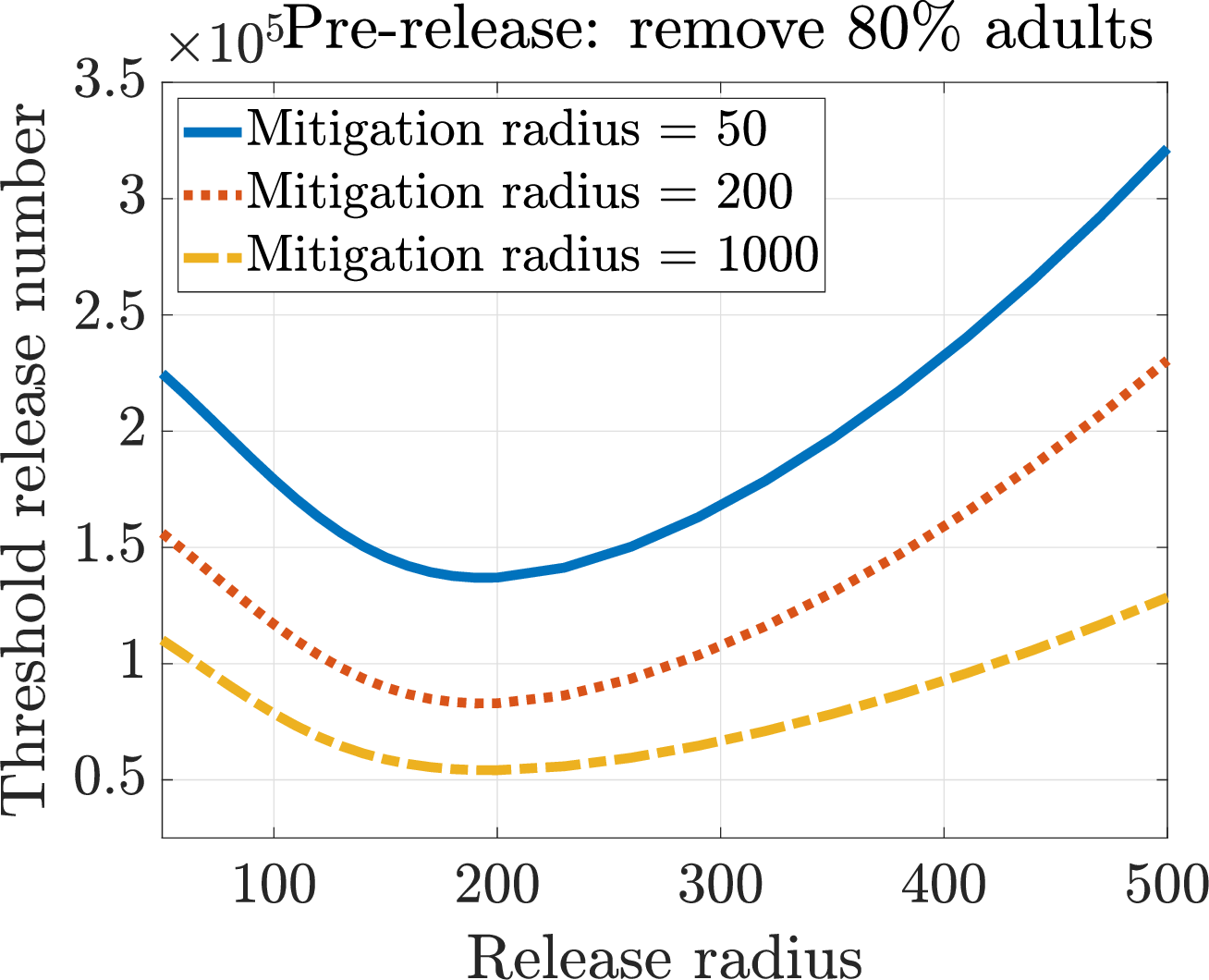}{B}\hfill
\insertfigdefault{0.31}{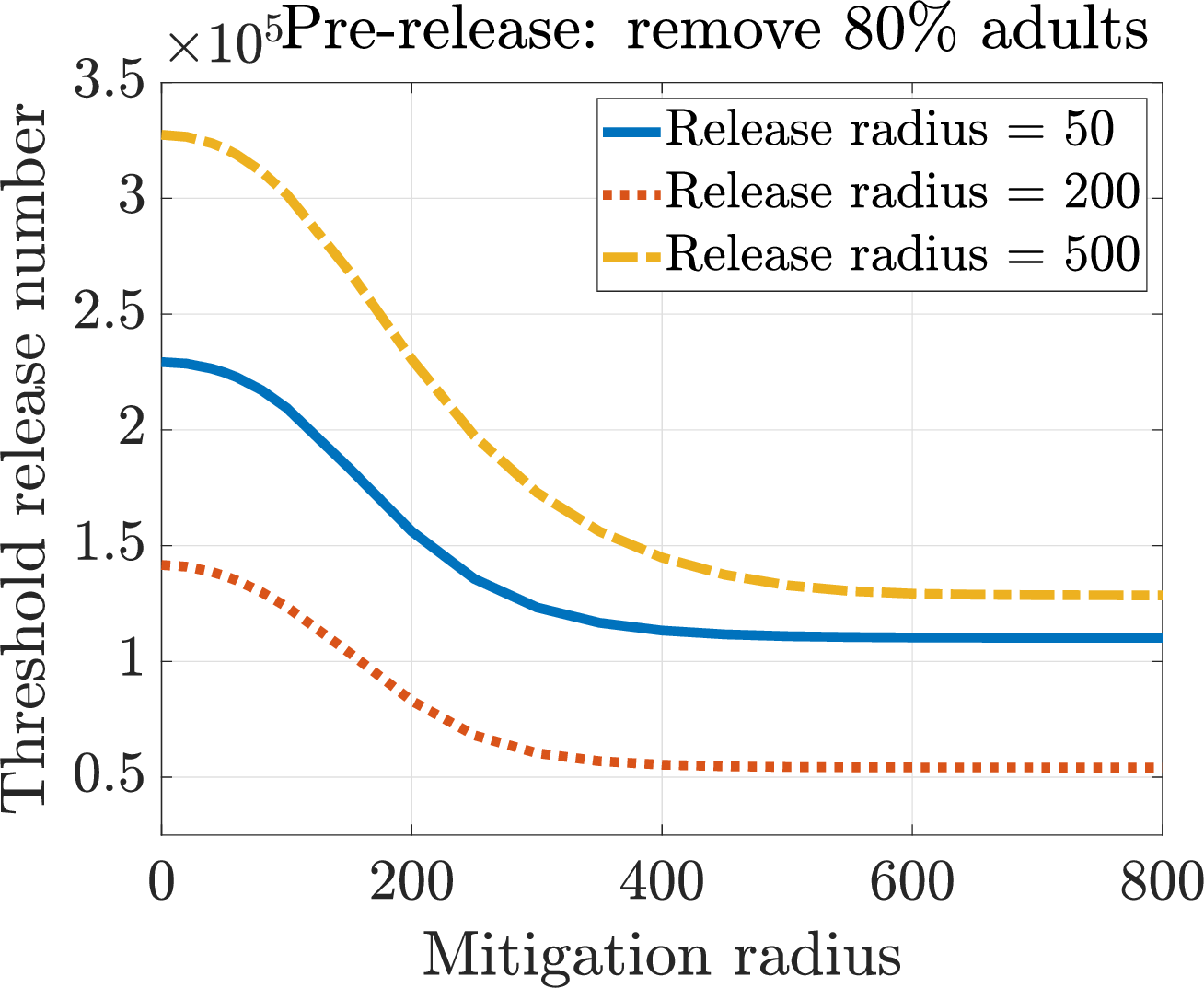}{C}\\
\insertfigclose{0.325}{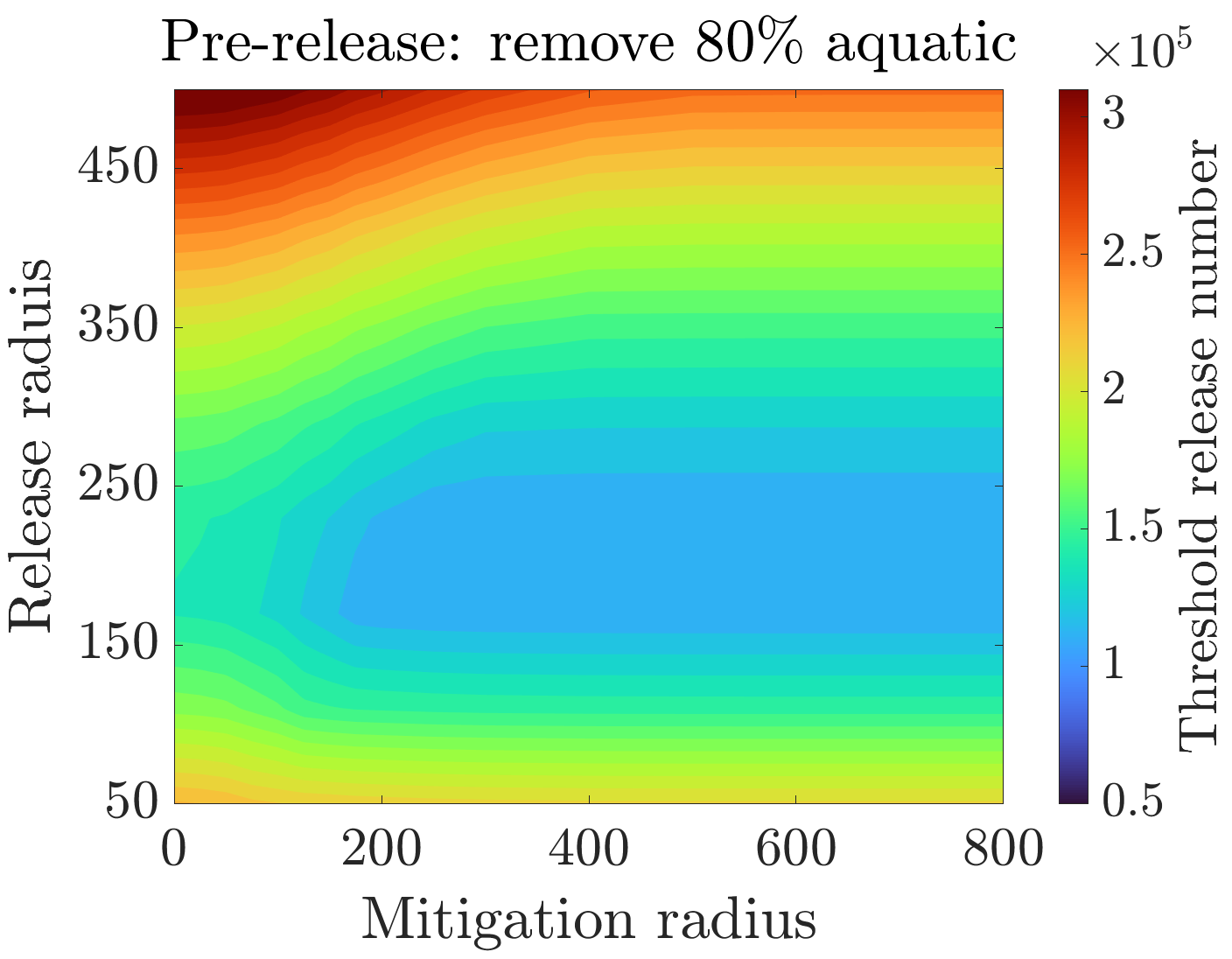}{D}\hfill
\insertfigclose{0.325}{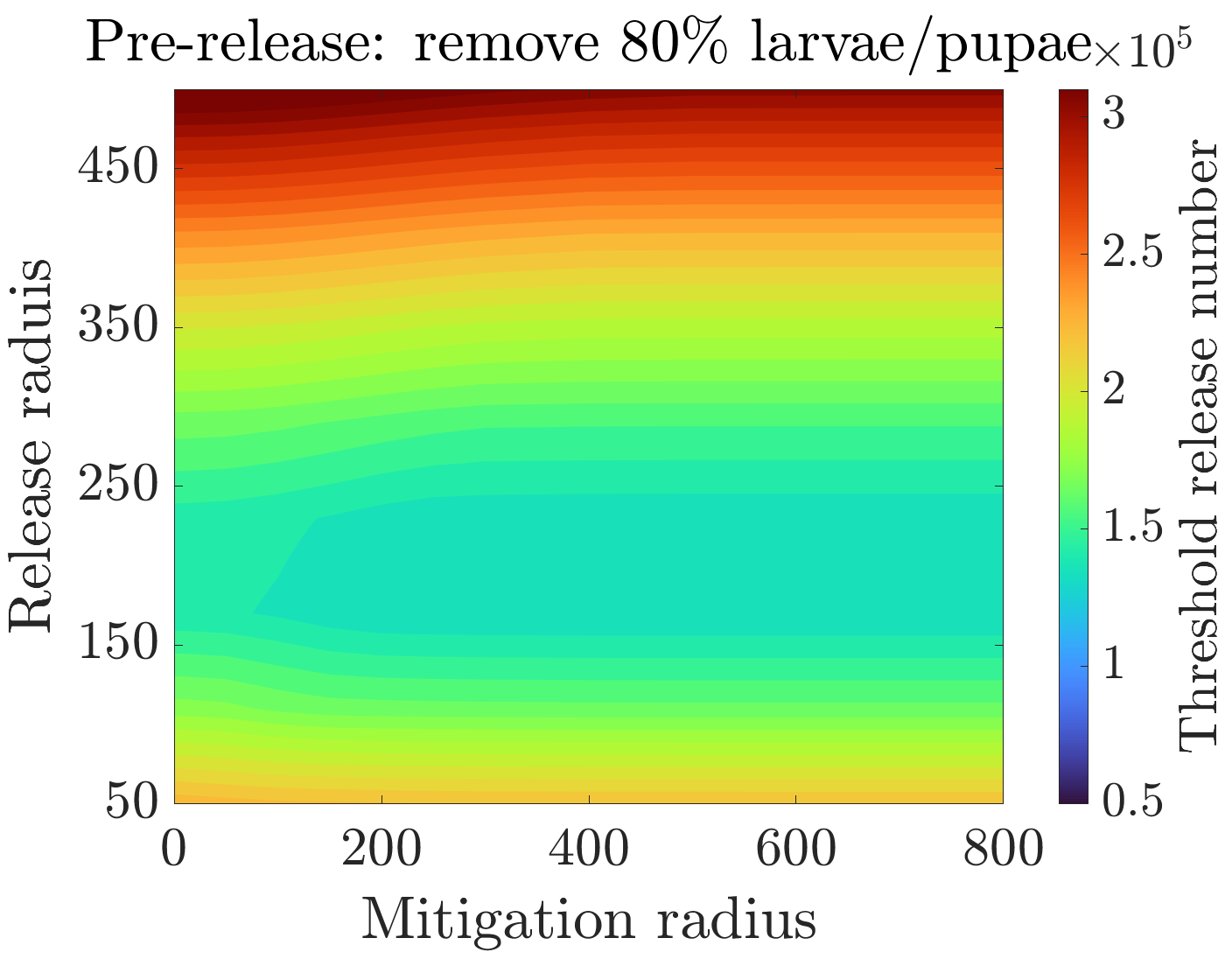}{E}\hfill
\insertfigdefault{0.3}{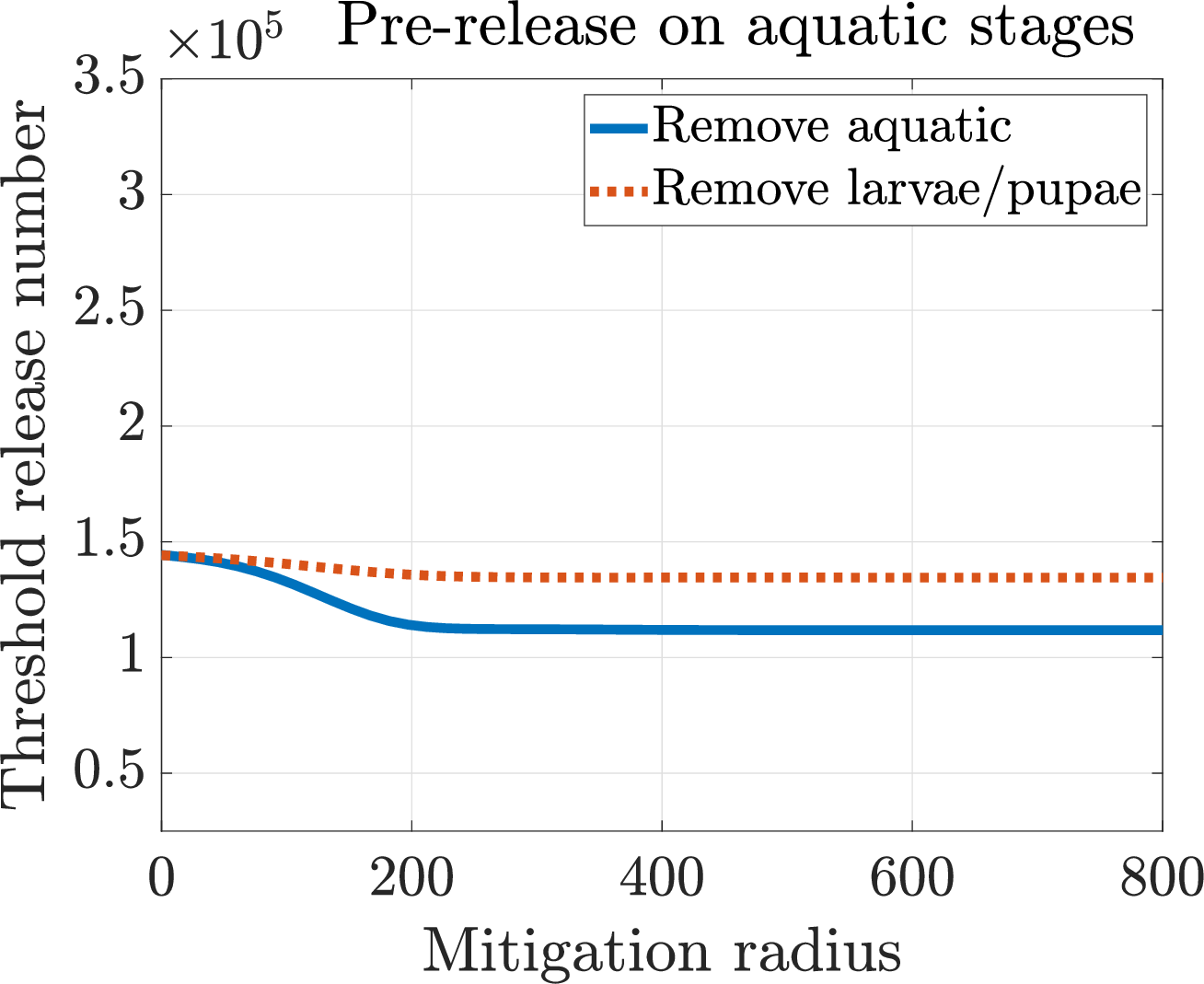}{F}\\
\insertfigclose{0.325}{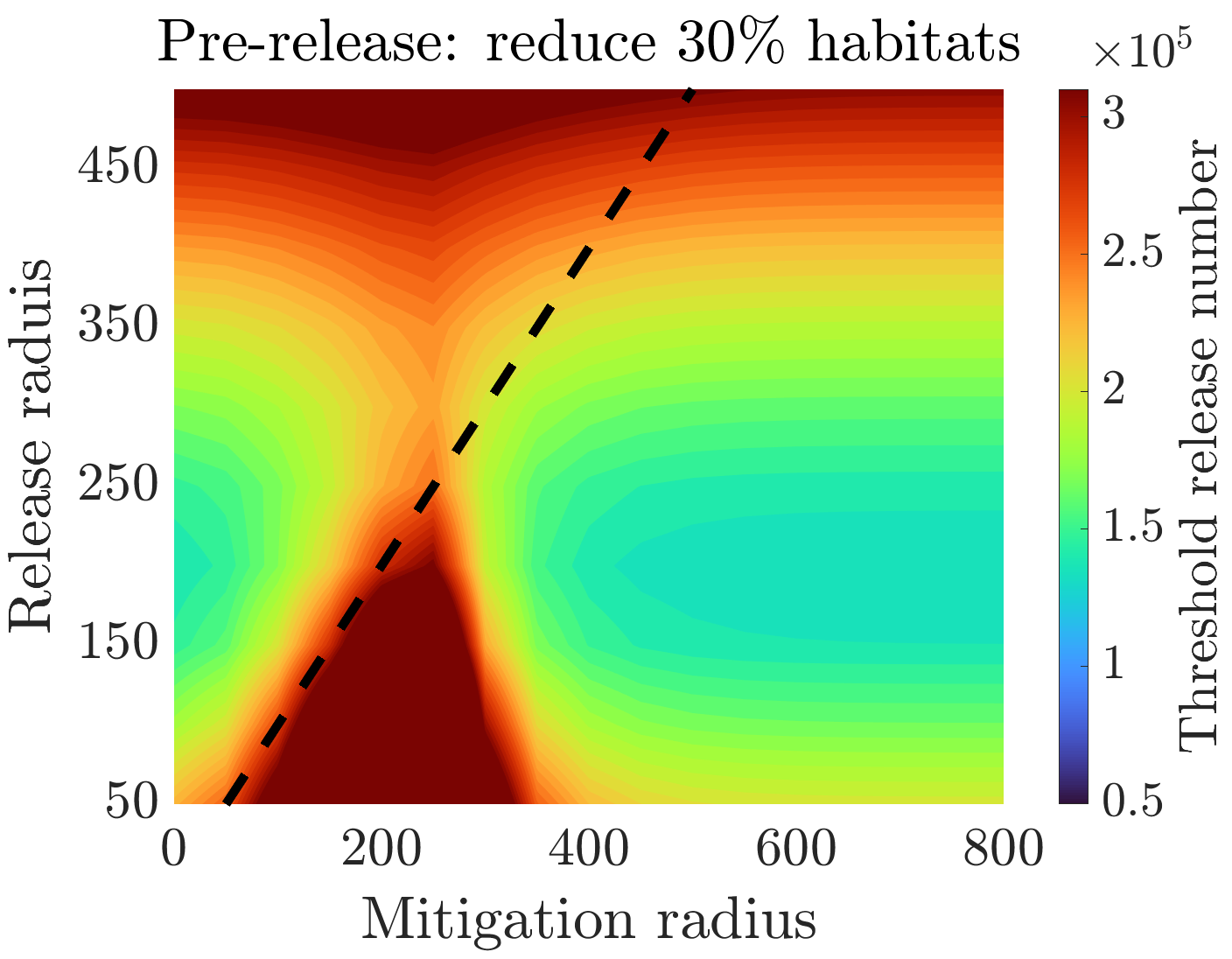}{G}\hfill
\insertfigclose{0.325}{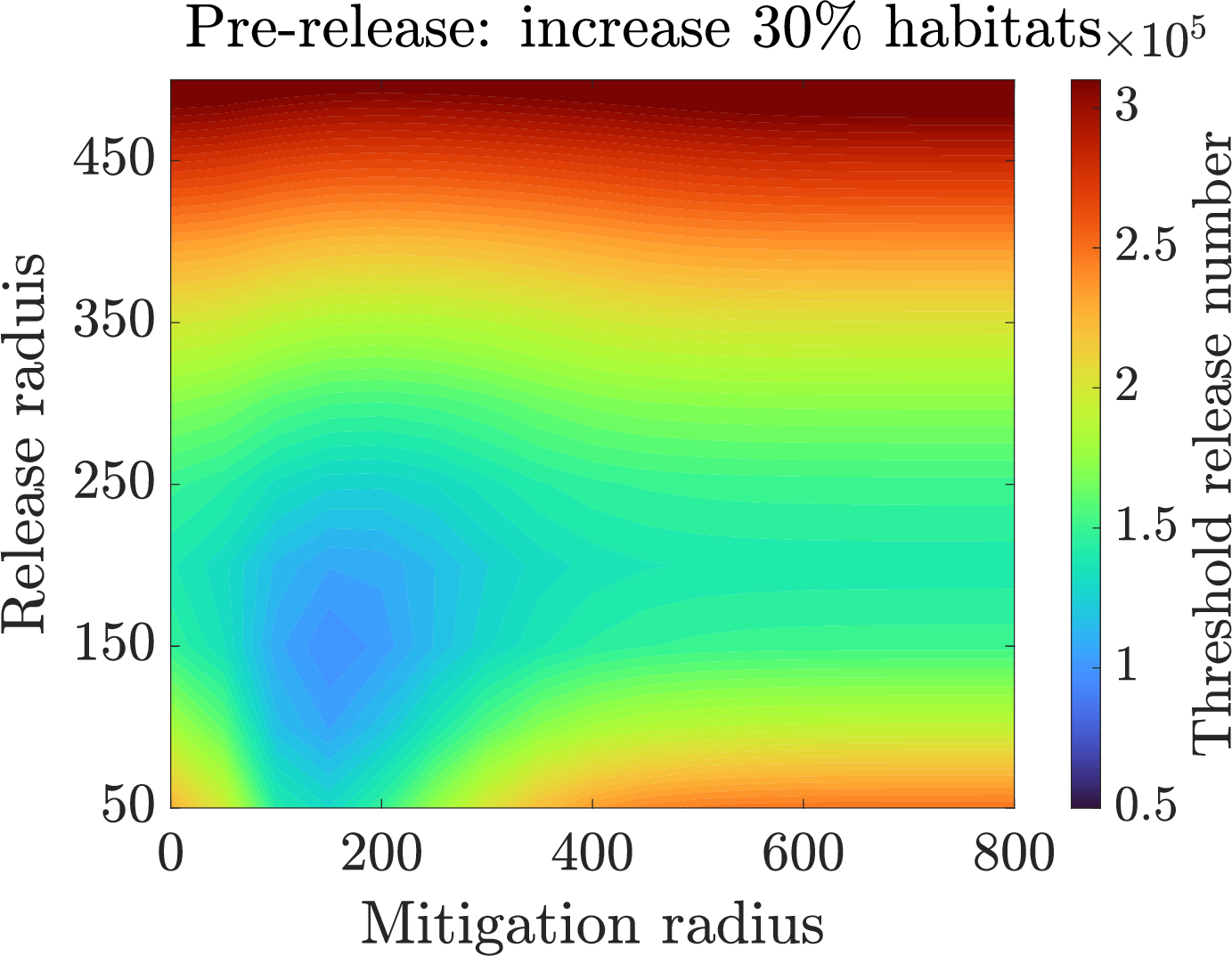}{H}\hfill
\insertfigdefault{0.3}{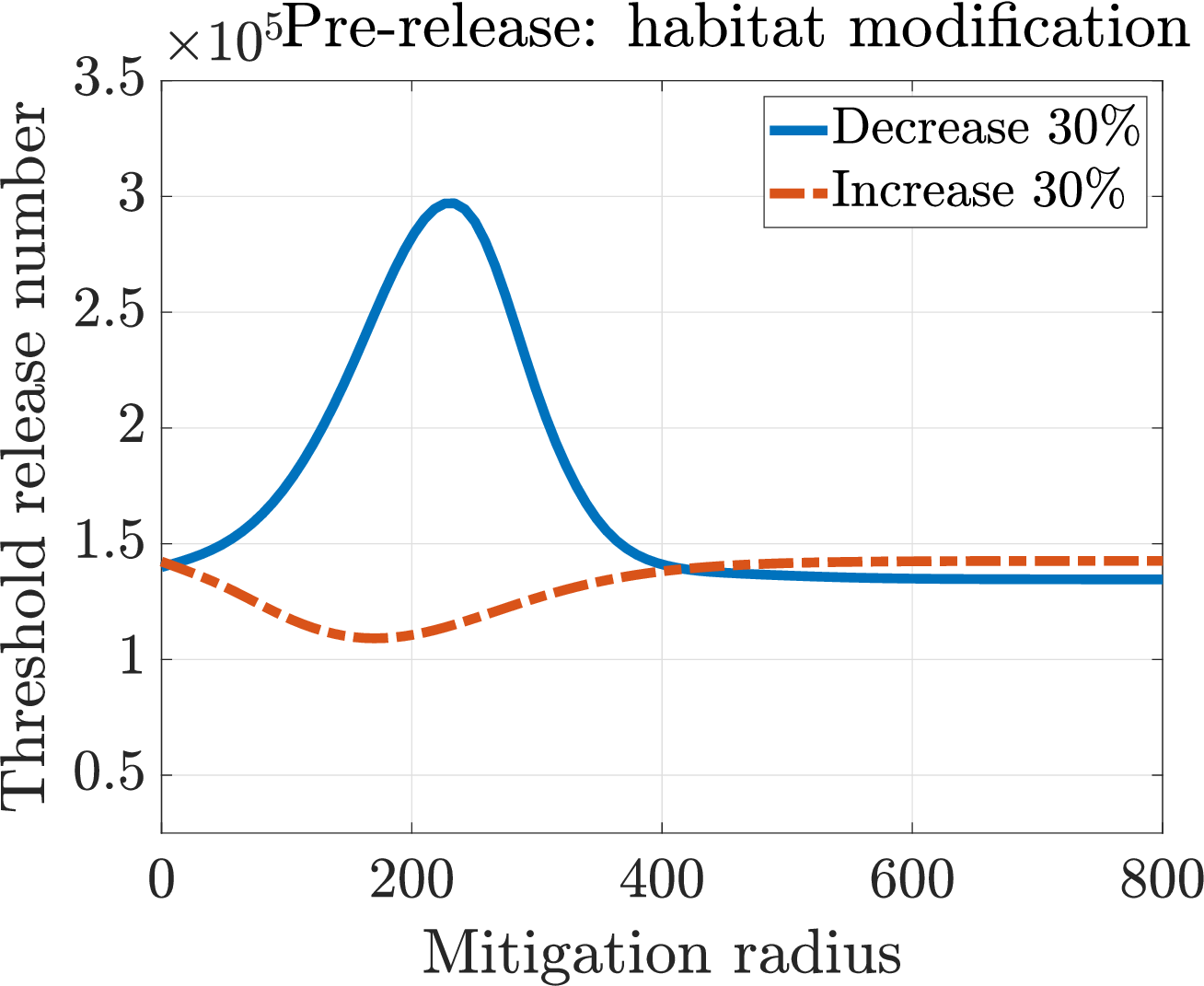}{I}
\caption{Impact of pre-release mitigation on threshold release number for \W establishment.\\
\textbf{(A)--(C)} Mitigation by removing $80\%$ of adult mosquitoes (males and females): (A) Heatmap showing release number as a function of mitigation radius and \W release radius. (B) Vertical slices of the heatmap (A) at fixed mitigation radii, indicating an optimal release radius of around 200 meters. (C) Horizontal slices of the heatmap (A) at fixed release radii, showing a sigmoidal decay in release numbers. \\
\textbf{(D)--(F)} Mitigation targeting aquatic stages:
(D) Heatmap for removing $80\%$ of aquatic-stage mosquitoes (eggs and larvae/pupae). (E) Heatmap for removing $80\%$ of larvae/pupae. (F) Horizontal slices of the heatmap (D) and (E) at a release radius of 200 meters. \\
\textbf{(G)--(I)} Mitigation through habitat modifications:
(G) Threshold release number when $30\%$ of breeding sites are removed. The dashed line indicates where the mitigation radius equals the release radius. (H) Threshold release number when  $30\%$ of the breeding sites are added. (I) Horizontal slices of the heatmap in (G) and (H) at a release radius of 200 meters.
}
\label{fig:heatmap_mitigation}
\end{widepage}
\end{figure}

Pre-release mitigation targeting adults (efficacy $\theta = 0.8$) substantially reduces the threshold release number needed to establish a stable infection (\cref{fig:heatmap_mitigation}A). For a given mitigation radius (\cref{fig:heatmap_mitigation}B), the threshold release number is consistently minimized at a release radius of around 200 meters. For a given release radius (\cref{fig:heatmap_mitigation}C), increasing the pre-release mitigation radius reduces the threshold release number, following a sigmoidal trend: reductions are slow at small mitigation radii; but there exists an inflection point, where the reduction is the fastest. After this inflection point, the benefit of additional mitigation coverage diminishes, as the release number saturates. The inflection and the saturation points inform the optimal pre-release mitigation coverage: mitigation should extend to at least the radius at the inflection point, where reductions in the release number are most efficient per each additional meter of mitigation radius, but further coverage beyond the saturation point yields marginal additional benefit. 

\subsubsection{Pre-release mitigation against aquatic stages}
We evaluate two pre-release mitigation strategies targeting aquatic stages: (1) drainage and scrubbing of water containers, which reduce juvenile stages ($E_u^0$ and $L_u^0$), and (2) larviciding, which impacts larvae/pupae stages ($L_u^0$) only. In both cases, water containers remain available, so the carrying capacity is assumed not unaffected. Both strategies show similar trends when varying the mitigation and release radii (\cref{fig:heatmap_mitigation}, panel D and E). Mitigation targeting all the juvenile stages leads to a more noticeable reduction in threshold release number compared to larviciding alone (\cref{fig:heatmap_mitigation}F). However, both strategies are still much less effective than pre-release mitigation targeting adults (\cref{fig:heatmap_mitigation}C).

These results underscore that adult-stage mitigation is significantly more effective in reducing the threshold release number for \W establishment. This finding aligns with the mechanism of \Wns-based population replacement, which is primarily driven by the reproduction advantage during the adult stages through the CI phenomenon that sterilizes wild-type mosquitoes. In contrast, the fitness costs associated with \W during the aquatic stages disadvantage \Wns-infected mosquitoes, slowing the population replacement progress. Therefore, targeting the adult stages via pre-release mitigation is a more efficient strategy to enhance the replacement process. 

Interestingly, this result is different from findings in \cite{florez2022modeling}, where an ODE model suggested comparable impacts on the invasion threshold for targeting adult versus aquatic stages. The discrepancy likely arises from the spatial diffusion dynamics among adults captured in our model, and it facilitates the spread of infection and enables the \Wns-infected adults to quickly fill gaps created by the pre-release mitigation upon their release. 

\subsubsection{Pre-release mitigation via habitat modifications}
Another common pre-release mitigation strategy involves habitat modifications that affect the aquatic-stage mosquitoes and alter the breeding sites. We considered two scenarios: (1) reducing breeding sites (removing water containers), which reduces all juvenile stages ($E_u^0$ and $L_u^0$) and carrying capacity $K_l$; and (2) increasing breeding sites (adding containers with fresh water), which increases the carrying capacity $K_l$.

When breeding sites are reduced ($\theta = 30\%$), the threshold release number increases significantly (\cref{fig:heatmap_mitigation}G). This is because reducing the carrying capacity increases the competition for resources among mosquitoes, leading to premature death of the infected offspring that otherwise could used to replace the native mosquitoes. This effect is exacerbated with greater reduction efficacy ($\theta = 0.5, 0.7$, \cref{fig:heatmap_app} in appendix). Notably, the threshold release number increases dramatically when the mitigation radius is near or slightly exceeds the release radius (dashed line). The reduced carrying capacity creates a gradient of uninfected mosquitoes diffusing into the release area at the boundary of the mitigation zone, disrupting the establishment of \W infection. While increasing breeding sites ($\theta = -30\%$) can moderately lower the threshold release number for small release radii and limited mitigation radius (\cref{fig:heatmap_mitigation}, panel H and I),
designing a pre-release habitat modification strategy for population replacement \W releases is not straightforward and can increase the threshold and reduce the overall efficiency.

\subsection{Release infection in wet or dry regions}
\begin{figure}[htbp]
\begin{widepage}
\centering
\insertfigcloser{0.325}{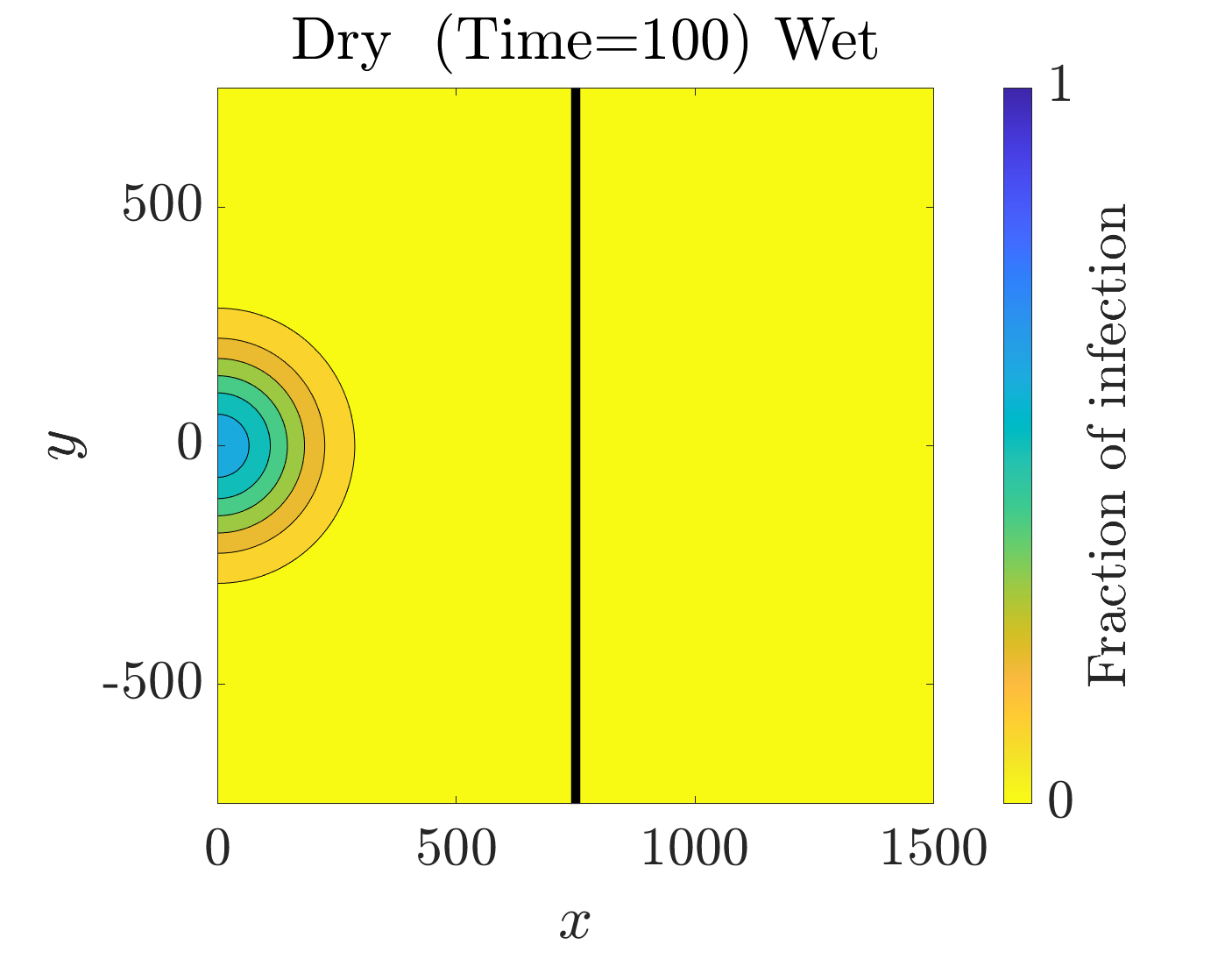}{A}
\insertfigcloser{0.325}{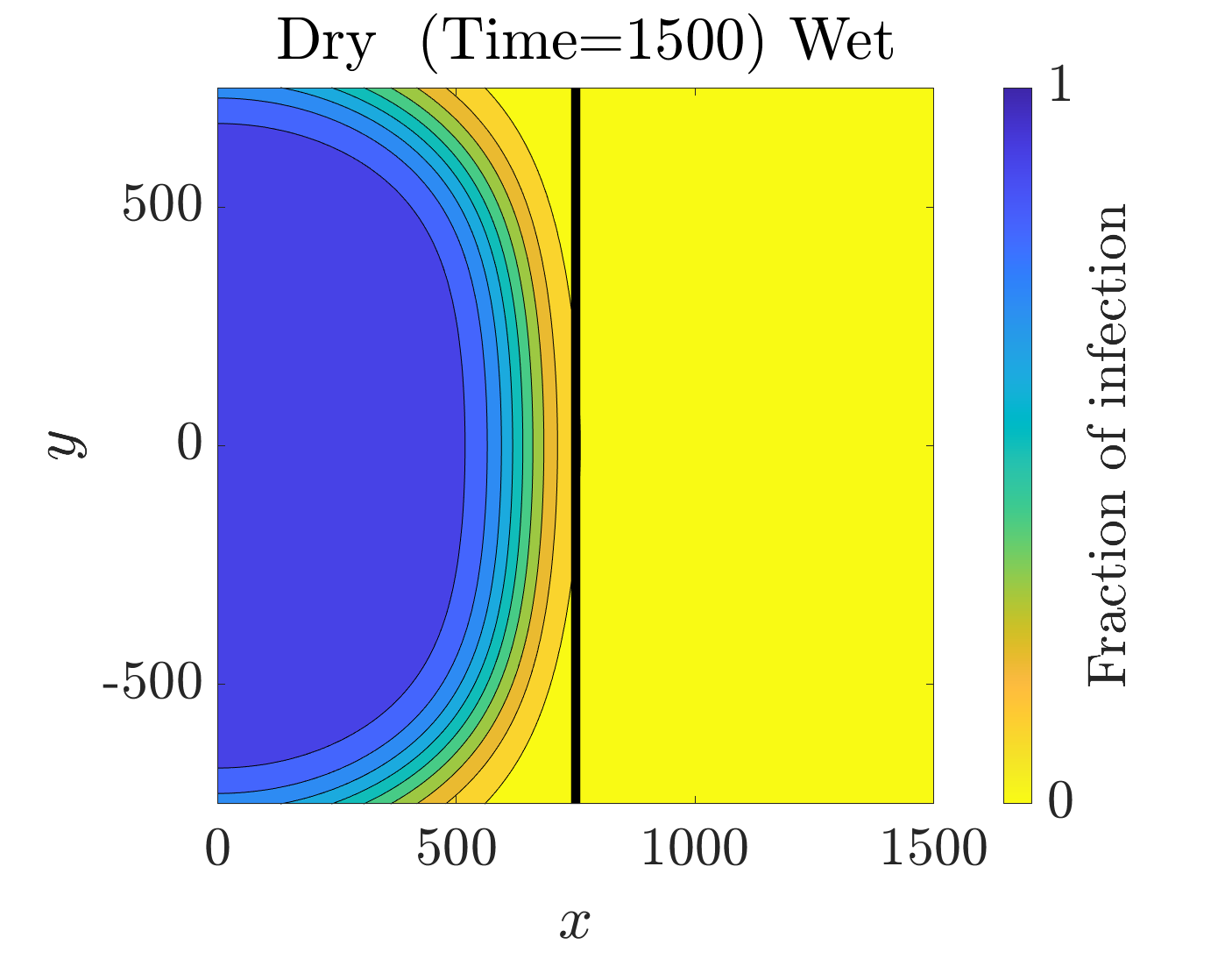}{B}
\insertfigcloser{0.325}{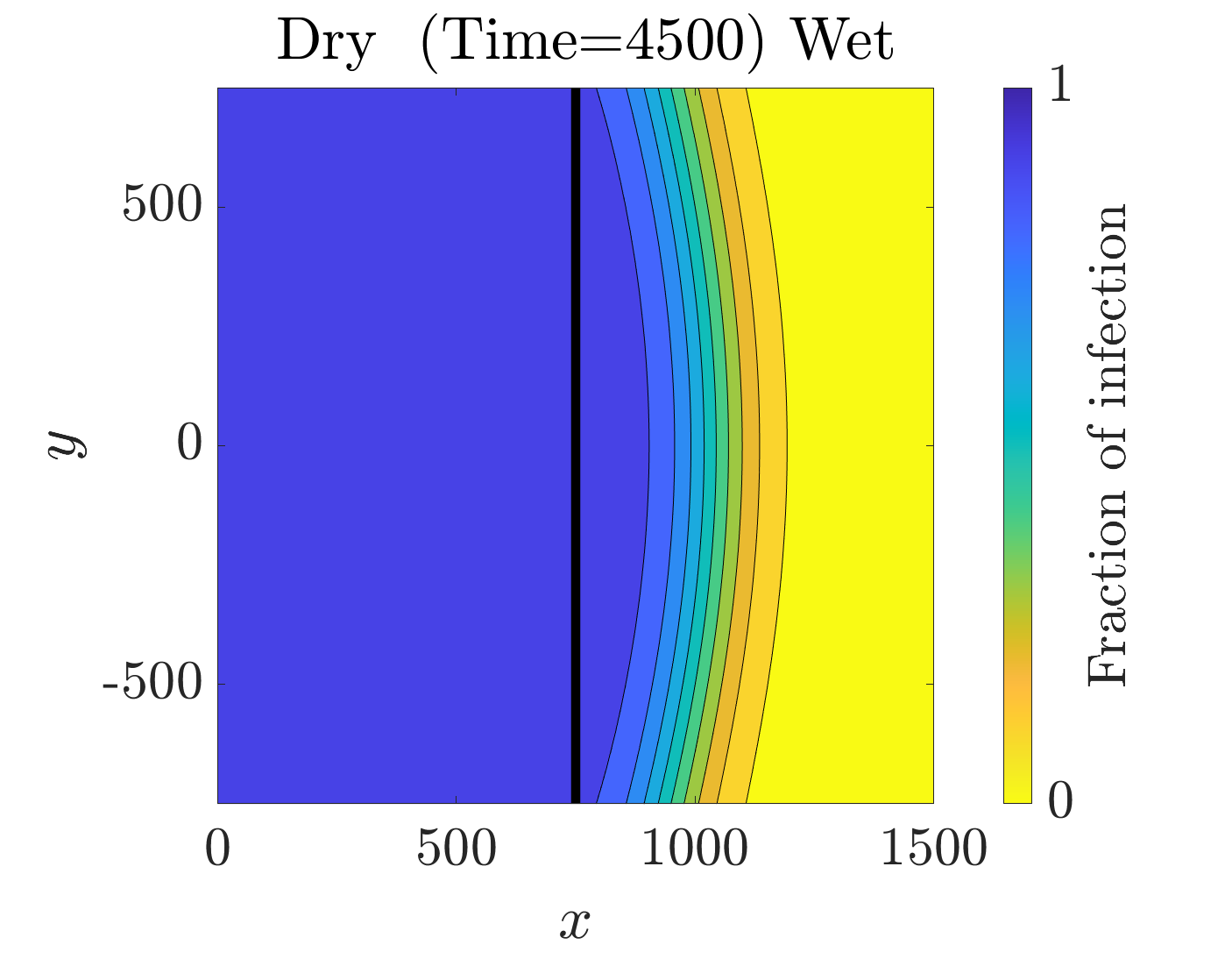}{C}\\
\insertfigcloser{0.325}{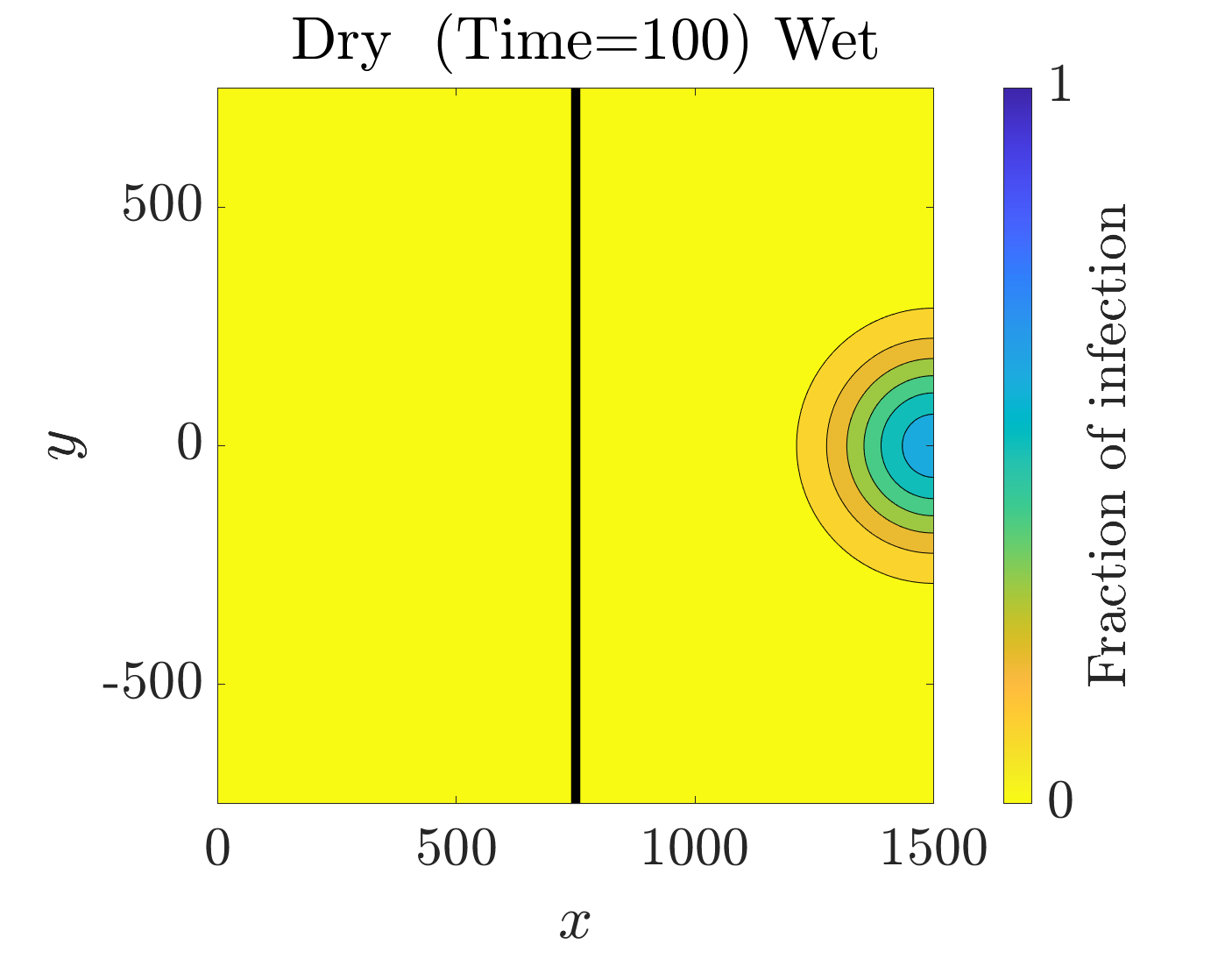}{D}
\insertfigcloser{0.325}{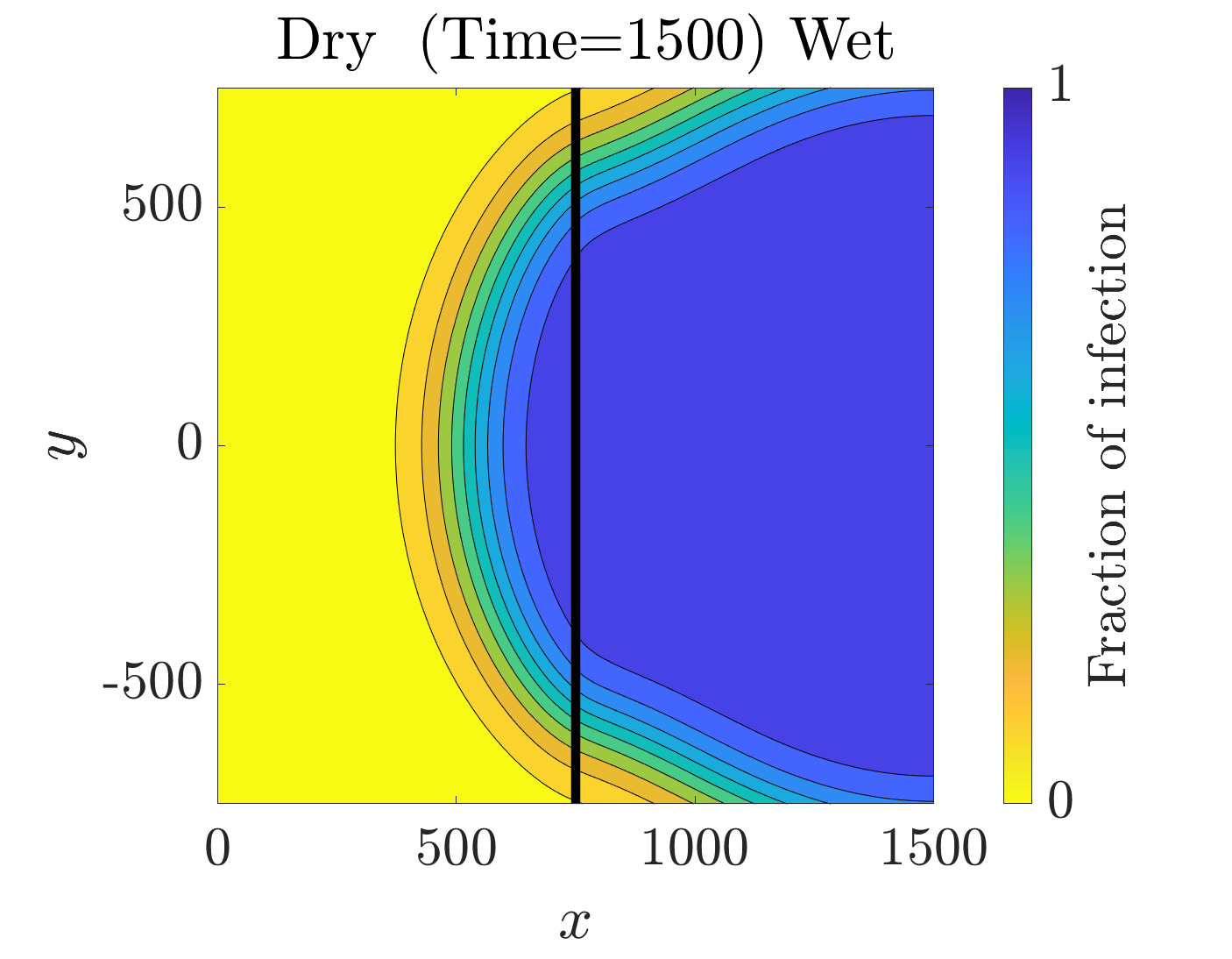}{E}
\begin{tikzpicture}
\node[anchor=north west,inner sep=0pt] at (0,0){\includegraphics[width= 0.33 \linewidth]{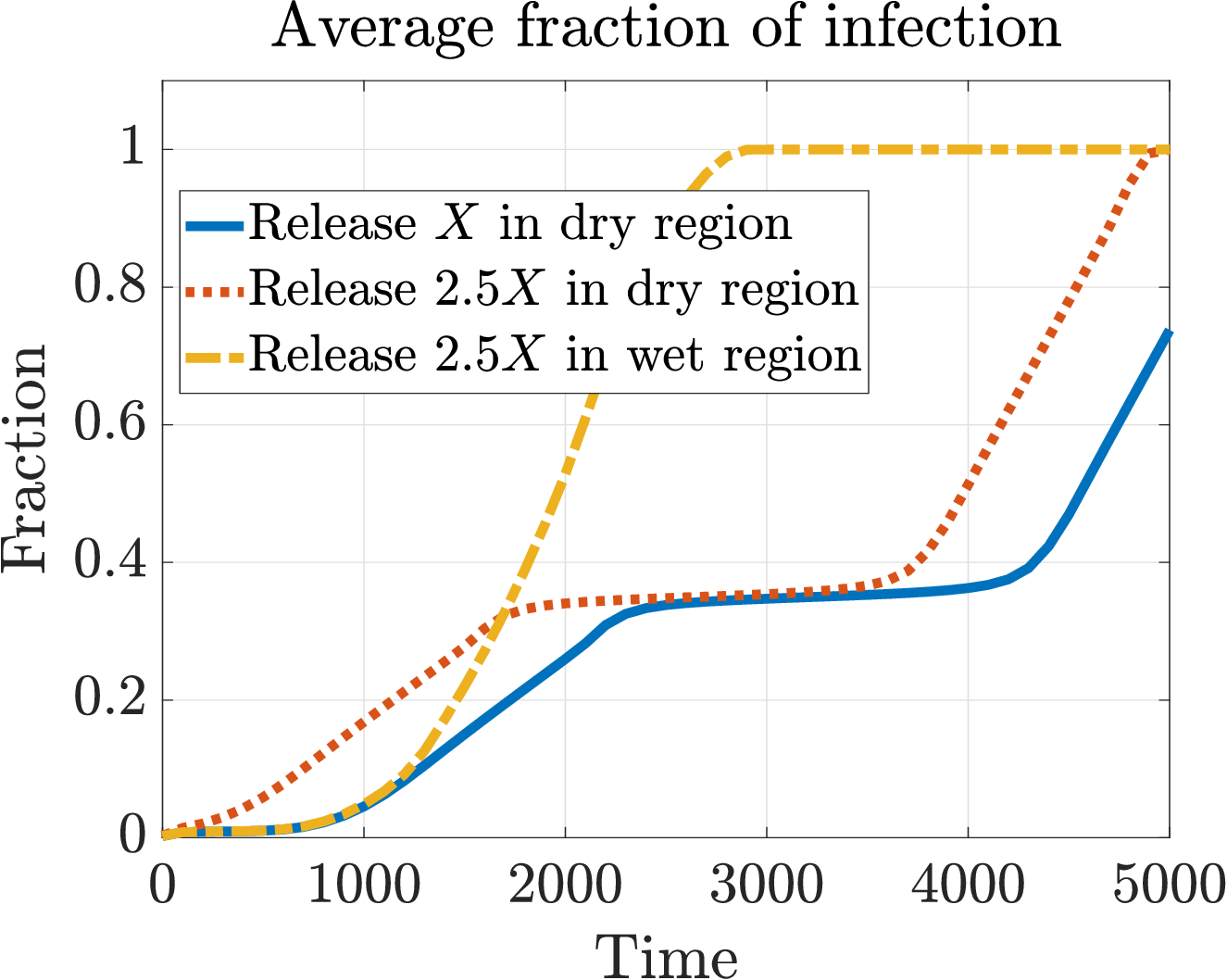}};
\node[font=\sffamily\bfseries\large] at (4ex,1ex) {F};
\end{tikzpicture}
\caption{Simulations of releasing infections in dry vs. wet regions. The dry-wet carrying capacity ratio is $1:2.5$. (A, B, C): release in the dry region, using threshold release level ($X \approx 6.6/m^2$ for both females and males) at point (0,0) with radius 100. (D, E): release in the wet region, using the threshold release level ($2.5X$) at point (1500, 0) with a radius of 100. Note that the threshold release level is 2.5 more in the wet region than in the dry region, proportional to the carrying capacity. (F): The average fraction of infection in the entire domain in time for different release scenarios. }
\label{fig:wetdry}
\end{widepage}
\end{figure}

We examine the selection of release sites in regions with disparities in carrying capacities by comparing releasing in dry and wet regions. When infection is released in the dry region (\cref{fig:wetdry}A--C, dry-to-wet carrying capacity ratio $1:2.5$), the infection wavefront slows substantially upon crossing into the wet region (around $t= 1500$) and enters the wet region around $t= 4500$. This delay is caused by competition from the much higher mosquito abundance in the wet region, proportional to the carrying capacity. The delay is further illustrated in \cref{fig:wetdry}F, where the average fraction of infection over the entire domain shows a plateau between $t\approx 2500$ and $t\approx 4500$ (solid curve). If the carrying capacity ratio between the dry and wet regions increases to 1:5, creating a stronger heterogeneity, the wavefront may even stop at the interface (\cref{fig:wetdry_app}, appendix) as infected mosquitoes are overwhelmed by the larger uninfected population in the wet region. Increasing the initial release level speeds up the local establishment of the infection (\cref{fig:wetdry}F, dotted curve) but does not affect the wavefront propagation once established.

Releasing infection in the wet region, by contrast, makes it easier to spread infection into the dryer area (\cref{fig:wetdry}D and E), due to the advantage of higher population abundance in the wet region. The infection wavefront accelerates slightly when crossing the dry-wet interface (around $t\approx 1500$). However, establishing stable infection in the wet region requires a much larger initial release level (\cref{fig:wetdry}F, dash-dotted curve).

\subsection{Impact of seasonality}
\begin{figure}[htbp]
\centering
\insertfigdefault{0.48}{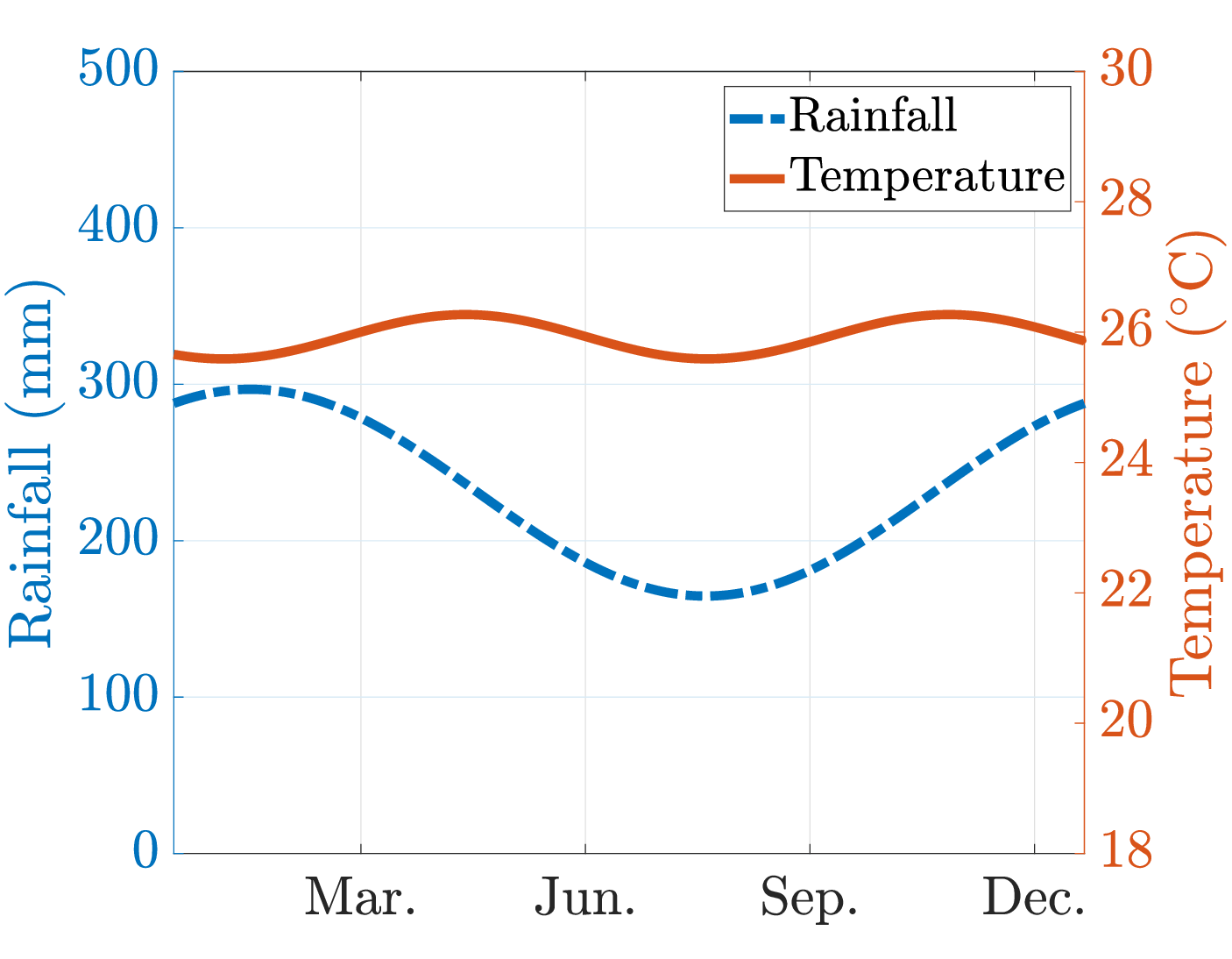}{A}
\insertfigdefault{0.48}{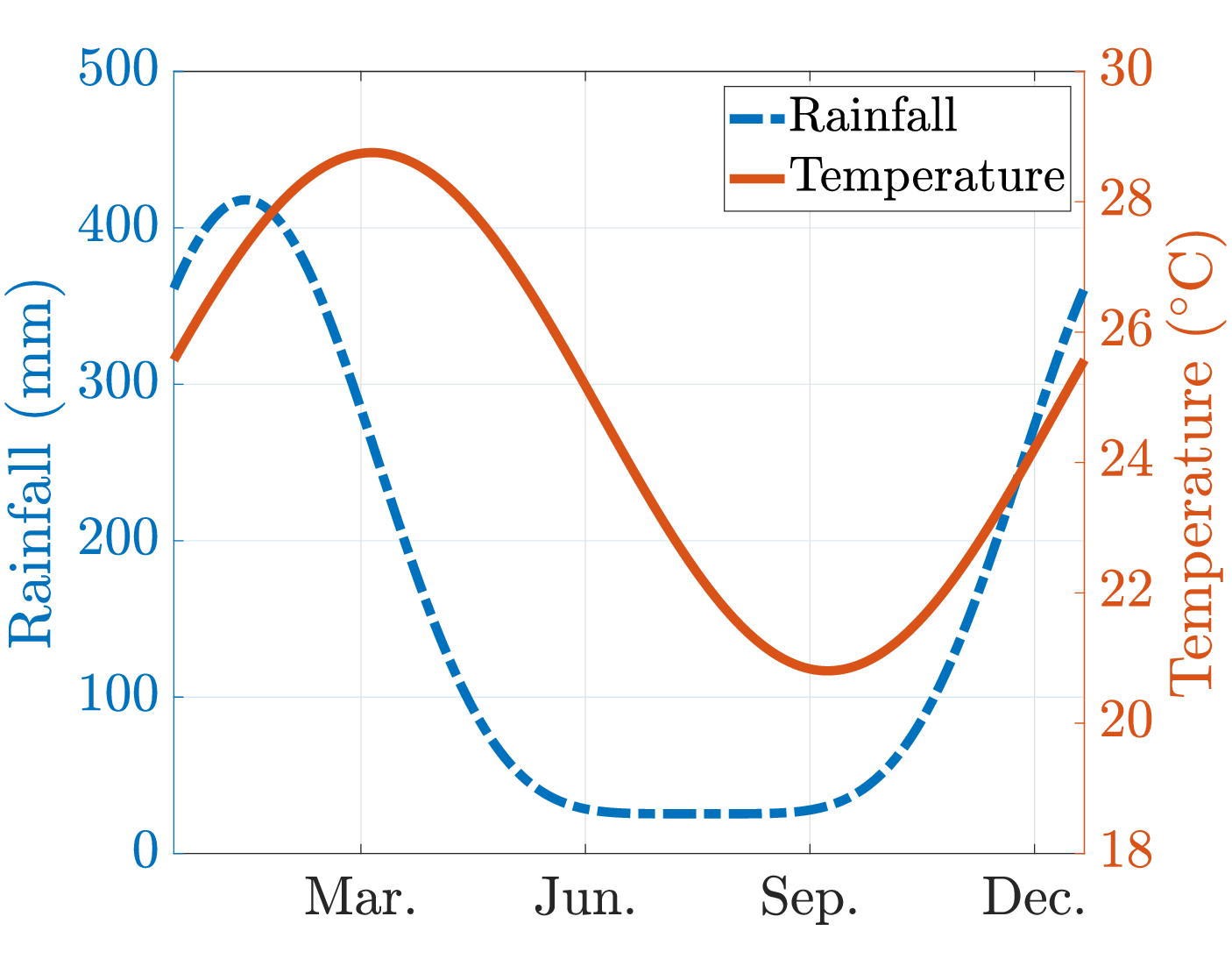}{B}
\insertfigdefault{0.48}{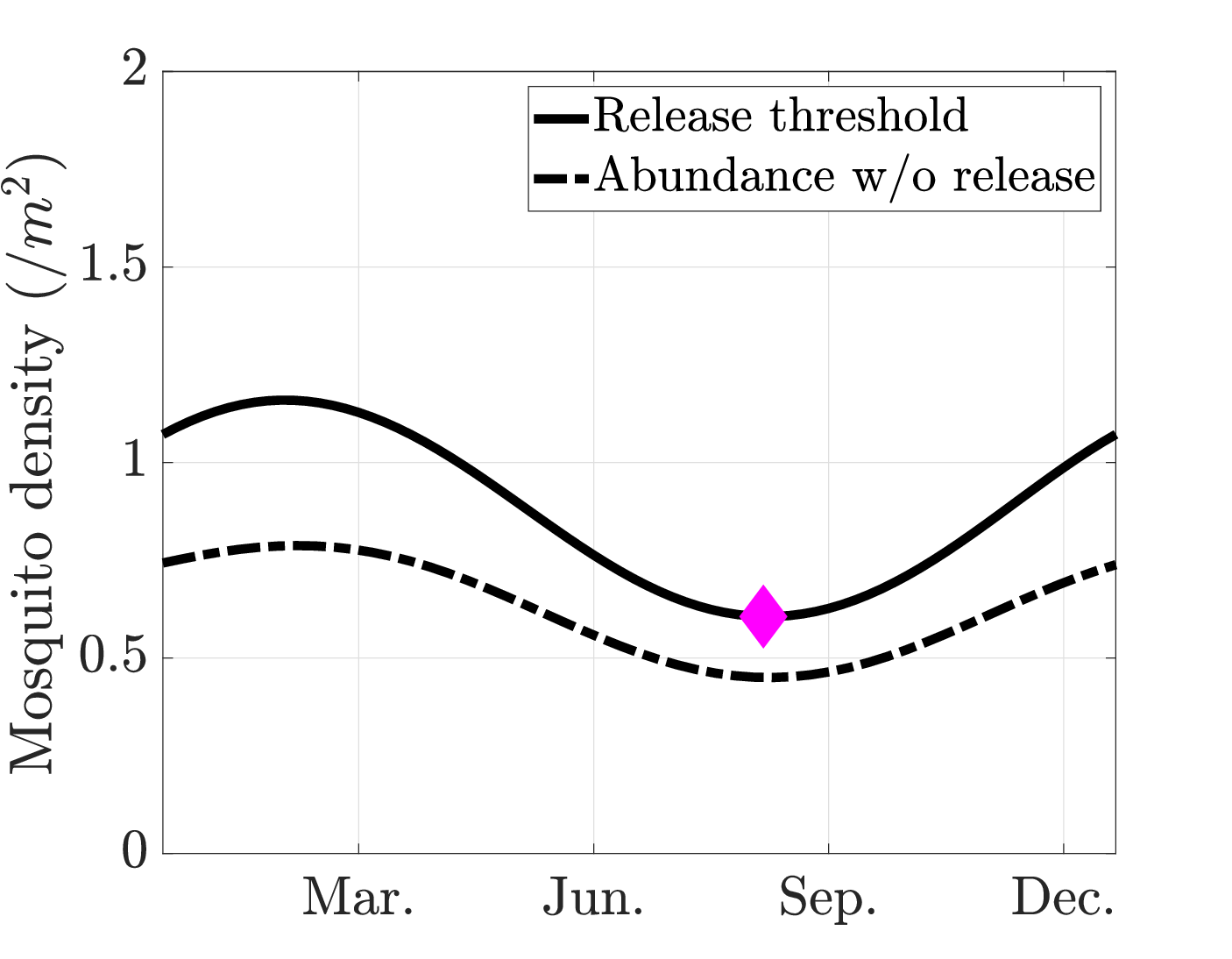}{C}
\insertfigdefault{0.48}{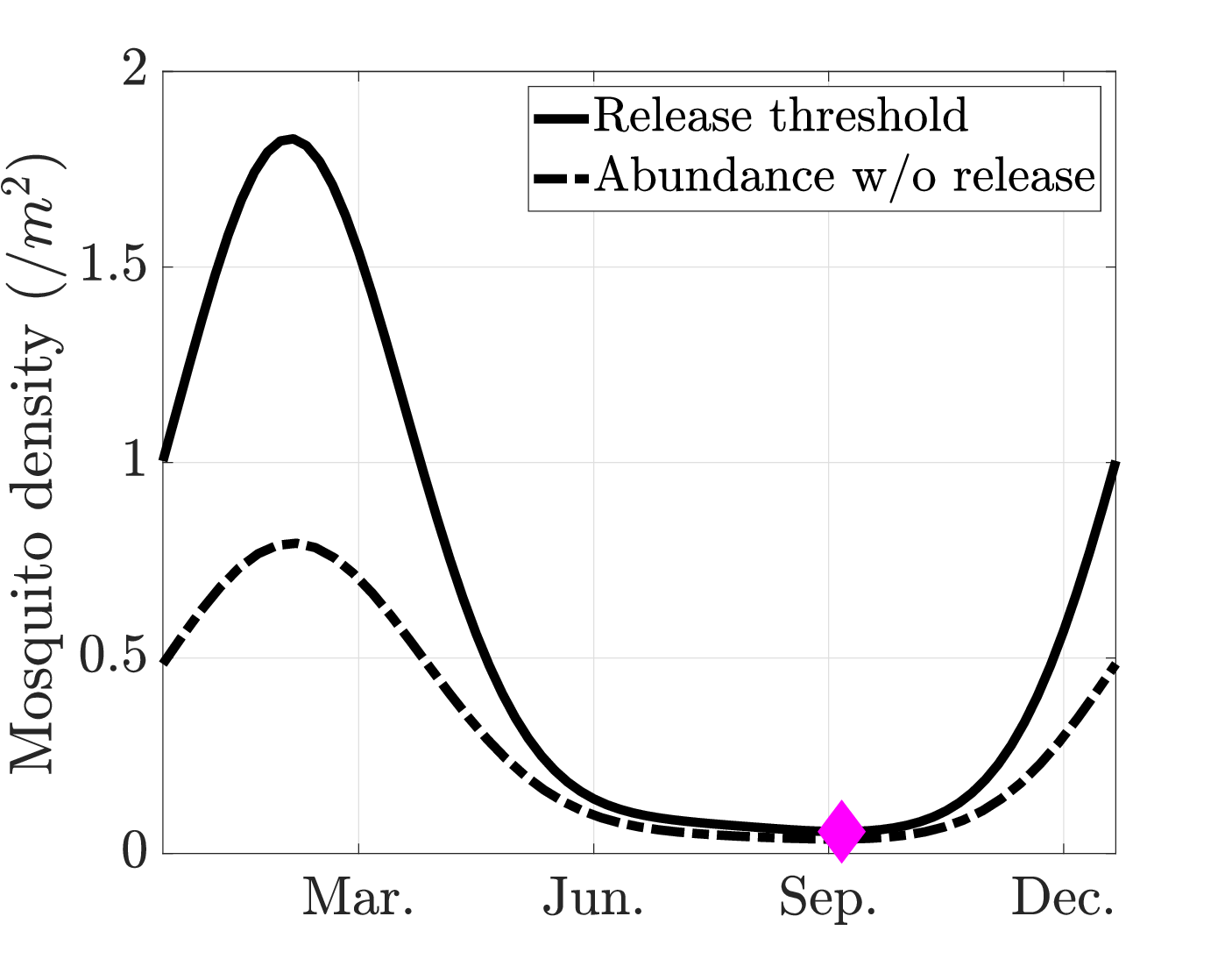}{D}
\caption{Impact of seasonality on \W population replacement threshold. \textit{Top row}: seasonal profiles for Cairns, Australia (left) and Yogyakarta, Indonesia (right). \textit{Bottom row:} Given the climate profile above, \W threshold when being deployed at different times of the year (solid curve) and the natural abundance of mosquitoes (dashed curve). The minimal threshold level is indicated using a magenta diamond marker.}
\label{fig:seasonality}
\end{figure}

The prior simulations assumed constant model parameters, representing minimal or no variation in environmental conditions. However, mosquito life traits are highly sensitive to environmental factors, such as temperature and rainfall \cite{reinhold2018effects,foster2002mosquitoes}. To explore how these variations may influence \W releases, we incorporate seasonality into the model using time-varying parameters based on lab and environmental data (details provided in \cref{sec:method_season}). We focus on two case studies: Cairns, Australia, and Yogyakarta, Indonesia, where \W releases have been successfully implemented \cite{hoffmann2011successful,indriani2020reduced}. These locations feature distinct seasonal profiles (\cref{fig:seasonality}, top row), offering complementary scenarios to evaluate the impact of seasonality on release strategies. 

Yogyakarta has a tropical monsoon climate with moderate rainfall variation, influenced by the monsoon, and little temperature variation, averaging around 26$^\circ$C year-round (\cref{fig:seasonality}A). The corresponding seasonal variation in natural mosquito abundance (without \W release) follows a similar trend (\cref{fig:seasonality}C, dashed curve), with low mosquito season correlated with the low rainfall season. Simulations of different release timings (\cref{fig:seasonality}C, solid curve) indicate that the optimal release window for minimizing the \W release threshold occurs around September (indicated by magenta diamond marker) when natural mosquito abundance is at its lowest.

Cairns also has a tropical monsoon climate but with more pronounced seasonal variations in both temperature and rainfall (\cref{fig:seasonality}B). Notably, it features a long dry season from June to September with minimal rainfall. The mosquito abundance curve reflects this pattern (\cref{fig:seasonality}D, dashed curve). Simulations of release timing (\cref{fig:seasonality}D, solid curve) show that the optimal release timing occurs at the end of the low mosquito season and just before the natural mosquito abundance begins to rise. This timing minimizes the competition from wild-type mosquitoes and takes advantage of the increasing mosquito abundance trend, which creates a favorable breeding condition for mosquitoes in general and will provide an opportunity for the \Wns-infected mosquitoes to efficiently outgrow the wild-type cohort. 

\section{Discussion}
Our multistage spatial model captures the complex maternal transmission dynamics of \W among mosquitoes across life stages as well as the spatial dispersal dynamics of adult mosquitoes. Based on the parameterization for the wMel strain of \W in \textit{Aedes aegypti} mosquitoes, we numerically determine the critical threshold, termed the ``critical bubble'', in a 2-D spatial field necessary for the establishment of \W infection upon a local release. Sensitivity analysis of various metrics related to the \W threshold and wave propagation reveals the influence of mosquito life traits at different stages (\cref{fig:SA_PRCC}), and simulations on pre-release mitigation and release strategies inform efficient \W release design. 

Among the parameters analyzed, the maternal transmission rate is identified as the most impactful parameter for both \W establishment and propagation. Reproduction parameters, including female fecundity and lifespan, and egg survivorship) also play an important role in determining the threshold level at the release center and the propagation velocity of the infection wave once established. Meanwhile, the diffusion coefficients and life traits of male mosquitoes primarily influence the shapes of the threshold bubble and the propagating wave after establishment. 

Our results on integrating \W with various pre-release mitigation approaches suggest that the most effective strategy targets the adult-stage mosquitoes, such as through adult spraying (\cref{fig:heatmap_mitigation}). This approach can substantially reduce the release number needed for \W invasion. In contrast, pre-release mitigation targeting aquatic-stage mosquitoes has a much smaller impact. This highlights that, as a population replacement strategy, \Wns-infected mosquitoes out-compete the wild cohort primarily due to their reproduction advantage during the adult stages (via CI). Removing competition from wild adult mosquitoes therefore efficiently facilitates the replacement process. 

Increasing the mitigation radius targeting adult mosquitoes reduces the threshold release number, however, its effectiveness eventually saturates. Moreover, there is a consistent optimal release radius of approximately 200 meters under our parameterization, regardless of the mitigation radius, assuming the same mitigation and release centers.

We also found that pre-release habitat modifications reducing the carrying capacity for aquatic-stage mosquitoes can increase the threshold for \W establishment. While such mitigations are often effective for mosquito population suppression programs, they may have adverse effects in the context of \W population replacement. Reducing breeding sites increases competition among mosquitoes, which can hinder the establishment of \Wns-infected cohorts. Conversely, increasing breeding sites may lower the threshold by providing an opportunity for \Wns-infected mosquitoes to establish dominance over the wild-type population.

Our simulation results also show that in the presence of spatial heterogeneity, with uneven carrying capacity across regions, releasing \W in the dry region lowers the threshold compared to releases in the wet region, where wild mosquito abundance is higher. However, the spatial propagation of the established \Wns-infection wave may slow down or even stagnate when crossing the dry-wet transition zone, depending on the disparity in carrying capacities.

Seasonal climate profiles can greatly impact the threshold release number for \W establishment. To account for this, we incorporated the seasonal variations in temperature and rainfall by using time-dependent parameters. We examined two case studies: Yogyakarta, Indonesia with mild seasonality, and Cairns, Australia with strong seasonality. In both scenarios, the optimal release time for minimizing the release threshold occurs just before the wet season starts, when competition from wild mosquitoes is low. Additionally, as natural abundance starts growing afterward, which indicates that the environment becomes more favorable for mosquito breeding, it creates an opportunity for the released mosquito to take over.

Our modeling results rely on several underlying assumptions. First, we assumed equal and constant diffusion coefficients for female and male mosquitoes, based on the field studies, where mark-release-recapture methods found no statistically significant difference in the mean distance traveled (MDT) per day between females and males \cite{kay1998aedes,harrington2005dispersal}. However, other studies suggest males may exhibit longer daily MDT than females \cite{tsuda2001movement}. For example, in \cite{juarez2020dispersal}, male \textit{Aedes aegypti} showed higher dispersal than females, and unfed females dispersed slightly farther than gravid females. Moreover, the reported MDT values vary widely in the literature and are strongly influenced by environmental factors, such as the spatial distribution of houses near the release site \cite{tsuda2001movement}, wind patterns \cite{garrett-jones1950dispersion}, and seasonal climatic conditions during collection \cite{mains2019localizeda}. These complexities introduce uncertainty in parameterizing mosquito dispersal. A more detailed model could account for varying diffusion coefficients for males and females, as well as incorporate environmental heterogeneity to better capture the dynamics of mosquito movement. 

Our model employs a simple diffusion formulation to describe mosquito dispersion dynamics. This corresponds to a Gaussian dispersal kernel, where flight distances follow a Gaussian distribution. As a result, mosquitoes rarely make long-distance flights. However, non-Gaussian dispersal distributions may better capture the spatial dynamics of mosquito movement, and different spatial dispersal patterns could impact the outcomes. For instance, existing studies have explored the effects of fat-tailed dispersal distributions \cite{schofield2002spatially, turelli2017deploying}. They observed a substantial impact on wave velocity. In \cite{chan2013modelling}, a slow-fast approach was implemented to switch between dispersal modes to match the observations for mark-recapture studies. Our modeling framework could be adapted to incorporate these more realistic spatial formulations, potentially improving its accuracy and applicability.

\section{Detailed parameterizations and methods}\label{sec:param_detail}
\subsection{Mosquito-related parameters} 
All model parameter descriptions, baseline estimates, and ranges are summarized in \cref{tab:parameter_all}. To parameterize the model, we referenced biological studies on the life traits of \textit{Aedes aegypti} mosquitoes, the impact of \W infection on these traits, and recent \W field releases (primarily in Australia). Recognizing that certain parameters are highly influenced by local environmental conditions, we incorporated data from both laboratory experiments and field trials for more informed and context-specific estimates.

\paragraph{Adult lifespan}
In laboratory settings, the lifespan of female \textit{Aedes aegypti} mosquitoes can reach up to 70 days \cite[Figure S6]{walker2011wmel}. However, for field conditions, we adopted a more conservative estimate of approximately two weeks (17.5 days) \cite{nationalenvironmentagency2020aedes,zettel2009yellowa}. We assumed that the male lifespan to be about 1.5 weeks, roughly 60\% of the female lifespan \cite{walker2011wmel,styer2007mortality}.

For \Wns-infected adults, a reduction in longevity of approximately 10\% has been reported for infected females \cite[Supplementary Figure S6]{walker2011wmel}. Accordingly, we set the infected female lifespan to $1/\mu_{fw}=0.9/\mu_{fu}$, while assuming no difference in male longevity, i.e., $\mu_{mu}=\mu_{mw}$.

\paragraph{Fecundity}
In laboratory cage experiments, no significant differences in mean fecundity were observed between infected and uninfected cohorts, with both producing approximately 105 eggs per female \cite[Figure S4]{walker2011wmel}. However, from the field release estimates \cite{hoffmann2014stability}, there is a reduction in the number of eggs laid by 
\Wns-infected females compared to uninfected females, with uninfected females producing approximately 65 eggs per female and infected females producing 55 eggs per female.

Using these field estimates and dividing by the corresponding female lifespans, we estimate egg-laying rates of approximately 3.7 eggs/day for uninfected females (65/17.5) and 3.5 eggs/day for the infected females (55/15.8). These values are consistent with ranges observed in laboratory experiments \cite{styer2007mortality}.

\paragraph{Eggs}
Egg death rates and hatching rates (number of eggs hatched per time unit) are not directly reported in the biological literature. Instead, we rely on estimates of egg viability (the fraction of eggs that become larvae, $\delta/(\delta+\mu_{E}$), and the egg hatching period ($1/\delta$).

In laboratory studies, wMel-infected females mated to both uninfected and wMel-infected males showed embryo hatch rates of approximately 90\% \cite[Table S1]{walker2011wmel}, with no significant differences observed between the mean hatch rates of infected and uninfected lines \cite[Figure S5]{walker2011wmel}. However, in field releases, egg hatching rates were reported as 85\% for uninfected eggs (range: 75\% $\sim$ 86\%) and 73\% for infected eggs (range: 60\% $\sim$ 75\%) \cite{hoffmann2014stability}, suggesting that $\mu_{eu}\neq \mu_{ew}$. Using these values, we establish the following relations:
$$
\frac{\delta_u}{\delta_u+\mu_{eu}}=0.85,\quad \frac{\delta_w}{\delta_w+\mu_{ew}}=0.73.
$$ 
The egg hatching period is approximately 2 days \cite{soares-pinheiro2016eggs}, with a range of $1 \sim 4$ days, depending on temperature \cite{foster2002mosquitoes}. Thus, we assume $\delta_u=\delta_w=\delta=1/2$, and using the relation above, we estimate egg death rates as $\mu_{eu} = 0.088$ (range: $0.08 \sim 0.17$), and $\mu_{ew}=0.185$ (range: $0.17 \sim 0.33$).

Overall, the total number of viable eggs produced is approximately 55.25 eggs/female for uninfected mosquitoes (65 eggs/female $\times 85\%$) and 40.15 eggs/female for infected mosquitoes (55 eggs/female $\times 73\%$). This represents a fitness cost in infected female fecundity of about $27\%$, consistent with estimates in the literature (24\%, \cite{walker2011wmel,hoffmann2014stability}).

\paragraph{Larvae}
The mortality rate during the larvae/pupae stage was estimated as $\mu_l\sim 0.12$/day in \cite{yang2011follow}. The wMel infection does not appear to affect the development time \cite{walker2011wmel}, so we assume the development time from larvae to adults is approximately 10 days under field conditions \cite{yang2011follow}, with a broad range of $8 \sim 30$ days, depending on food availability and temperature \cite[Table S2]{walker2011wmel,foster2002mosquitoes}.

\paragraph{\W transmission}
We assumed a complete cytoplasmic incompatibility (CI) effect \cite[Table S1]{walker2011wmel,hoffmann2014stability} and perfect maternal transmission, $v_w=1$, with a range of $0.89 \sim 1$ based on reported values \cite{walker2011wmel,hoffmann2014stability}.

\paragraph{Mating}
We assumed equal male mating competitiveness between the infected and uninfected males \cite{segoli2014effect} and further assumed a homogeneous mixing of males at a given location and time. Under these assumptions, the probability of a female mating with an uninfected male is given by $M_u/(M_u+M_w)$.

\subsection{Estimate on diffusion coefficients} 
The diffusion coefficients for male and female mosquitoes are assumed to be the same, as their daily dispersion distances are not statistically different \cite{kay1998aedes}. Consider the 2-D diffusion equation $u_t = D\Delta u (D>0)$ and its solution to the initial value problem:
$$
u(\pmb{x},t)= \int_{\mathbb{R}^2}\frac{1}{4\pi D t}e^{-\frac{|\pmb{x}-\pmb{y}|^2}{4Dt}}u(\pmb{y},0)d\pmb{y},~~ \pmb{x}\in \mathbb{R}^2,~~ t>0,
$$
which can be rewritten as
$$
u(\pmb{x},t+\Delta t) = \int_{\mathbb{R}^2} k(\pmb{x}-\pmb{y},\Delta t)u(\pmb{y},t) d\pmb{y},
$$
where
$$
k(\pmb{x},\Delta t) = \frac{1}{4\pi D\Delta t }e^{-\frac{|\pmb{x}|^2}{4D\Delta t}} \sim \mathcal{N}_2\left(0,\begin{bmatrix}
2D\Delta t & 0 \\
0 & 2D\Delta t
\end{bmatrix}\right).
$$
We assume mosquito dispersal per unit time $\Delta t$ follows the heat kernel $k(\pmb{x},\Delta t)$ \cite{turelli2017deploying,silva2020modeling}, 
which is a two-dimensional Gaussian distribution with standard deviation $\sigma_{\Delta t} = \sqrt{2D \Delta t}$. To estimate the diffusion coefficient $D$, we match the mean absolute deviation (MAD) of the distribution, $E_{\Delta t} = \sigma_{\Delta t}\sqrt{\pi/2} =\sqrt{\pi D \Delta t}$, to the mean distance traveled (MDT) measured in biological studies \cite{morris1991measuringa}.

We refer to two field studies in Australia: one in Pentland \cite{kay1998aedes}, a rural inland town in Australia with a drier climate, and another in Cairns \cite{russell2005markreleaserecapture}, a coastal city with more humid weather. Both studies use the standard mark-release-recapture methodology. In the Pentland study, the daily MDT was estimated to be 25m/day. Using $E_1=\sqrt{\pi D}=25$, we calculate $D = 200$ m$^2$/day. Using the weekly MDT estimates of 60m/week, $E_7=\sqrt{\pi D\times 7}=60$, gives $D=160$m$^2$/day. 

In the Cairns study, daily MDT estimates were not available, but MDT values were reported at different time points, $E_5 = 70~(D\approx 310)$, $E_8 = 55~(D\approx120)$, $E_{11} = 140$ $(D\approx 570)$, $E_{15} = 100~(D\approx210)$, with the capture number decreasing in time. The overall MDT (within 15 days) was estimated to be $E_{15}^{all} = 78~(D\approx 130)$. The lower MDT in Cairns compared to Pentland may reflect shorter flights in the humid environment, where mosquitoes are less pressured to seek distant oviposition sites. 

Daily dispersal gives a more reliable characterization of mosquito movement, as it is less likely to be impacted by mosquito lifespan. Based on the Pentland data and MDT values from early time points in Cairns, we adopt $D = 200$ m$^2$/day as the baseline value. To capture the significant variability in the mosquito dispersal due to environmental and climatic factors \cite{kay1998aedes}, we use a broad range of $100\sim 300$ m$^2$/day for $D$ in our sensitivity analyses.

\subsection{Seasonality} \label{sec:method_season}
We incorporate the impact of seasonality by introducing time-varying parameters for mosquito life trait and carrying capacity $K_l$, which depends on the seasonal temperature ($T$) and rainfall ($R$). The governing equations are as follows:
\begin{equation}
\begin{aligned}\label{eq:PDE8_seasonal}
(E_u)_t&= \phi_u(T)\frac{M_u}{M_u+M_w} F_u + (1-v_w)\phi_w(T)F_w - \delta(T) E_u - \mu_{eu}(T) E_u, \\
(E_w)_t&= v_w \phi_w(T) F_w - \delta(T) E_w - \mu_{ew} (T)E_w,\\
(L_u)_t&= \delta(T)E_u \left(1-\frac{L_u+L_w}{K_l(R)}\right) - \psi(T) L_u - \mu_l(T) L_u,\\
(L_w)_t&= \delta(T) E_w\left(1-\frac{L_u+L_w}{K_l(R)}\right) - \psi(T) L_w - \mu_l(T) L_w,\\
(F_u)_t&= b_f \psi(T) L_u -  \mu_{fu}(T) F_u+D_1\Delta F_u,\\
(F_w)_t&=b_f\psi(T) L_w - \mu_{fw}(T)F_w+D_2\Delta F_w,\\
(M_u)_t&= b_m\psi(T) L_u - \mu_{mu}(T)M_u+D_3\Delta M_u,\\
(M_w)_t&= b_m \psi(T) L_w - \mu_{mw}(T) M_w+D_4\Delta M_w.
\end{aligned}
\end{equation}
The seasonal profiles for temperature and rainfall are presented in \cref{sec:climate}, while the functional dependencies of parameters on these climate factors are detailed in \cref{sec:climate2}.

\subsubsection{Climate profiles}\label{sec:climate}
\begin{figure}[htbp]
\centering
\includegraphics[width=0.5\linewidth]{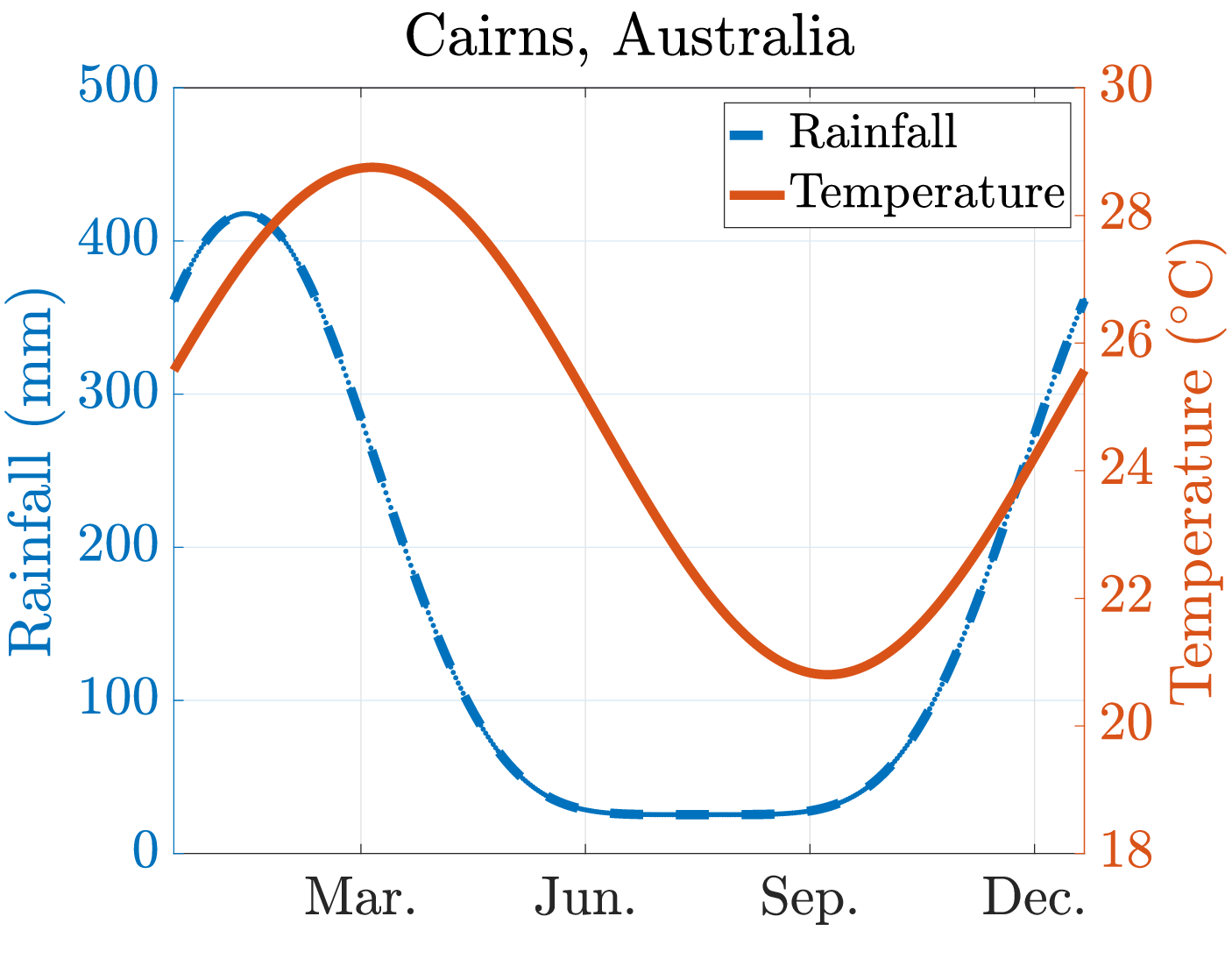}\includegraphics[width=0.5\linewidth]{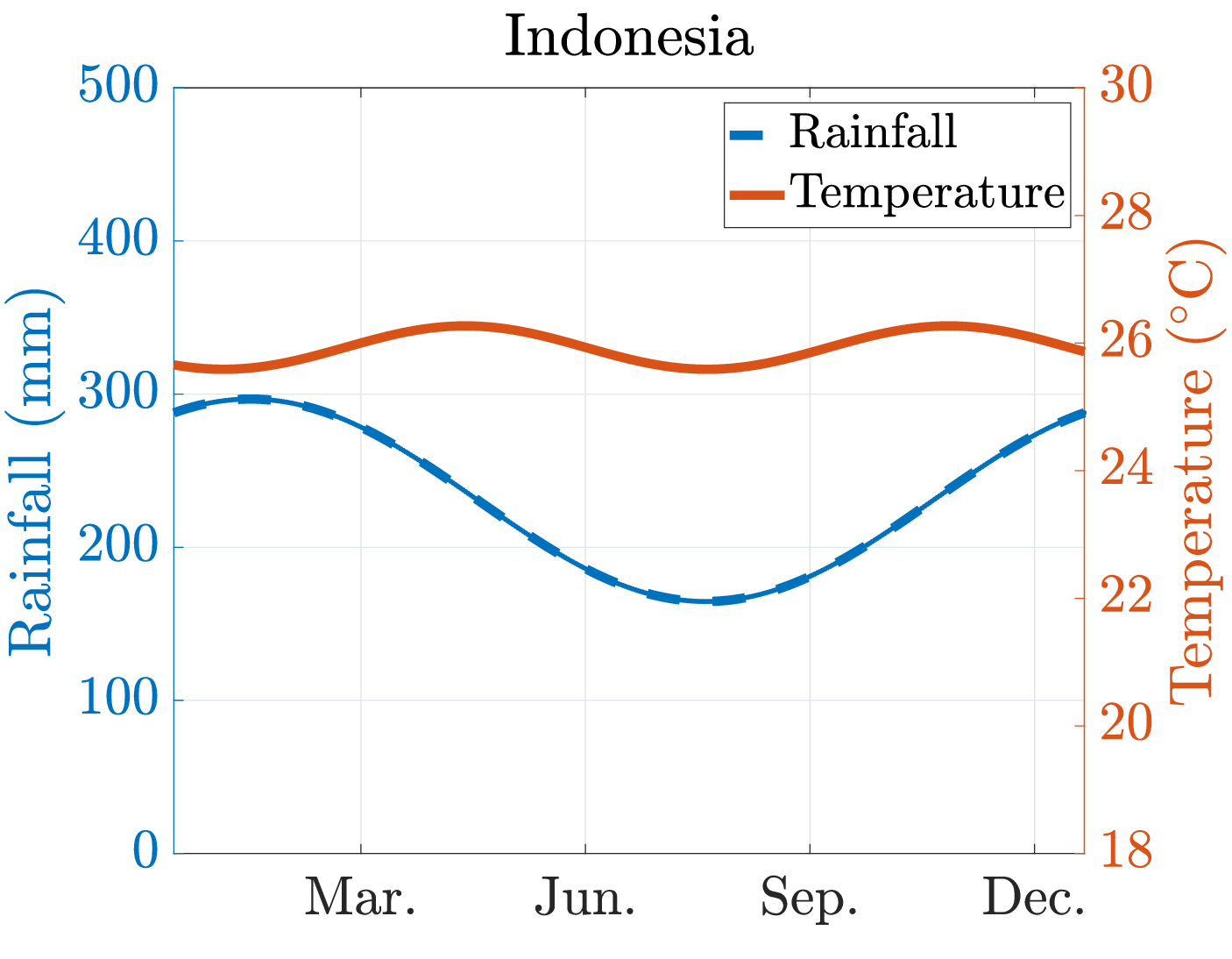}
\caption{Fitted seasonal rainfall and temperature curves for Cairns, Australia (left) and Yogyakarta, Indonesia (right). The corresponding fitted functions are provided in \cref{tab:seasonal_climate}.}
\label{fig:seasonality_fit}
\end{figure}

We analyze climate profiles from two regions, Cairns, Australia, and Yogyakarta, Indonesia. For Cairns, data was obtained from the National Oceanic and Atmospheric Administration (NOAA)'s Climate Data Online \cite{nationaloceanicandatmosphericadministrationnoaanoaa}, using records from the CAIRNS POST OFFICE, AS, weather station. The data includes the average hourly temperature and precipitation for the city, which were aggregated into monthly measurements for curve fitting. For Indonesia, climate data was sourced from the Climate Change Knowledge Portal (CCKP) \cite{worldbankgroup2021climate}, using monthly average values over the period 1991 - 2022. We fitted periodic functions with a one-year period to the temperature and rainfall data using a nonlinear least square algorithm (\texttt{fminsearch} in MATLAB). The resulting fitted functions are given in \cref{tab:seasonal_climate} and plotted in \cref{fig:seasonality_fit}.

\begin{table}[htbp]
\begin{widepage}
\centering
\caption{Fitted functions for seasonal climate profiles. The corresponding plots are presented in \cref{fig:seasonality_fit}.}
\begin{tabular}{lll}
\toprule
& Cairns, Australia & Indonesia \\
\midrule
Temperature &  $3.98\sin(2\pi t/365+0.20)+24.78$&  $-0.34\sin(2\pi 1.88 (t/365+0.08))+25.93$ \\
Rainfall & $0.03((1.47\sin(2\pi(t/365+0.17))+1.47)^{2.54}+1)$ & $-0.07\sin(2\pi(t/365+0.67))+0.23$\\
Carrying capacity & $0.06((1.47\sin(2\pi(t/365+0.17))+1.47)^{2.54}+1)$ & $-0.40\sin(2\pi(t/365+0.67))+1.40$\\
\bottomrule
\end{tabular}
\label{tab:seasonal_climate}
\end{widepage}
\end{table}

\subsubsection{Impact on mosquito life trait parameters}\label{sec:climate2}
\begin{figure}[htbp]
\centering
\insertfigdefault{0.48}{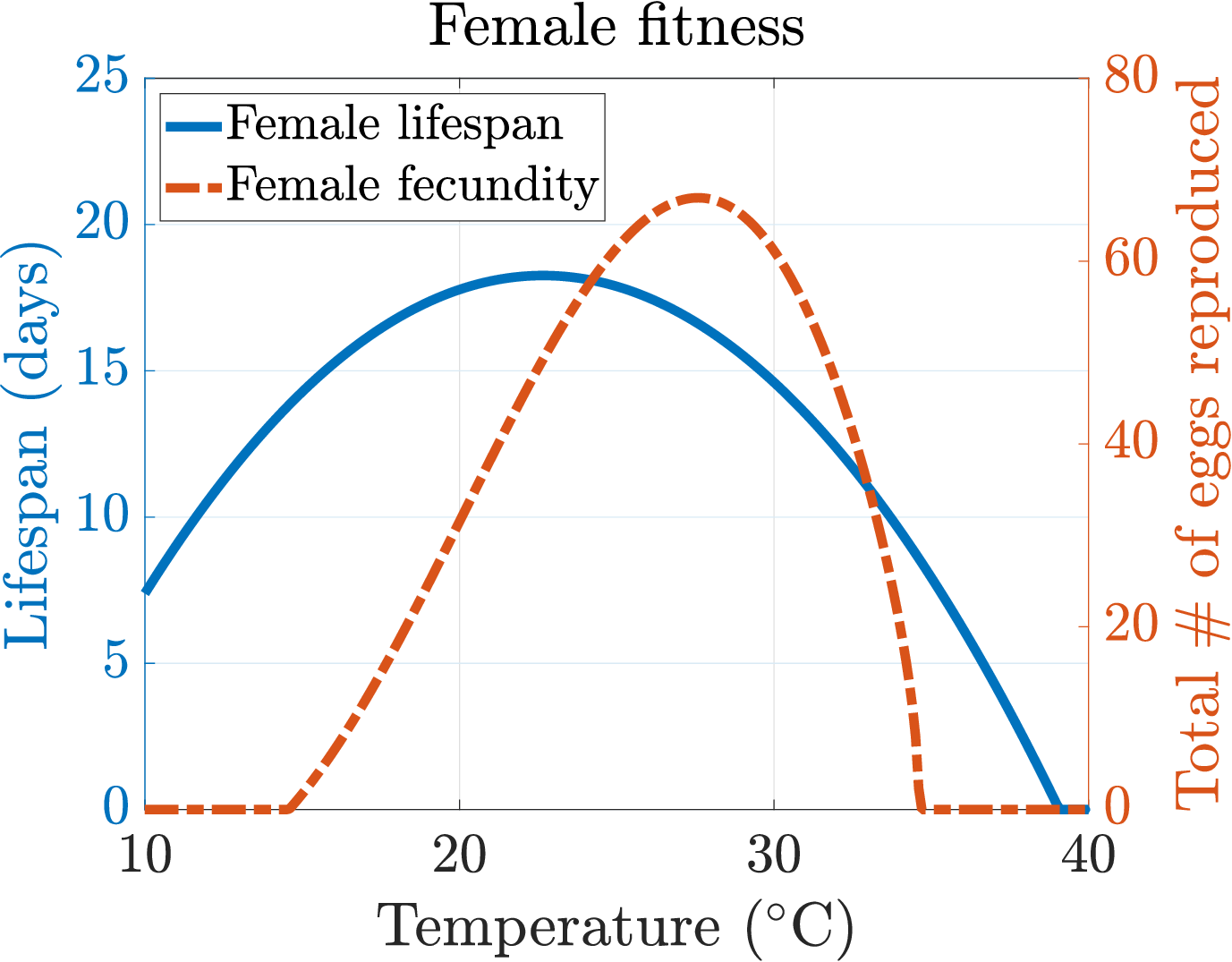}{A}
\insertfigdefault{0.48}{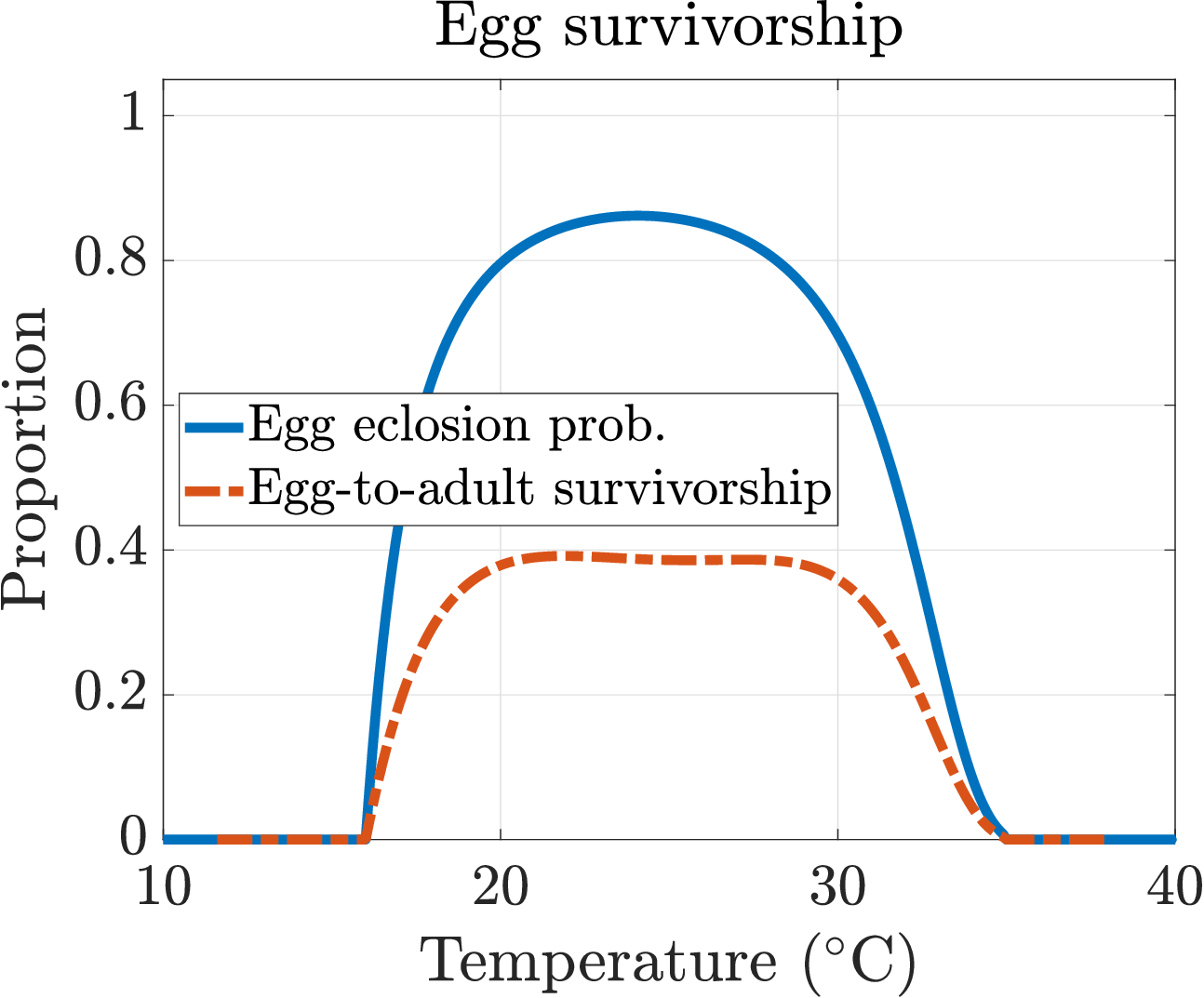}{B}
\insertfigdefault{0.48}{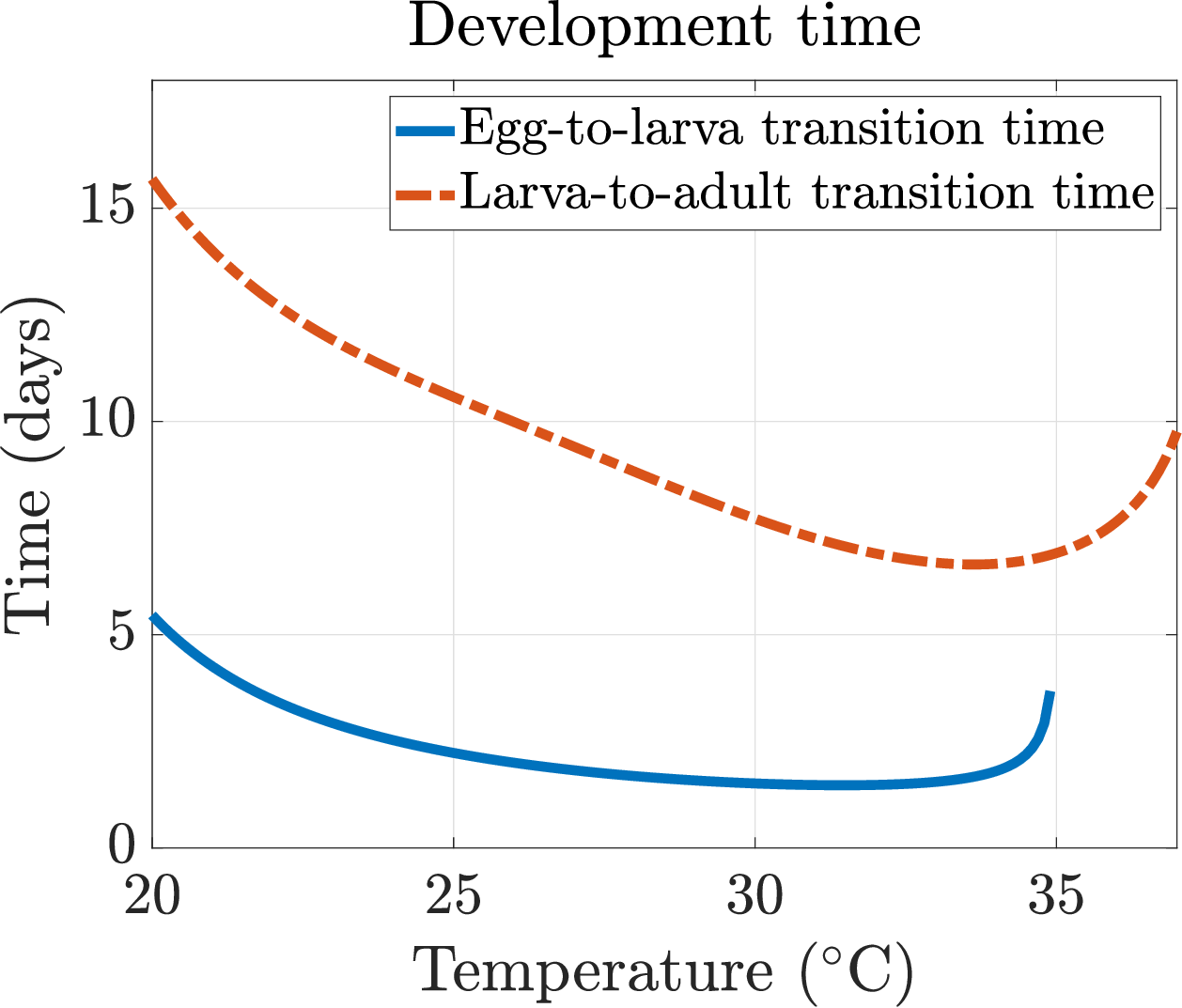}{C}
\insertfigdefault{0.48}{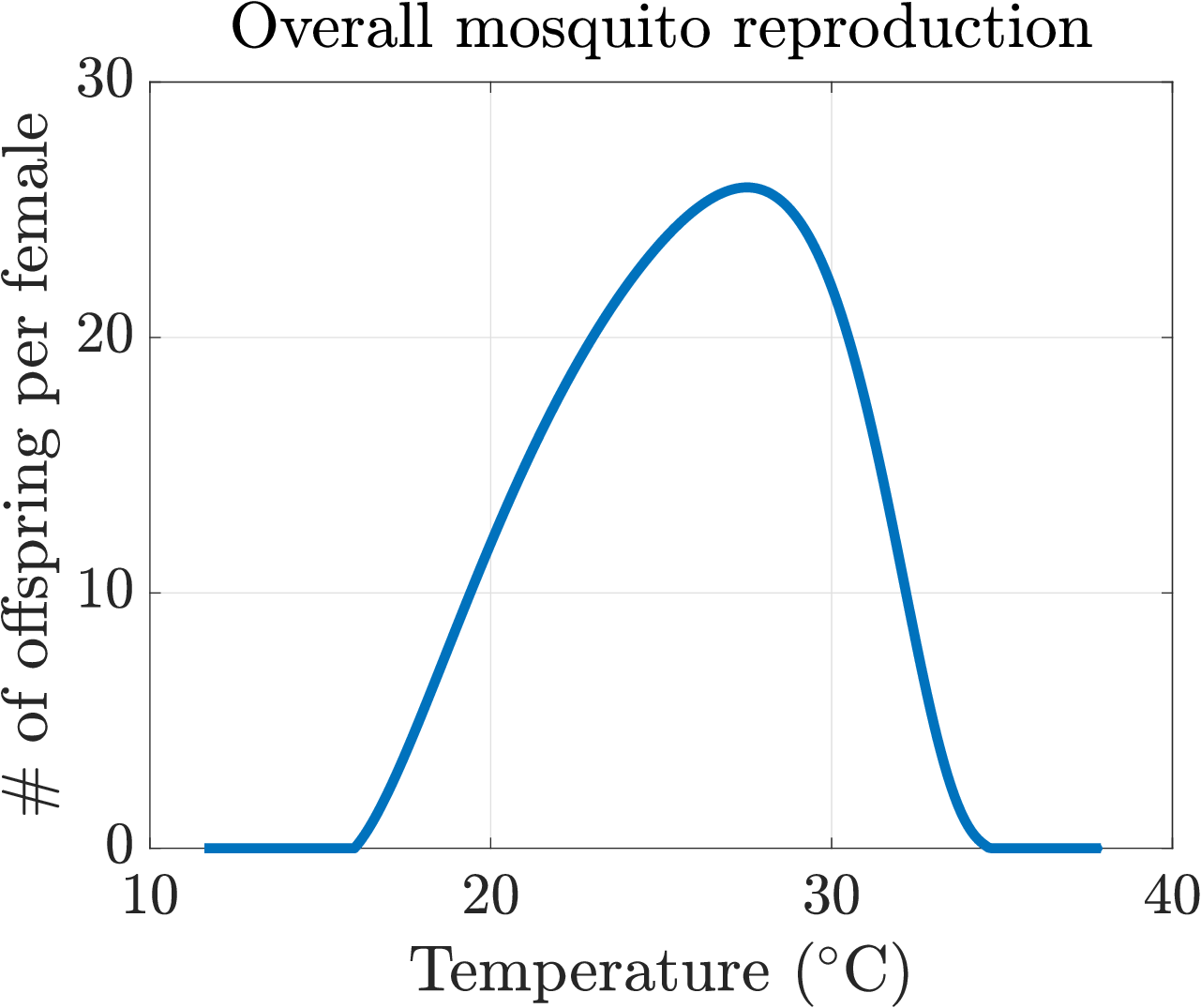}{D}
\caption{Temperature-dependent mosquito life traits. (A) Female fitness: lifespan in days ($1/\mu_{fu}$) and number of eggs reproduced per female ($\phi_u/\mu_{fu}$). (B) Egg survivorship: proportion of eggs that eclose ($\frac{\delta}{\delta+\mu_{eu}}$), and proportion of eggs surviving to adulthood ($\frac{\delta\psi}{(\delta+\mu_{eu})(\psi+\mu_l)}$). (C) The development time for juvenile stages: egg-to-larva transition time ($1/\delta$) and larva-to-adult transition time ($1/\psi$). (D) Overall mosquito reproduction, defined as the number of offspring that survive to adulthood per female mosquito: 
$$\frac{\phi_u}{\mu_{fu}} \frac{\delta\psi}{(\delta+\mu_{eu})(\psi+\mu_l)}.$$ }
\label{fig:seasonal_temp}
\end{figure}

For mosquito life traits, we utilized temperature-dependent measurements from the literature \cite{mordecai2017detecting,farnesi2009embryonic,yang2011follow} (top section of \cref{tab:params_season}). These quantities were transformed into the model parameters and rescaled to align with baseline estimates at $26^\circ$C (\cref{tab:parameter_all}), the bottom section of \cref{tab:params_season}. The biological relevant quantities are visualized in \cref{fig:seasonal_temp}.

\begin{table}[htbp]
\begin{widepage}
\caption{Temperature dependence functions for mosquito life traits, transformed into model parameters and rescaled to match the baseline values in \cref{tab:parameter_all} at $26^\circ$C.
$^\dag$ For each temperature dependence function $f(x)$, the actual value used is $\max\{f(x),0\}$, to ensure positivity.}
\label{tab:params_season}
\centering
\begin{tabular}{llp{8cm}ll}
\toprule
& Parameters & Temperature dependence$^\dag$ & Baseline & Reference\\
\midrule
$s_{P\to A}$& Pupae $\to$ adult trans. rate & $1.32 \times 10^{-9}T^8 - 2.49\times 10^{-7}T^7 + 1.99\times 10^{-5} T^6 - 8.80 \times 10^{-4}T^5 + 0.024 T^4 - 0.39 T^3 + 3.88 T^2 - 21.33 T +  49.51$ & & \cite{yang2011follow}\\
$\mu_P$ & Pupae mortality rate & $4.40\times10^{-7}T^3+ 7.06\times10^{-4}T^2 -0.033T + 0.43 $& & \cite{yang2011follow}\\
$\mu_L$ & Larvae mortality rate & $7.503\times 10^{-6}T^4-7.54\times 10^{-4}T^3+2.74\times 10^{-2}T^2-0.42T+2.32$ &  & \cite{yang2011follow}\\
$s_{L\to P}$&  Larvae $\to$ pupae trans. rate& $1.55 \times 10^{-9}T^7 - 2.62 \times 10^{-7}T^6 + 1.80 \times 10^{-5}T^5 - 6.46 \times 10^{-4}T^4 + 0.013T^3 - 0.15T^2 + 0.83T - 1.85
$&  & \cite{yang2011follow}\\
$\tau_F$ & Unadjusted female lifespan & $-0.11T^2+4.87 T-26.25$ & & \cite{mordecai2017detecting}\\
$\phi^*$ & Unadjusted female egg-laying rate &  $0.01 (T-14.58) T (34.61\, -T)^{0.51}$& & \cite{mordecai2017detecting}\\
$\delta^*$ & Unadjusted hatching rate for eggs & $6.43\times 10^{-4}x(x-16)(35-x)^{0.34}$ & & \cite{farnesi2009embryonic}\\
$\mu_{E}$ & Unadjusted egg mortality rate & $1.15 \times 10^{-4}x(x-12)(36-x)^{3.08}$ & & \cite{farnesi2009embryonic}\\
\midrule
\multicolumn{5}{c}{Transformed and rescaled parameters (used in the main text for seasonality results)} \\
\midrule
$\phi_u$& Egg-laying rate for uninf. females & $\phi^* \times 0.5 $ & 3.7 &\\
$\phi_w$ & Egg-laying rate for infected females &  $\phi^* \times 0.47 $& 3.5 & \\
$\delta$ & Hatching rate for eggs  & $\delta^*\times 1.41 $ & 0.5&\\
$\psi$ & Larvae to adult develop. rate & $1/(1/s_{L\to P}+1/s_{P\to A})\times 0.75$ & 0.1 & \\
$\mu_{eu}$& Death rate for uninf. eggs & $\mu_{E}\times 4.47$ & 0.088 & \\
$\mu_{ew}$& Death rate for infected eggs & $\mu_{E}\times 9.40 $ & 0.185 & \\
$\mu_l$ & Death rates for larvae/pupae stages & $(\psi(s_{L\to P}+\mu_L)(s_{P\to A}+\mu_P)/(s_{L\to P}s_{P\to A})-\psi) \times 1.35$  & 0.12 \\
$\mu_{fu}$ & Death rate for uninf. females & $1/\tau_F\times 1.59$& 1/17.5\\
$\mu_{fw}$ & Death rate for infected females & $1/\tau_F\times 1.76 $& 1/15.8\\
$\mu_{mu}$ & Death rate for uninf. males & $1/\tau_F\times 2.64 $& 1/10.5\\
$\mu_{mw}$ & Death rate for infected males & $1/\tau_F\times 2.64 $& 1/10.5\\
\bottomrule
\end{tabular}
\end{widepage}
\end{table}

\subsubsection{Impact on mosquito breeding sites}
We incorporate the impact of seasonal rainfall variation into the carrying capacity, $K_l(R)$. More sophisticated implementations could also incorporate temperature dependence, as explored in studies such as \cite{valdez2018impact, tran2013rainfall}. In particular, we rescaled the rainfall curves fitted in \cref{sec:climate} so that the maximum values match the baseline carrying capacity (one mosquito per square meter) for each rainfall function. The resulting functions are provided in \cref{tab:seasonal_climate} (row 3).

\section*{Acknowledgment}
ZQ was partially supported by the National Science Foundation award DMS-2316242. The funder had no role in study design, data collection and analysis, decision to publish, or preparation of the manuscript.

\bibliographystyle{plos2015}
\makeatletter
\renewcommand{\@biblabel}[1]{\quad#1.}
\makeatother

\bibliography{References}

\newpage 
\appendix
\setcounter{figure}{0} 
\setcounter{equation}{0} 
\setcounter{table}{0}
\renewcommand\thefigure{\thesection.\arabic{figure}}   
\renewcommand\theequation{\thesection.\arabic{equation}}  
\renewcommand\thetable{\thesection.\arabic{table}}  
\section{Critical bubble under different release shapes}\label{sec:appendixB}
\begin{figure}[htbp]
\centering
\includegraphics[width=0.32\textwidth]{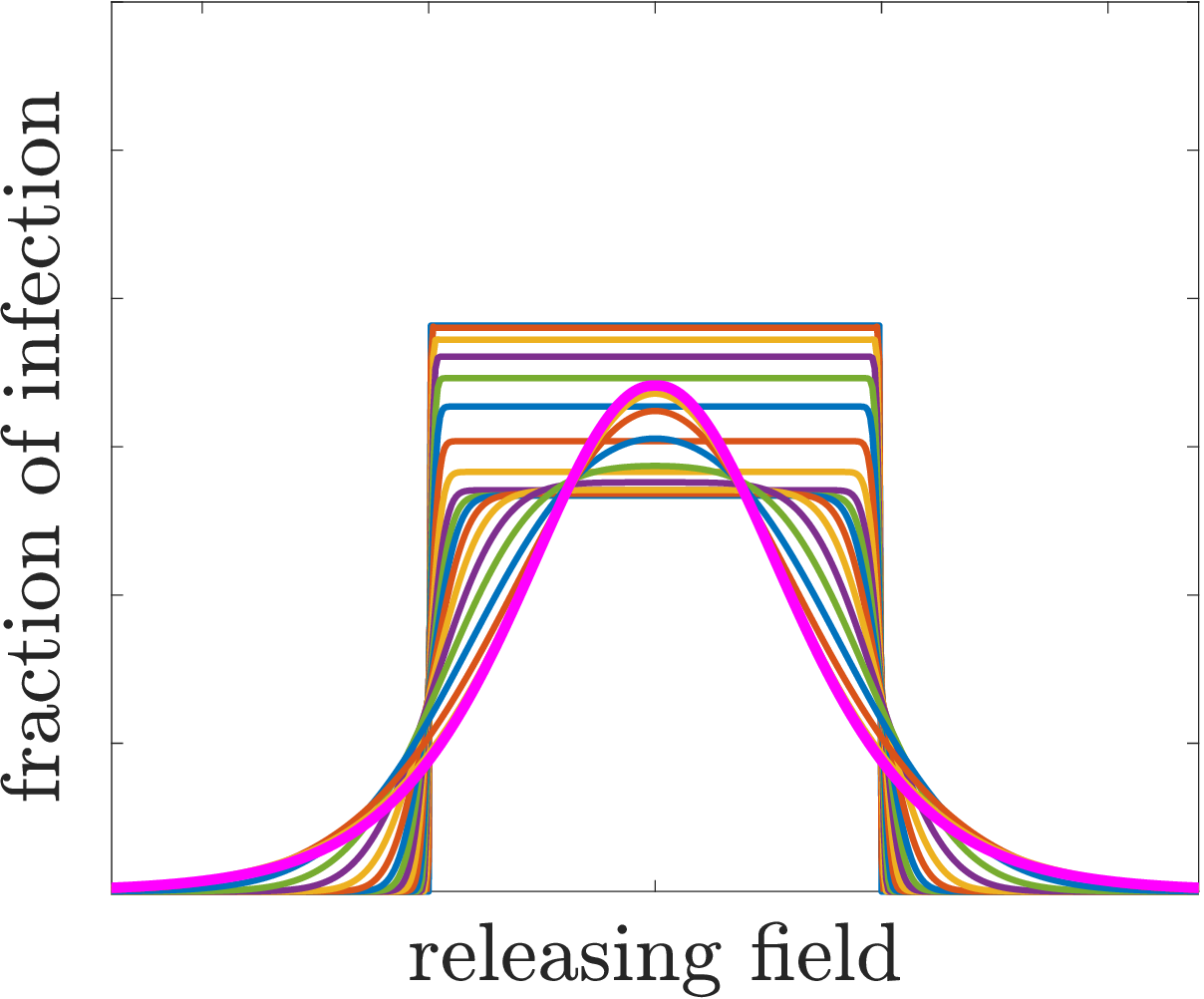}\hfill
\includegraphics[width=0.32\textwidth]{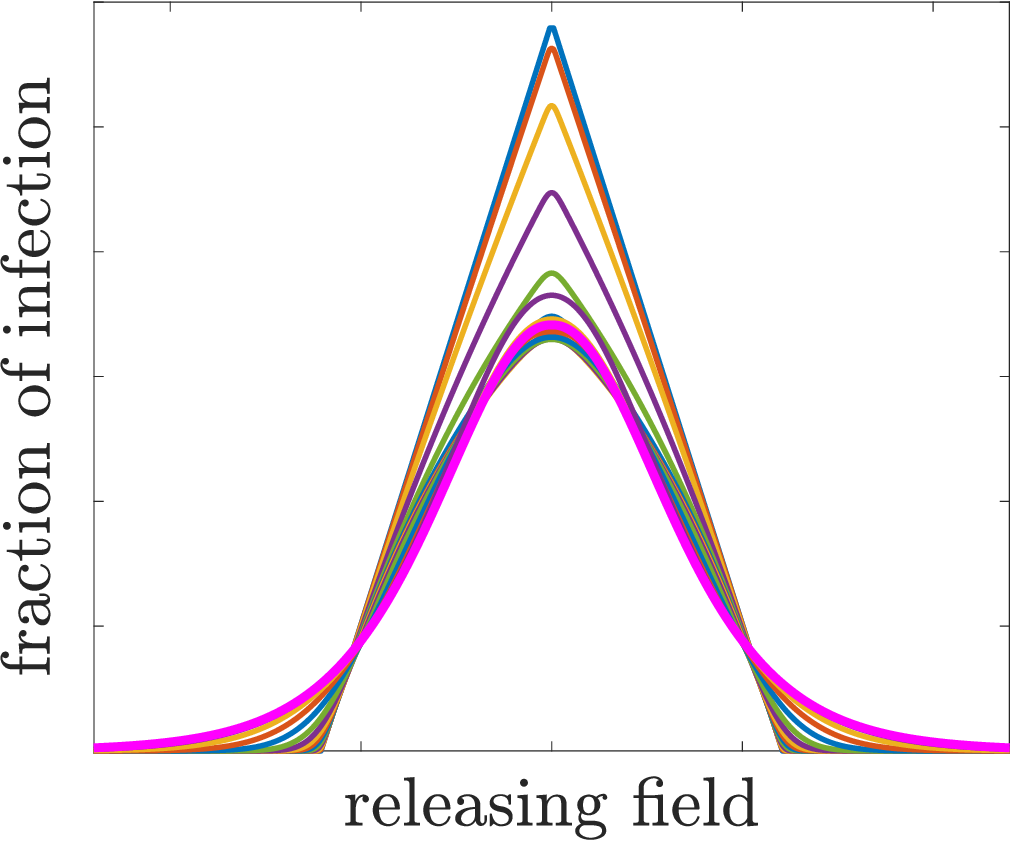}\hfill
\includegraphics[width=0.32\textwidth]{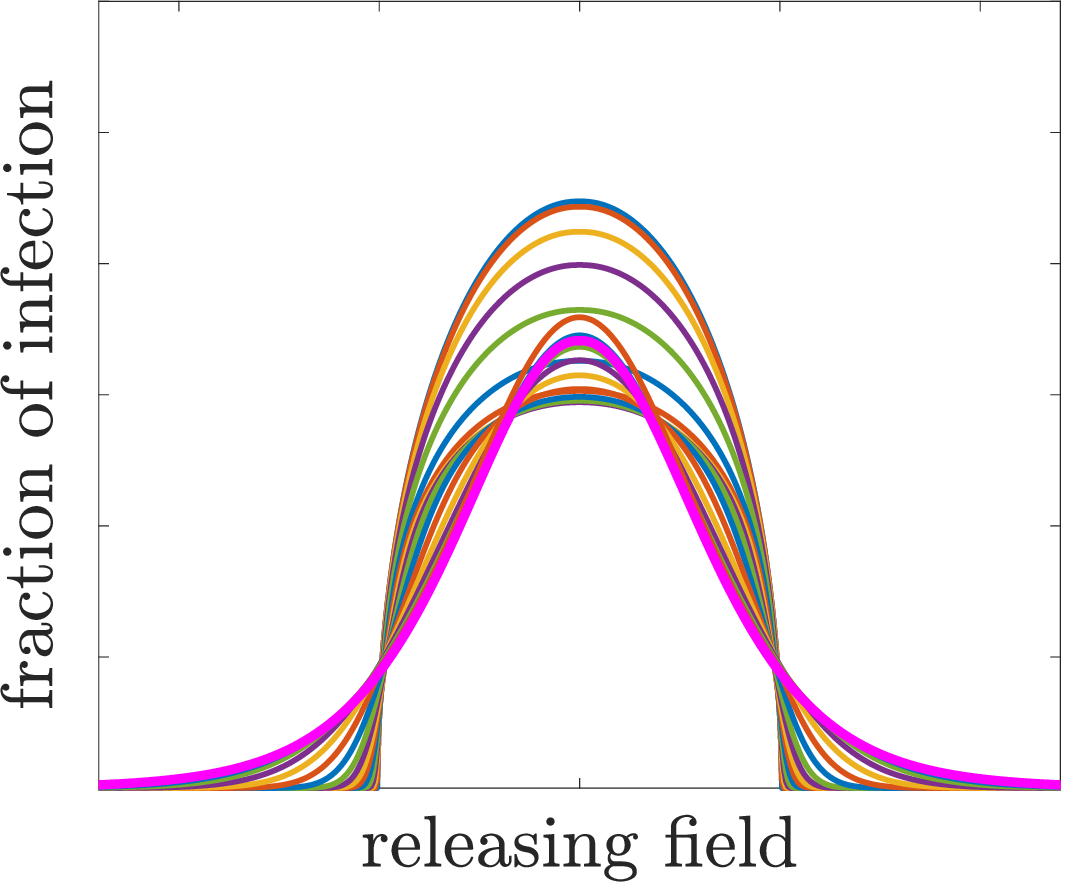}
\caption{Threshold for \W releases with different initial release distributions evolve to the same critical bubble.}\label{fig:bubble_ICs}
\end{figure}

The threshold identification algorithm described in the main text (\cref{sec:threshold2}) can be applied to different initial release distributions, and the resulting critical bubble remains consistent. \Cref{fig:bubble_ICs} illustrates the critical bubbles obtained using various release shapes, including step, triangular, and elliptical distributions. At the corresponding threshold release level, identified by the algorithm, all configurations evolve into the same critical bubble.

\section{Details on sensitivity analysis}\label{sec:appendixC}
Below we provide details on the quantities of interest (QOIs) and the sampling methods for the sensitivity analysis in \cref{sec:SA} of the main text.

\begin{figure}[ht]
\centering
\insertfigclose{0.48}{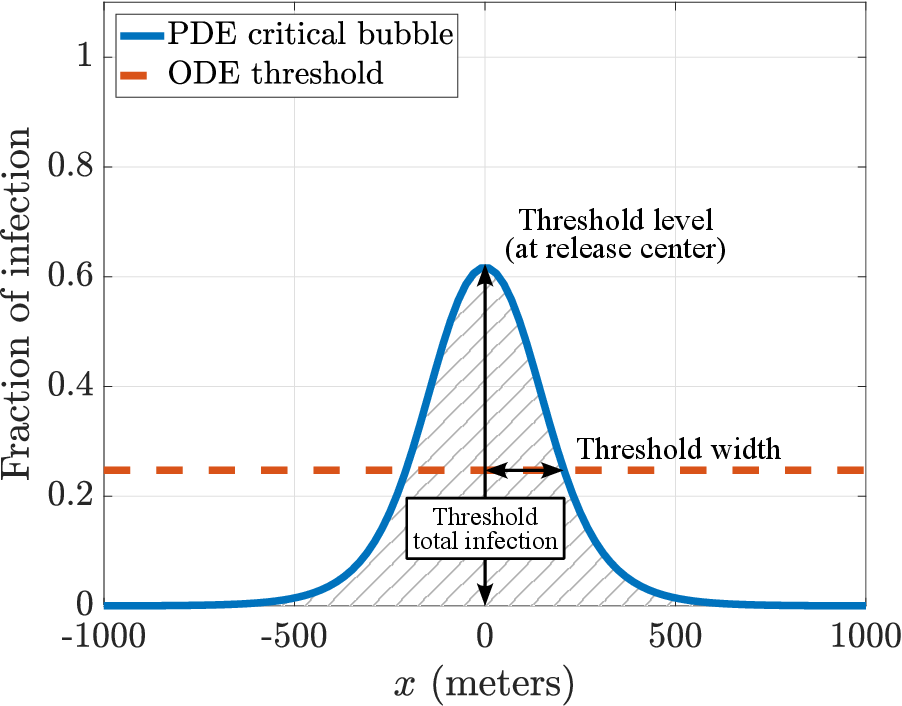}{A}\hfill
\insertfigclose{0.48}{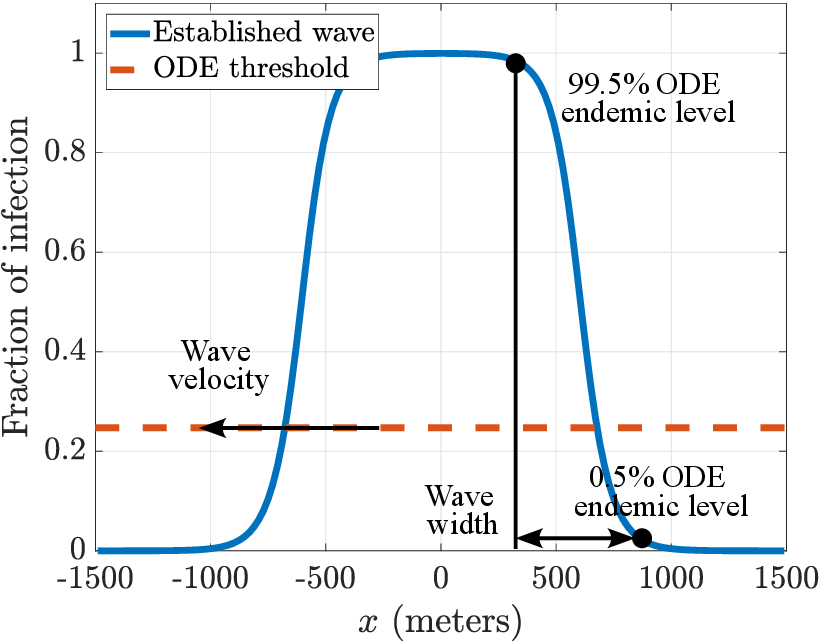}{B}
\caption{Quantities for sensitivity analysis. (A) Threshold-related quantities, including the threshold level, threshold width, and threshold total infection (the shaded area under the curve). (B) Wave-related quantities, including wave velocity (assuming outward propagation) and wave width.}\label{fig:QOIs_SA}
\end{figure}

\subsection{Quantities of interest}\label{sec:app_QOI}
\paragraph{Threshold bubble quantities:} For quantities related to the threshold bubble, we define the fraction of infection in female mosquitoes as: 
$$
p^*(x,y) = \frac{F_w^*(x,y)}{F_u^*(x,y)+F_w^*(x,y)},
$$
where $F_u^*$ and $F_w^*$ are the solutions of uninfected and infected female populations, respectively, of the critical bubble. The infection profile is illustrated in \cref{fig:QOIs_SA}A.
\begin{enumerate}
\item Threshold total infection: This measures the overall fraction of infection within the bubble's coverage, $\Omega \in \mathbb{R}^2$:
$$
\text{QOI}=\int_{\Omega} p^*(x,y) dx dy
$$
\item Threshold level: The fraction of infection at the release center of the critical bubble, QOI $=p^*(0,0)$.
\item Threshold width: This measures the spread of the critical bubble. It is calculated as the radial distance from the release center, (0, 0), to the inflection point of the bubble front (\cref{fig:QOIs_SA}A). At the infection point, the fraction of infection matches the ODE threshold level $p_{ODE}$, derived in \cite{florez2022modeling}, 
\begin{equation}\label{eq:EE}
p_{ODE} = \frac{r_{wu}\frac{1}{\mu_{fw}}}{\frac{1}{\mu_{fu}}+r_{wu}\frac{1}{\mu_{fw}}},     
\end{equation}
where $r_{wu}$ is the smaller root for the following equation
\begin{equation}\label{eq:EEr}
\frac{v_u}{v_w}\frac{\delta+\mu_{ew}}{\delta+\mu_{eu}}r_{wu}^2+\left(\frac{v_u}{v_w}\frac{\delta+\mu_{ew}}{\delta+\mu_{eu}}-1\right)r_{wu}+\frac{1-\mathcal{R}_0}{\mathcal{R}_0}=0,    
\end{equation}
and $\mathcal{R}_0$ is the basic reproduction number as defined in \cref{eq:R0} in the main text. Due to the radial symmetry of the bubble, there are infinitely many inflection points located on a circle. For simplicity, we select the point on the positive $x$-axis, that is $p^*(x_{inf},0)$ and define the threshold width $\text{QOI} = x_{inf}$.
\end{enumerate}

\paragraph{Infection wave quantities:} For the infection wave, we consider the fraction of infection in female mosquitoes at a large final time, when an asymptotic propagating wave has formed:
$$
p^{**}(x,y) = \frac{F_w^{**}(x,y)}{F_u^{**}(x,y)+F_w^{**}(x,y)}.
$$
The solution is a smooth transition front connecting two stable ODE steady states (\cref{fig:QOIs_SA}B).
\begin{enumerate}
\item Wave velocity: This is the velocity of wave propagation (assuming expanding coverage, $c>0$, and radial symmetry):
$$
c(x_{inf})\approx \frac{-p_t^{**}(x_{inf},0)}{p_x^{**}(x_{inf},0
)},
$$
where $x_{inf}$ is the inflection point calculated as in the threshold width, $p_t$ is evaluated based on the right-hand side of the equations, and $p_x$ is approximated using numerical derivatives. We use the asymptotic wave velocity as a proxy and define wave velocity as $\text{QOI} = c(x_{inf})$.

\item Wave width: This measures how spread out the wavefront is, calculated as the horizontal distance between 99.5\% of the ODE endemic infection rate and 0.5\% infection (\cref{fig:QOIs_SA}B). The ODE endemic infection level is determined using \cref{eq:EE}, where $r_{wu}$ is the larger root of \cref{eq:EEr}. Further details on ODE steady states can be found in \cite{florez2022modeling}.
\end{enumerate}

\subsection{Parameter samplings}\label{sec:app_sample}
When implementing the global sensitivity analysis, we adjusted the parameter ranges so that all parameters of interest (POIs) are varied by a comparable relative amount. The updated parameter ranges are listed in \cref{tab:parameter_SA}.

To ensure biological relevancy, we enforced the following assumptions across all samples: a reduced reproduction rate for \Wns-infected females ($\phi_w<\phi_u$), increased death rates for infected females ($\mu_{fw}>\mu_{fu}$) and infected eggs ($\mu_{ew}>\mu_{eu}$). To incorporate these fitness costs, we introduce the independent fitness cost parameters (e.g. $c_\phi$) as the POIs for sampling. The dependent parameters are then back-calculated (e.g. $\phi_w = \phi_u(1-c_\phi)$).

Additionally, we maintain the model assumptions that \W infection does not impact male death rates ($\mu_{mw}=\mu_{mu}$) or mosquitoes dispersion ($D_1 = D_3$, $D_2 = D_4$). As a result, $\mu_{mu}$, $D_1$, and $D_3$ were treated as independent POIs.

\begin{table}[htbp]
\begin{widepage}
\caption{Model parameters involved in the sensitivity analysis (SA) and their baseline values. All the rates have dimension $day^{-1}$ and are estimated based on the wMel strain of \textit{Wolbachia} and \textit{Aedes aegypti} mosquitoes, see \cref{tab:parameter_all}. The ranges for SA have been adjusted, so that for a baseline value $X\in [0.75X, 1.25X]$, subject to biological constraints. $^\S$ indicates new fitness cost parameters for sampling purposes.}
\label{tab:parameter_SA}
\begin{tabular}{llllll}
\toprule
\multicolumn{2}{l}{Parameters} & Baseline & Range for SA & Reference & Notes\\ \midrule
$\phi_u$ &Per capita egg-laying rate for $F_{u}$ &3.7  & $2.775\sim 4.625$ & \cite{hoffmann2014stability,styer2007mortality,yang2011follow} & \\
$c_{\phi}$ & Fitness cost: reduction in egg-laying rates$^\S$ &  0.054 &  $0.041 \sim 0.068$ & \cite{hoffmann2014stability,styer2007mortality} & $\phi_w = \phi_u(1-c_\phi)$\\
$\delta$ &Per capita hatching rate for eggs & 0.5 & 0.375 $\sim$ 0.625 &\cite{soares-pinheiro2016eggs,foster2002mosquitoes} & \\
$\psi$ &Per capita development rate & 0.1 & 0.075$\sim$ 0.125 & \cite{foster2002mosquitoes,walker2011wmel} & \\
$\mu_{eu}$& Death rate for $E_u$ & 0.088 & 0.066 $\sim$ 0.11& \cite{hoffmann2014stability,soares-pinheiro2016eggs} & \\
$c_{\mu e}$& Fitness cost: increase in egg death rates$^\S$ & $1.102$ &  $0.827 \sim 1.378\%$ &\cite{hoffmann2014stability,soares-pinheiro2016eggs}  & $\mu_{ew} = \mu_{eu}(1+c_{\mu e})$\\
$\mu_{l}$& Death rate for $L_u$ and $L_w$ & 0.12 & 0.09 $\sim$ 0.15 & \cite{yang2011follow} &\\
$\mu_{fu}$ &Death rate for $F_u$ & 1/17.5 & 1/23.33 $\sim$ 1/14& \cite{qu2018modeling,zettel2009yellowa,nationalenvironmentagency2020aedes} & \\
$c_{\mu f}$ & Fitness cost: increase in female death rates$^\S$ & 0.108 & $0.081 \sim 0.135$ & \cite{walker2011wmel} & $\mu_{fw} = \mu_{fu}(1+c_{\mu f})$\\
$\mu_{mu}$ &Death rate for $M_{u}$ &1/10.5 & 1/14 $\sim$ 1/8.4 &\cite{styer2007mortality} & $\mu_{mw}=\mu_{mu}$ \\
$v_w$ & Maternal transmission rate & 1 &  0.89 $\sim$ 1 & \cite{walker2011wmel,hoffmann2014stability} & biological constraint \\
$D_1$ & Diffusion coefficients (m$^2$/day)& 200 & $150 \sim 250$&
\cite{kay1998aedes,russell2005markreleaserecapture,turelli2017deploying} & $ D_1 = D_2 = D_3 = D_4$\\
\bottomrule
\end{tabular}
\end{widepage}
\end{table}

\section{Additional plots for results}\label{sec:appendixD}
\begin{figure}[htbp]
\begin{widepage}
\centering
\insertfigclose{0.32}{pics/1_pic-2_1.png}{A}
\insertfigclose{0.32}{pics/1_pic-2_2.png}{B}
\insertfigclose{0.32}{pics/1_pic-2_3.png}{C}
\insertfigclose{0.31}{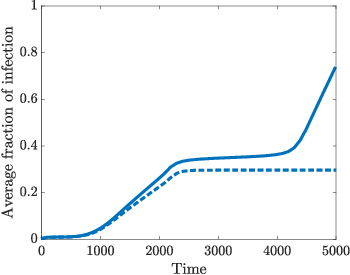}{D}
\insertfigclose{0.32}{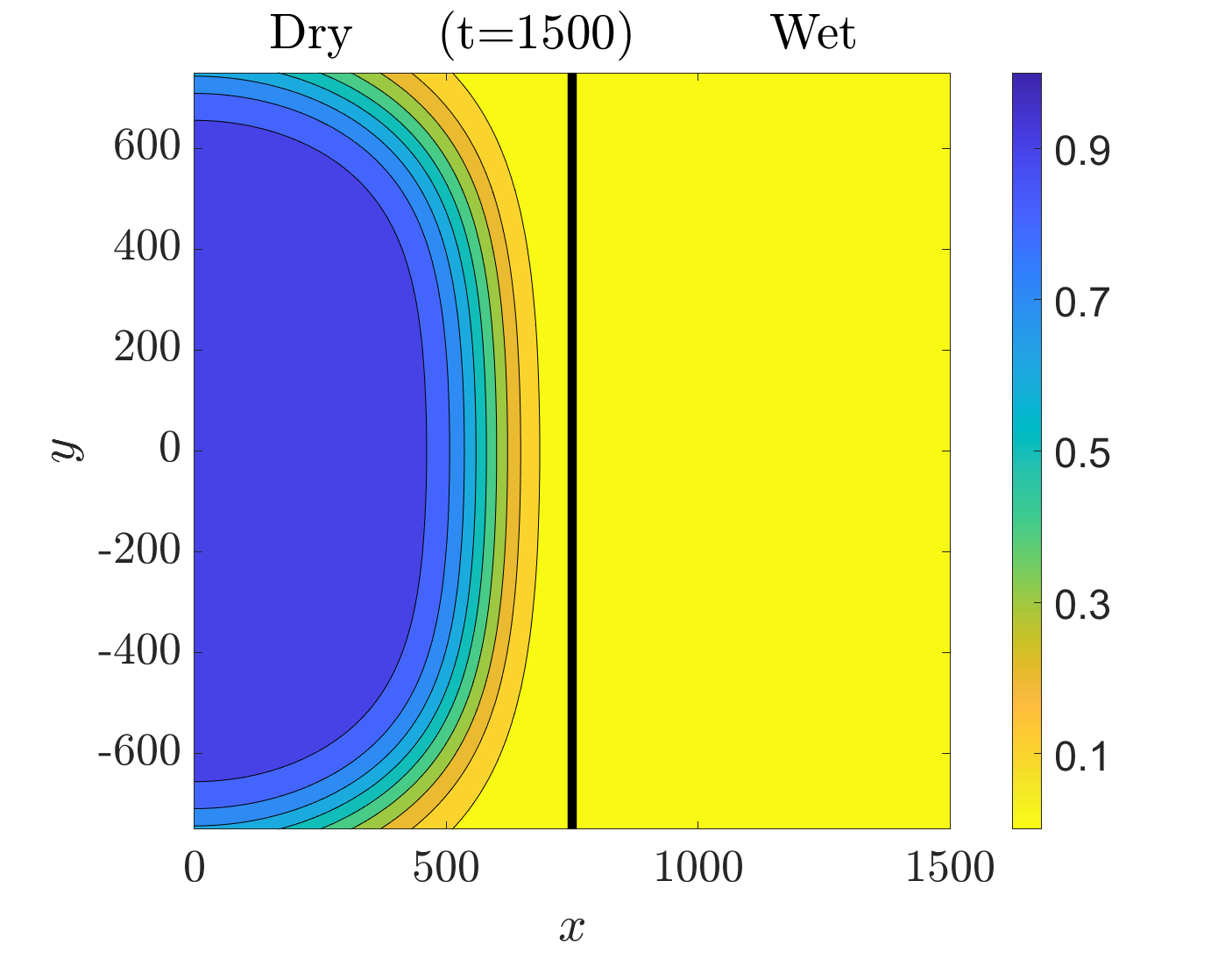}{E}
\insertfigclose{0.32}{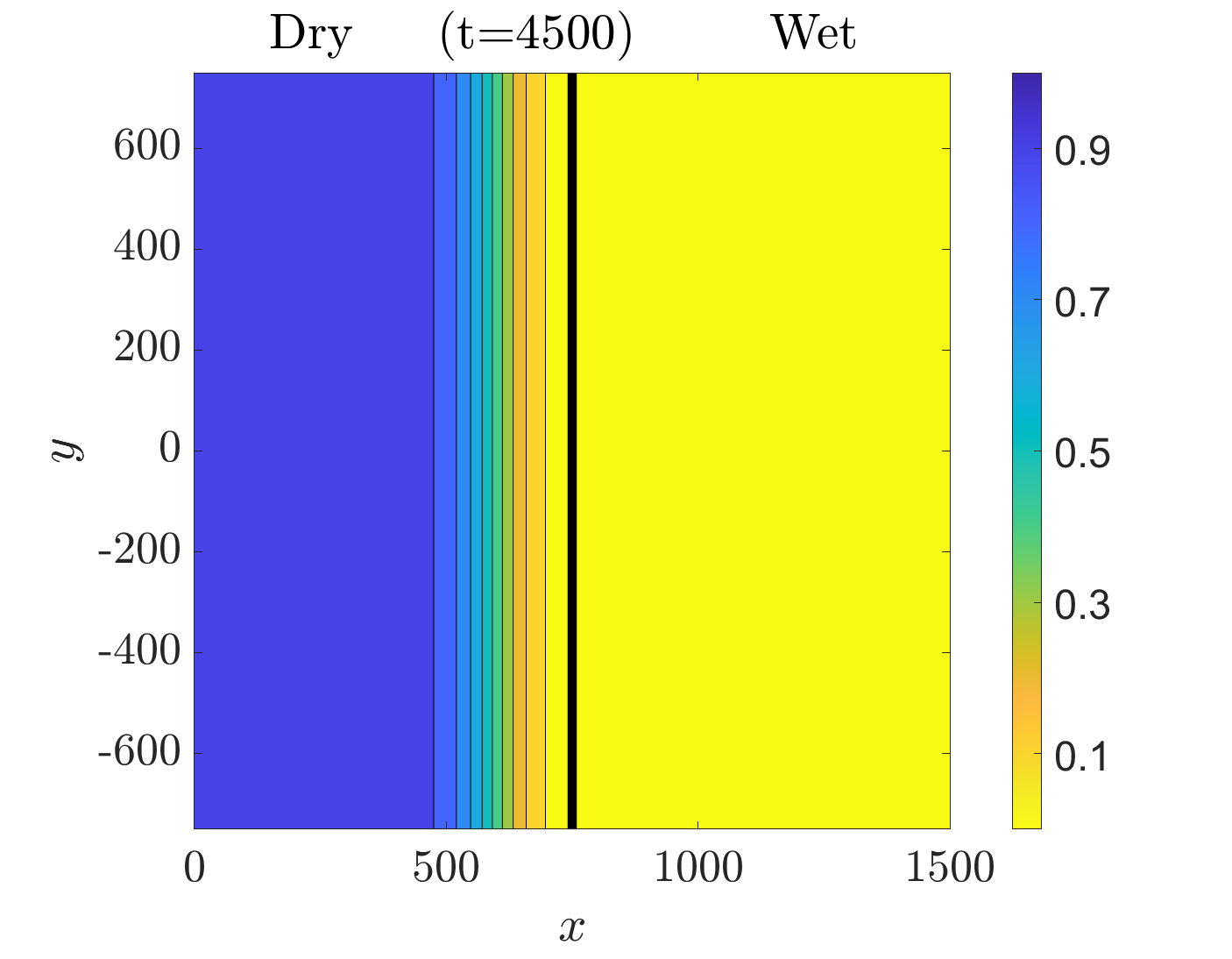}{F}
\caption{Release infection in the dry vs. wet regions. (A)--(C) Results for a dry-to-wet carrying capacity ratio of $1:2.5$. (E)--(F) Results for a dry-to-wet carrying capacity ratio of $1:5$. (D) The average fraction of infection across the entire domain for both scenarios: solid curve for the $1:2.5$ ratio and dashed curve for the $1:5$ ratio, where the infection wave stops near the dry-wet interface.}
\label{fig:wetdry_app}
\end{widepage}
\end{figure}

\begin{figure}[htbp]
\begin{widepage}
\centering
\insertfigclose{0.31}{pics/heatmap_Prerelease_Kl_EL_K3.eps}{A}
\insertfigclose{0.31}{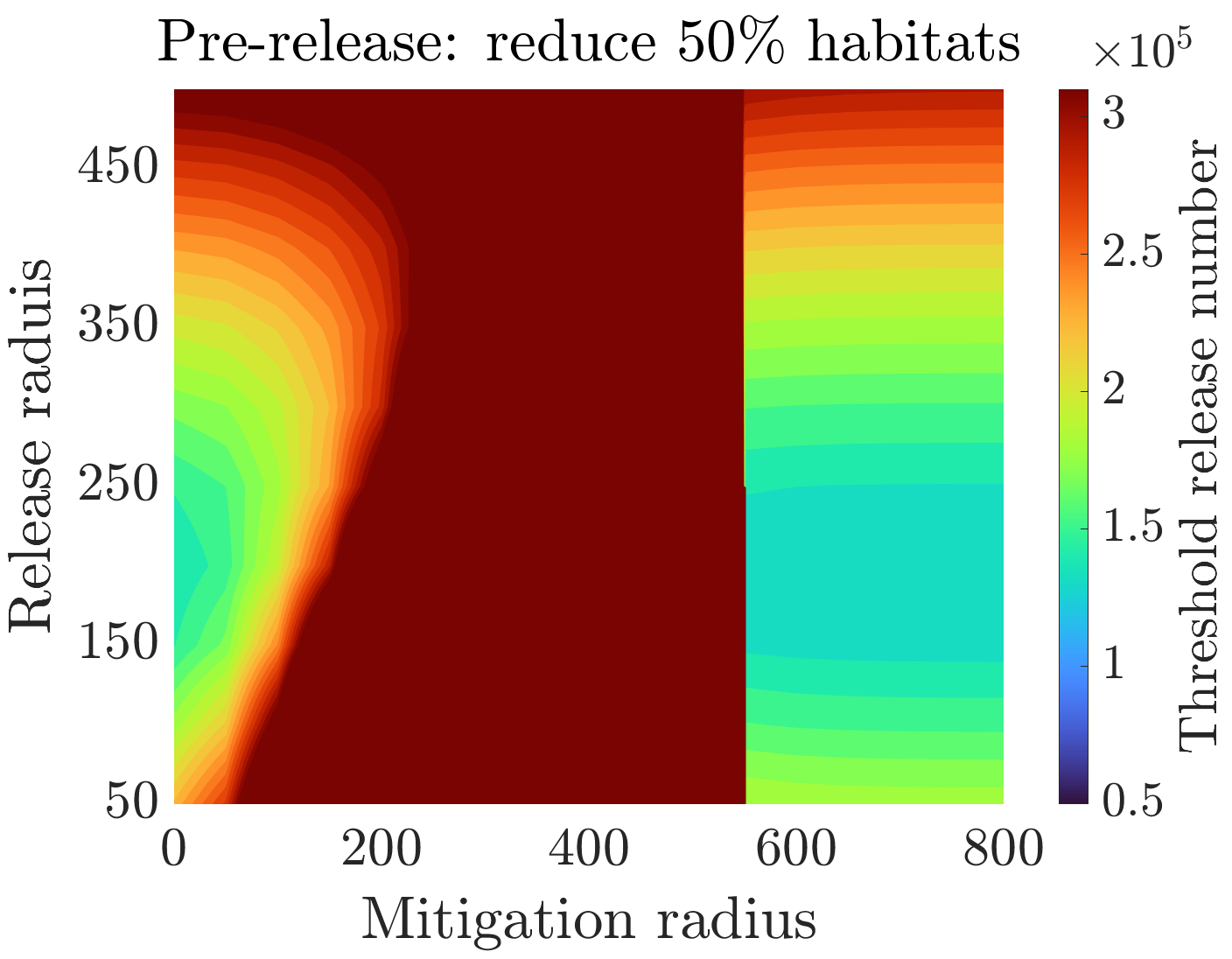}{B}
\insertfigclose{0.31}{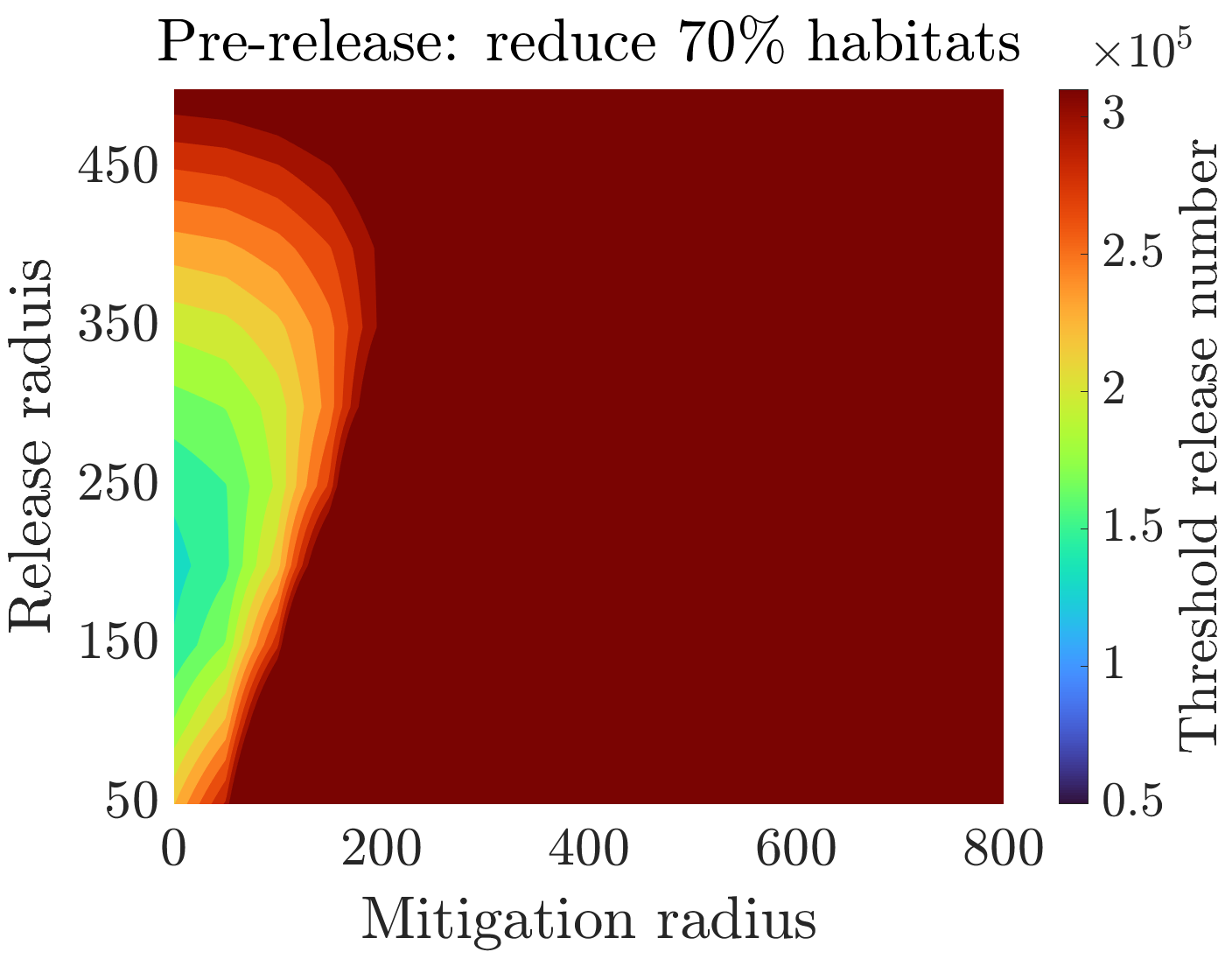}{C}
\caption{Pre-release mitigation through habitat modifications, removing (A) $30\%$ (same as \cref{fig:heatmap_mitigation}G), (B) $50\%$, and (C) $70\%$ of the breeding sites. Regions with a threshold release number exceeding $3\times 10^5$ are shown in dark red and are considered non-achievable. As the reduction fraction increases, the non-achievable regions expand.}
\label{fig:heatmap_app}
\end{widepage}
\end{figure}



\end{document}